\documentclass[twocolumn,twocolappendix]{aastex7}
\usepackage{graphicx}
\usepackage{amsmath}
\usepackage{scalerel}
\usepackage{apjfonts}
\usepackage{tikz}
\usetikzlibrary{svg.path}
\usepackage{comment}
\usepackage{array,multirow,color,tablefootnote}
\usepackage{xcolor}
\usepackage[version=4]{mhchem}
\usepackage{adjustbox}
\usepackage{placeins}
\definecolor{orcidlogocol}{HTML}{A6CE39}
\tikzset{
	orcidlogo/.pic={
		\fill[orcidlogocol] svg{M256,128c0,70.7-57.3,128-128,128C57.3,256,0,198.7,0,128C0,57.3,57.3,0,128,0C198.7,0,256,57.3,256,128z};
		\fill[white] svg{M86.3,186.2H70.9V79.1h15.4v48.4V186.2z}
		svg{M108.9,79.1h41.6c39.6,0,57,28.3,57,53.6c0,27.5-21.5,53.6-56.8,53.6h-41.8V79.1z M124.3,172.4h24.5c34.9,0,42.9-26.5,42.9-39.7c0-21.5-13.7-39.7-43.7-39.7h-23.7V172.4z}
		svg{M88.7,56.8c0,5.5-4.5,10.1-10.1,10.1c-5.6,0-10.1-4.6-10.1-10.1c0-5.6,4.5-10.1,10.1-10.1C84.2,46.7,88.7,51.3,88.7,56.8z};
	}
}

\newcommand\orcidicon[1]{\href{https://orcid.org/#1}{\mbox{\scalerel*{
				\begin{tikzpicture}[yscale=-1,transform shape]
					\pic{orcidlogo};
				\end{tikzpicture}
			}{|}}}}

\begin{document}

\title{Digging into the Massive Protostar S255IR NIRS3: A Study of Nitrogen-Bearing Molecules and Their Prebiotic Chemistry}
	
\correspondingauthor{Tapas Baug and Arijit Manna}
\author[orcid=0000-0001-9133-3465]{Arijit Manna}
\altaffiliation{Authors contributed equally to this work.}
\affiliation{S. N. Bose National Centre for Basic Sciences, Block-JD, Sector-III, Salt Lake, Kolkata 700106, India}
\email{amanna.astro@gmail.com}
	
\author[orcid=0000-0003-2325-8509]{Sabyasachi Pal}
\altaffiliation{Authors contributed equally to this work.}
\affiliation{Department of Physics and Astronomy, Midnapore City College, Kuturia, Bhadutala, Paschim Medinipur 721129, India}
\email{sabya.pal@gmail.com}
	
\author[orcid=0000-0003-0295-6586]{Tapas Baug}
\altaffiliation{Authors contributed equally to this work.}
\affiliation{S. N. Bose National Centre for Basic Sciences, Block-JD, Sector-III, Salt Lake, Kolkata 700106, India}
\email{tapasbaug@bose.res.in}
	
\author[orcid=0009-0003-6633-525X]{Ariful Hoque}
\affiliation{S. N. Bose National Centre for Basic Sciences, Block-JD, Sector-III, Salt Lake, Kolkata 700106, India}
\email{ariful.hoque@bose.res.in}
	
\author[orcid=0000-0003-0818-7474]{Sandip Dutta}
\affiliation{Department of Applied Mathematics, Dinabandhu Andrews Institute of Technology and Management, Kolkata, West Bengal 700094, India}
\email{duttasandip.mathematics@gmail.com}
	
\author[orcid=0009-0005-2662-0245]{Sekhar Sinha}
\affiliation{Department of Physics, Sidho Kanho Birsha University, Ranchi Road, Purulia 723104, India}
\email{secr.sina3@gmail.com}
	
\author[orcid=0000-0003-4353-8487]{Sushanta Kumar Mondal}
\affiliation{Department of Physics, Sidho Kanho Birsha University, Ranchi Road, Purulia 723104, India}
\email{skm.phy@skbu.ac.in}
	
\begin{abstract}
The study of complex nitrogen (N)-bearing molecules is essential for probing the physical and chemical evolution of star-forming regions. In this paper, we present the identification of rotational emission lines from several complex N-bearing species such as methyl cyanide (\ce{CH3CN}), ethyl cyanide (\ce{C2H5CN}), vinyl cyanide (\ce{C2H3CN}), cyanamide (\ce{NH2CN}), and formamide (\ce{NH2CHO}) toward the high-mass protostar S255IR NIRS3 using ALMA band 4 observations. In addition, the vibrationally excited transitions of cyanoacetylene (\ce{HC3N}, $\nu_{7}$ = 2) were detected. The column densities and excitation temperatures of these molecules were derived through LTE spectral modelling, yielding excitation temperatures in the range of 175--220 K. The high excitation temperatures (175--220 K) indicate that the identified N-bearing molecules arise from the warm inner regions ($T \geq 100$ K) of the source. The fractional abundances were further estimated relative to \ce{H2}, \ce{CH3OH}, and \ce{CH3CN}. A Pearson correlation heat map of the abundances reveals a strong positive correlation ($r > 0.7$) among three molecules in the cyanide family, such as \ce{CH3CN}, \ce{C2H3CN}, and \ce{C2H5CN}, suggesting that these N-bearing molecules may be chemically linked. Comparison with three-phase warm-up chemical models shows that the observed abundances of \ce{CH3CN}, \ce{C2H5CN}, \ce{C2H3CN}, \ce{NH2CN}, \ce{NH2CHO}, and \ce{HC3N} ($\nu_{7}$ = 2) relative to \ce{H2} are consistent with model predictions within factors of 1.04, 0.67, 1.28, 0.76, 0.72, and 0.96, respectively. Finally, we discuss the potential formation pathways of the identified N-bearing molecules in the context of gas-grain chemistry within S255IR NIRS3.
\end{abstract}
	
\keywords{\uat{Astrochemistry}{75}; \uat{Interstellar molecules}{849}; \uat{Submillimeter astronomy}{1647}; \uat{Prebiotic astrochemistry}{2079}; \uat{Star formation}{1569}; \uat{Chemical abundances}{224}}
	
\section{Introduction}
\label{sec:intro} 
In the interstellar medium (ISM) or circumstellar shells, more than 340 molecular species were detected at (sub)millimeter wavelengths\footnote{\url{https://cdms.astro.uni-koeln.de/classic/molecules}}. Among the 340 molecular species, approximately 96 molecules have at least one nitrogen (N) atom, 12 molecules contain isocyanide (--NC), and 33 molecules have a functional group of cyanide (--CN). CN was the first detected N-bearing molecule observed in 1940 towards different stars \citep{mc40}. CN plays an important role in the formation of cyanopolynes (HC$_{2n+1}$N), hydrocarbons (C$_{n}$H$_{2n+2}$), and several prebiotic molecules \citep{her09, tan19, jo20, tan22}. The study of N-bearing molecules with CN bonds is crucial because CN bonds form between the \ce{-NH2} and --COOH groups, creating different types of amino acids \citep{gol10}. Previous studies show that N-bearing molecules are essential for the production of different types of complex biomolecules \citep{bal09}. Earlier, a wide variety of complex N-bearing and N-ion-bearing molecules have been detected across diverse Galactic environments, including high- and low-mass star-forming regions, comets, protoplanetary disks, and Saturn’s moon Titan \citep{jon77, ir84, kaw92, zi99, cor15, ob15, man24a, zi86, kaw94, th08, cer08, cer13, mar18, ag15, ag10}.
	
In the ISM, complex organic molecules form efficiently on dust grain surfaces, which act as catalytic sites for chemical reactions \citep{gar06}.  Nitriles heavier than HCN are classified as interstellar complex organic molecules (iCOMs), defined as species containing at least six atoms \citep{her09, ce17}.  Their formation and detectability are closely linked to the rise in temperature and density during star formation, particularly in high-mass environments. The formation of high-mass stars (\(M \geq 8 M_{\odot}\)) within dense clusters provides a key framework for chemical enrichment, offering insight into the astrochemical processes that may have preceded the formation of our Solar System \citep{car00, la03, riv13, ad10}. Based on observational results, the earliest phase can be characterized by a high-mass starless core (HMSC), which is dense ($n \geq 10^{5}$ cm$^{-3}$), cold ($T \sim 15-20$ K), and massive molecular condensation, but there is no evidence of star formation activity due to the gravitational instability \citep{tan13}. The subsequent evolutionary phase, often associated with high-mass protostellar objects (HMPOs), is thought to mark the onset of protostar formation within hot molecular cores (HMCs), characterized by temperatures $T \geq 100$ K and gas densities $n \geq 10^{7}$ cm$^{-3}$ \citep{kur00, fon07, wil14}. High-mass protostellar cores have been identified observationally \citep[see][and references therein]{olg23} and serve as important test cases for high-mass star formation (HMSF) theories. However, the existence and lifetime of this evolutionary phase are not yet firmly established and remain subjects of active investigation. At $T \geq 100$ K, molecules formed within icy grain mantles are released into the gas phase via thermal desorption or sputtering by shocks, UV photons, and cosmic rays \citep{gar06, gar13}. Once in the gas phase, these species undergo further reactions, producing more complex molecules such as \ce{CH3OH}, \ce{CH3OCHO}, \ce{CH3OCH3}, and \ce{C2H5OH} \citep{her09}. Several chemical models suggest that the lifetime of HMCs is on the order of $\sim10^{5}$ yr for a medium warm-up phase and up to $\sim10^{6}$ yr for a slow warm-up phase \citep{gar06, gar13}. In the warm-up phase, complex N-bearing iCOMs, including methyl cyanide (\ce{CH3CN}), vinyl cyanide (\ce{C2H3CN}), ethyl cyanide (\ce{C2H5CN}), and cyanoacetylene (\ce{HC3N}), become prominent spectral features \citep{gar13}. Using three-phase chemical modelling, \citet{gar17} demonstrated that \ce{HC3N} acts as a key precursor of both \ce{C2H3CN} and \ce{C2H5CN} on grain surfaces in hot cores. These three species are consistently detected in hot cores, hot corinos, and even in the atmosphere of Saturn’s moon Titan (see Table~\ref{tab:molecular_sources}). Their detection in both Galactic star-forming regions and Titan’s atmosphere suggests that these molecules may trace chemical pathways linking interstellar material to planetary atmospheres, highlighting their relevance across different stages of star and planet formation \citep{pal17, I20}. Similarly, in the ISM, amide-bearing molecules such as cyanamide (\ce{NH2CN}) and formamide (\ce{NH2CHO}) are recognized as important prebiotic species and have been identified in several Galactic sources (see Table~\ref{tab:molecular_sources}). Using gas-grain chemical networks, \citet{man24b} proposed that \ce{NH2CN} may act as a potential precursor of the simplest amino acid, glycine (\ce{NH2CH2COOH}), through both grain-surface and gas-phase pathways, highlighting its role in linking interstellar chemistry with molecular inventories relevant to the origins of life.

\begin{figure}
\centering
\includegraphics[width=0.48\textwidth]{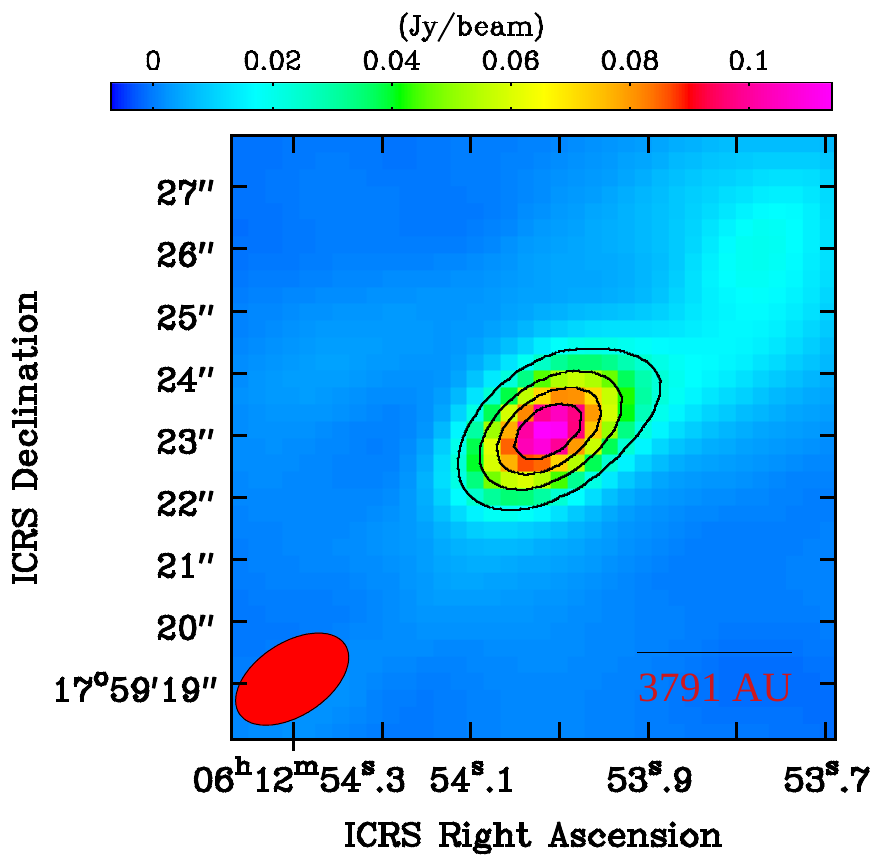}
\caption{Dust continuum emission image of NIRS3 at 142.49 GHz ($\lambda$ = 2.10 mm). The red circle represents the synthesized beam of the continuum emission image, whose size is 2.13$^{\prime\prime}$$\times$1.18$^{\prime\prime}$. The RMS of the continuum emission image is 2.35 mJy.}
\label{fig:continuum}
\end{figure}
	
\begin{table*}[t]
\centering
\caption{Dust continuum emission properties of NIRS3.}
\label{tab:continuum}
\begin{tabular}{cccccccc}
\hline
Frequency&Integrated flux density&Peak intensity&Synthesized beam size& Position angle&RMS&Source size\\
(GHz) & (mJy) & (mJy beam$^{-1}$) & ($^{\prime\prime}$) & ($^{\circ}$) & (mJy) & ($^{\prime\prime}$) \\
\hline
128.82 & 105.7$\pm$5.3 & 92.5$\pm$2.7 & 2.30$\times$1.28 & 124$\pm$28 & 2.72 & 2.44$\times$1.35 \\
130.56 & 108.8$\pm$5.9 & 92.3$\pm$2.9 & 2.29$\times$1.28 & 125$\pm$23 & 2.91 & 2.46$\times$1.39 \\
141.07 & 125.1$\pm$6.1 & 104.3$\pm$3.0 & 2.07$\times$1.19 & 122$\pm$12 & 2.67 & 2.28$\times$1.30 \\
142.49 & 139.7$\pm$5.1 & 116.6$\pm$5.6 & 2.13$\times$1.18 & 128$\pm$27 & 2.35 & 2.27$\times$1.29 \\
\hline
\end{tabular}
\end{table*}
	
The goal of this study is to conduct a comprehensive investigation of N-bearing molecules, focusing on their spatial distributions and prebiotic formation chemistry toward the high-mass young stellar object (HMYSO) S255IR NIRS3 (hereafter NIRS3) using ALMA band 4 observations. This source is also referred to as S255IR-SMA1 or G192.600–0.048. We selected NIRS3 for this study because it hosts a chemically rich HMC \citep{wan11, zin15}. Previous studies by \citet{zin15} and \citet{liu20} identified rotational emission lines of numerous oxygen (O)- and sulphur (S)-bearing molecules toward NIRS3, but placed little emphasis on complex N-bearing species. Therefore, NIRS3 is an excellent target for detailed studies of both the physical properties and chemical evolution of N-bearing molecules.  NIRS3 is situated in the S255IR star-formation region at a distance of 1.78 kpc \citep{bur16}. The mass and bolometric luminosity of NIRS3 are $\sim$20 \(\textup{M}_\odot\) and $\sim$2.4$\times$10$^{4}$ \(\textup{L}_\odot\), respectively \citep{wan11, zin15}. Previous observations have shown that the molecular gas content in this source is approximately 300--400 \(\textup{M}_\odot\) \citep{wan11, zin15}. The maser emissions of water (\ce{H2O}) and methanol (\ce{CH3OH}) were detected towards NIRS3 \citep{hir21, be23}. The molecular outflows of carbon monoxide (CO), silicon monoxide (SiO), and carbon monosulfide (CS) were also found in NIRS3 \citep{zin15}. 
	
This paper is organized as follows: The observation and data analysis are discussed in Section~\ref{sec:obs}. The continuum emission and molecular line identification are shown in Section~\ref{sec:result}. The discussion and conclusion are shown in Sections~\ref{sec:dis} and \ref{sec:con}.
	
\begin{figure*}
\centering
\includegraphics[width=0.9\textwidth]{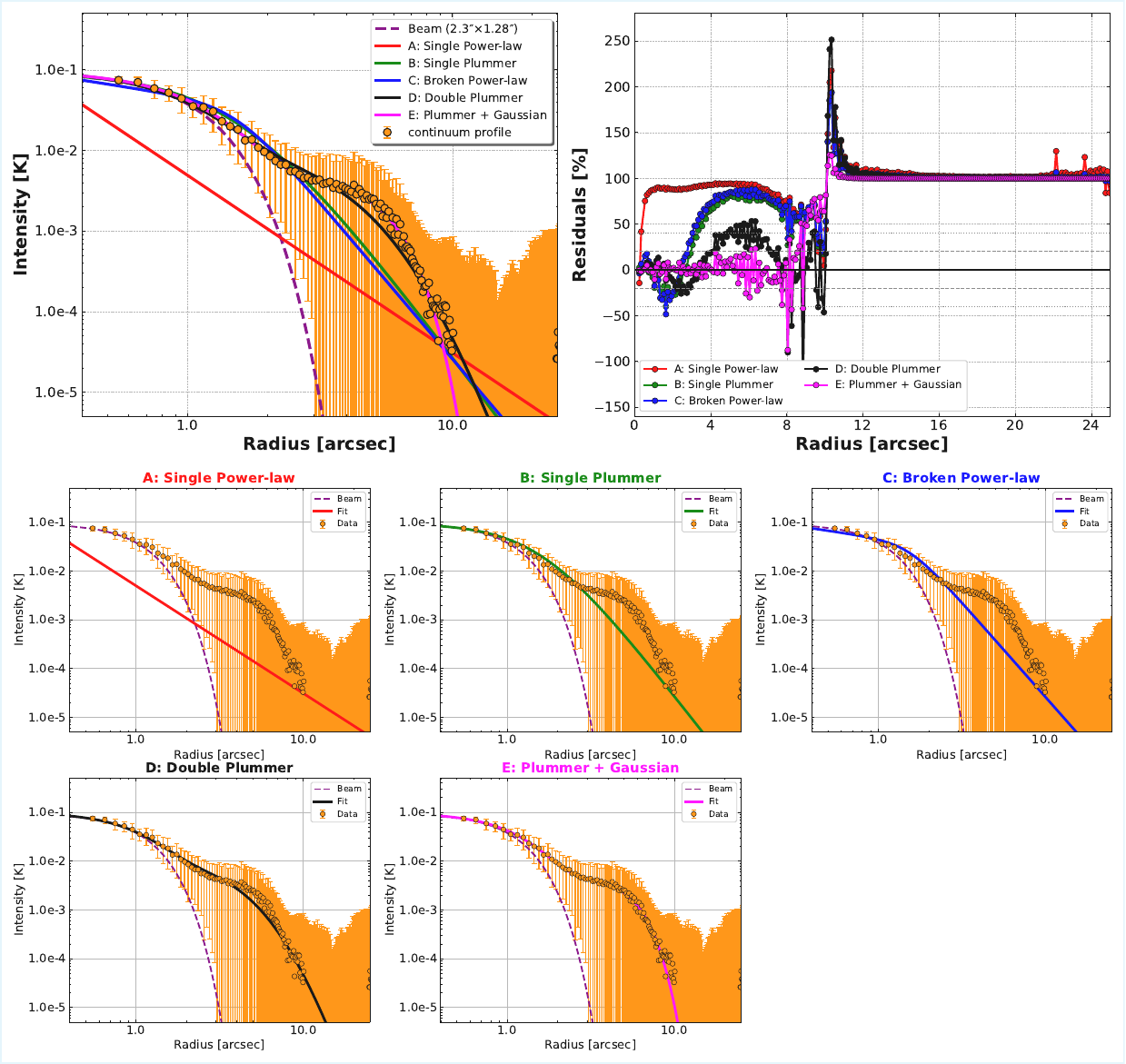}
\caption{\emph{Left panel:} Radial intensity profile of NIRS3 at 2.32 mm (orange circles with error bars showing the rms azimuthal scatter). The dashed purple curve shows the beam profile ($2.30^{\prime\prime}\times1.28^{\prime\prime}$). Colored lines represent the best-fit models: single power-law (red), single Plummer (green), broken power-law (blue), double Plummer (black), and Plummer + Gaussian (magenta). \emph{Right panel:} Residuals (data $-$ model) expressed as a percentage of the data for each model. \emph{Lower panel:} Individual fits of all five models.}
\label{fig:profile_fit}
\end{figure*}
	
\section{Observations and data reductions}
\label{sec:obs}
We used the archival data of NIRS3, observed using a high-resolution Atacama Large Millimeter/Submillimeter Array (ALMA) (PI: Cesaroni, Riccardo, ID: 2016.A.00008.T) with the 12 m array. This observation was carried out on December 14, 2016 (MJD 57736) as part of an ALMA Director's Discretionary Time (DDT) program aimed at monitoring the free–free continuum emission associated with the accretion burst in NIRS3. The primary objective of these observations was to investigate the connection between accretion and jet ejection processes. The on-source integration time was 5.04 minutes, which is relatively short, and the observations were not originally optimized for a detailed astrochemical study. However, the available spectral coverage enables the detection of several molecular transitions, which we exploit here to investigate the chemical properties of the source. The phase centre of NIRS3 was ($\alpha,\delta$)$_{\rm J2000}$ = 06:12:54.020, +17:59:23.100. For this observation, J0510+1800 was used as both a bandpass and flux calibrator. In addition, J0613+1708 was used as the phase calibrator. During the observation, 45 antennas were used with a minimum baseline of 25 m and a maximum baseline of 400 m. Observations were performed in the frequency ranges of 127.61--129.49 GHz, 129.56--131.44 GHz, 139.67--141.54 GHz, and 141.56--143.43 GHz, with spectral and velocity resolutions of 976.56 kHz and 2.06 km s$^{-1}$, respectively.
	
We used the Common Astronomy Software Application ({CASA 5.4.1}) with the ALMA data analysis pipeline for calibration and imaging of the data \citep{mc07}. We applied the Perley-Butler 2017 flux calibration model for each baseline for flux calibration using the SETJY task \citep{per17}. We also used the pipeline tasks HIFA\_bandpassflag and HIFA\_flagdata for flagging bad antenna data and channels, which were performed after the flux and bandpass calibration. After preliminary data reduction, we separated the target data (NIRS3) using the MSTRANSFORM task with all the available rest frequencies. We also used the UVCONTSUB task to subtract continuum emission from the UV plane of the calibrated data. We created the continuum emission map of NIRS3 using the TCLEAN task with a HOGBOM deconvolver for line-free channels.  We also generated spectral data cubes using the TCLEAN task with the cube spectral definition mode (SPECMODE) with a Briggs weighting of 0.5. Several rounds of self-calibration were applied using the GAINCAL and APPLYCAL tasks to reduce the RMS. We applied the task IMPBCOR to correct the primary beam pattern in the continuum images and spectral data cubes. 
	
\begin{figure*}
\centering
\includegraphics[width=1.0\textwidth]{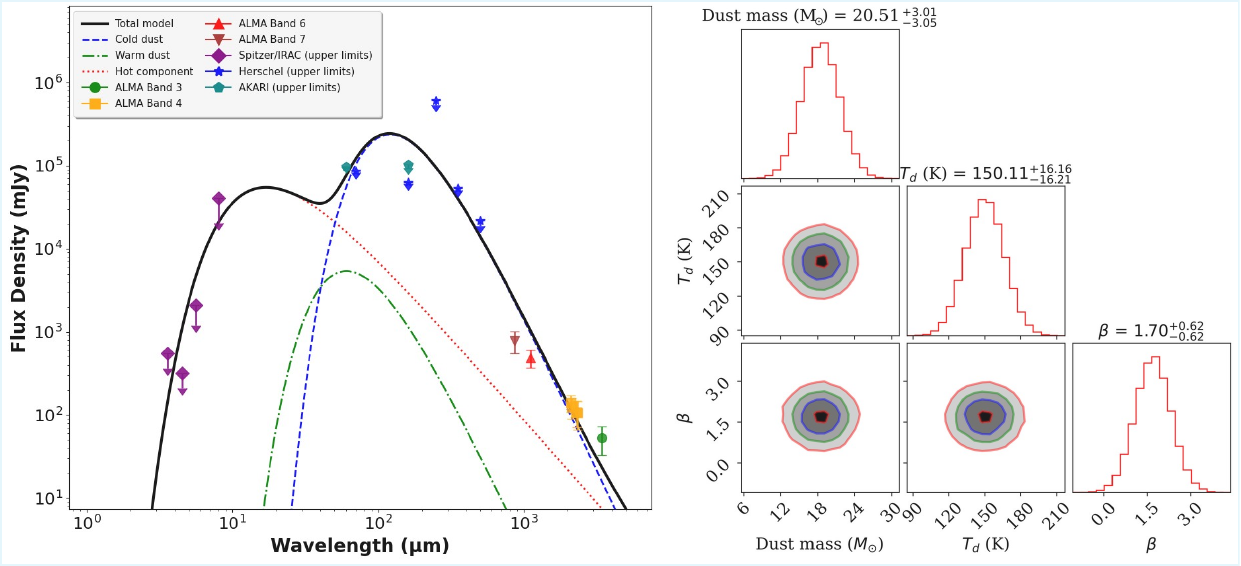}%\includegraphics[width=0.5\textwidth]{corner_sed.pdf}
\caption{SED of NIRS3 covering the wavelength range from 3.6 $\mu$m to 3443 $\mu$m. The black curve represents the best-fit SED obtained from the radiative transfer model of \cite{rob17}. The right panel shows a corner diagram demonstrating the results of our MCMC parameter estimation for the SED model. The 1-D histograms on the diagonal show the marginalized posterior densities of mass, dust temperature, and $\beta$. The off-diagonal panels indicate 2-D projections of the posterior probability distributions for each pair of parameters.}
\label{fig:sed}
\end{figure*}
	
\section{Result}
\label{sec:result}
\subsection{Dust continuum emission towards NIRS3}
We generated the dust continuum images of NIRS3 for line-free channels at frequencies centering on 128.82 GHz ($\lambda$ = 2.32 mm), 130.56 GHz ($\lambda$ = 2.29 mm), 141.07 GHz ($\lambda$ = 2.12 mm), and 142.49 GHz ($\lambda$ = 2.10 mm), respectively. The dust continuum image of NIRS3 at a wavelength of 2.10 mm is shown in Figure~\ref{fig:continuum}. To estimate the physical parameters of the continuum emission images, we fitted a 2-D Gaussian using the IMFIT task over the main core. The physical properties of the NIRS3 core are shown in Table~\ref{tab:continuum}. We found that the deconvolved source sizes are larger than the corresponding synthesized beam sizes by a factor of $\sim$1.05--1.10 along both the major and minor axis across all four frequencies, indicating that the emission is marginally resolved over the observable frequency ranges of band 4.
	
\subsubsection{Radial intensity profile}
The radial intensity profile was constructed by azimuthally averaging the 2.32 mm continuum emission image in concentric annuli binned at 0.1$^{\prime\prime}$ centered on the emission peak. To quantitatively characterize the radial intensity profile of NIRS3, we fitted five different analytical models to the radial intensity distribution data. The models include (A) a single power-law, (B) a single Plummer-like profile \citep{pl11}, (C) a broken power-law, (D) a double Plummer profile, and (E) a Plummer profile with an additional Gaussian component (Plummer + Gaussian). The mathematical formulations of these models are as follows.\\\\
\underline{\textbf{Model A: Single Power-law:}}
\begin{equation}
I(r) = I_0 \left(\frac{r}{r_0}\right)^{-\alpha}
\end{equation}
where $I_0$ is the intensity at the reference radius $r_0$, and $\alpha$ is the power-law index that determines the steepness of the profile. The variable $r$ denotes the projected angular radius from the emission peak position. Here, $r$ define as $r = \sqrt{(x-x_0)^2 + (y-y_0)^2}$, where $x_0$ and $y_0$ is the peak position of the continuum emission.\\\\
\underline{\textbf{Model B: Single Plummer-like Profile:}}
\begin{equation}
I(r) = \frac{I_0}{\left[1 + (r/r_{\text{flat}})^2\right]^{\alpha/2}}
\end{equation}
where $I_0$ is the central intensity, $r_{\mathrm{flat}}$ is the characteristic radius of the flattened core, and $\alpha$ is the spectral index of the asymptotic slope at large radius, with $I(r) \propto r^{-\alpha}$ for $r \gg r_{\mathrm{flat}}$.\\\\
\underline{\textbf{Model C: Broken Power-law:}}
\begin{equation}
I(r) = I_0 \left(\frac{r}{r_{\text{break}}}\right)^{-\alpha_1} \left[1 + \left(\frac{r}{r_{\text{break}}}\right)^s\right]^{(\alpha_1-\alpha_2)/s}
\end{equation}
where $I_0$ is the intensity at the break radius $r_{\mathrm{break}}$, $\alpha_1$ and $\alpha_2$ are the power-law indices inside and outside the break, respectively, and $s$ is the smoothness of the transition. This formulation provides a smooth connection between the inner and outer power-law regimes. The parameter `$s$' controls the sharpness or smoothness of the transition between the inner power-law (with index $\alpha_1$) and the outer power-law (with index $\alpha_2$). \\\\
\underline{\textbf{Double Plummer Profile:}}
\begin{equation}
I(r) = \frac{I_{0,1}}{\left[1 + (r/r_1)^2\right]^{\alpha_1/2}} + \frac{I_{0,2}}{\left[1 + (r/r_2)^2\right]^{\alpha_2/2}}
\end{equation}
where $r$ is the projected angular radius from the continuum peak position. The parameters $I_{0,1}$ and $I_{0,2}$ represent the central intensities of the two Plummer components, while $r_1$ and $r_2$ are their respective characteristic (scale) radius that define the extent of the flat inner regions. The parameters $\alpha_1$ and $\alpha_2$ describe the steepness of the outer radial decline for each component. This formulation allows the radial profile to be modeled as the superposition of two structurally distinct components, typically interpreted as a compact inner structure (potentially a disk) and a more extended envelope.\\\\
		
\begin{table*}
\centering
\scriptsize
\caption{Best-fit different model parameters (with $1\sigma$ uncertainties) and model comparison statistics for the radial intensity profile of NIRS3.}
\label{tab:fit_results}
\begin{adjustbox}{width=0.98\textwidth}
\begin{tabular}{l c c c c}
\hline
Model & \multicolumn{1}{c}{Parameters} & Reduced $\chi^2$ & AIC & $\Delta\mathrm{AIC}^{a}$ \\
\hline
					
A: Single Power-law & $I_0 = (3.30 \pm 3.0)\times10^{-2}$\,K, $r_0 = 2.22\arcsec \pm 1.75\arcsec$, $\alpha = 2.20 \pm 0.06$ &20.23 & 1010.4 & 129.2 \\
					
B: Single Plummer & $I_0 = (9.42 \pm 0.19)\times10^{-2}$\,K, $r_{\mathrm{flat}} = 2.58\arcsec \pm 0.07\arcsec$, $\alpha = 4.41 \pm 0.21$ &14.82 & 906.2 & 25.0 \\
					
C: Broken Power-law & $I_0 = (3.77 \pm 0.28)\times10^{-2}$\,K, $r_{\mathrm{break}} = 2.53\arcsec \pm 0.06\arcsec$, $\alpha_1 = 2.10 \pm 0.37$, $\alpha_2 = 3.88 \pm 0.17$ & 15.18 & 915.1 & 33.9 \\
					
D: Double Plummer &$I_{0,1} = (8.95 \pm 0.43)\times10^{-2}$\,K, $r_1 = 2.31\arcsec \pm 0.80\arcsec$, $\alpha_1 = 2.58 \pm 0.72$, $I_{0,2} = (2.04 \pm 0.29)\times10^{-2}$\,K, $r_2 = 7.15\arcsec \pm 2.15\arcsec$, $\alpha_2 = 3.26 \pm 0.35$ &14.02 & 890.6 & 9.3 \\
					
E: Plummer + Gaussian & $I_{0,\mathrm{p}} = (9.57 \pm 0.27)\times10^{-2}$\,K, $r_{\mathrm{flat}} = 2.38\arcsec \pm 0.67\arcsec$, $\alpha = 2.86 \pm 0.52$, $I_{0,\mathrm{g}} = (3.71 \pm 0.65)\times10^{-3}$\,K, $r_{\mathrm{g}} = 2.90\arcsec \pm 0.56\arcsec$, $\sigma_{\mathrm{g}} = 2.08\arcsec \pm 0.20\arcsec$ &13.64 & 881.2 & 0.0 \\
\hline
\end{tabular}
\end{adjustbox}\\
\medskip
\footnotesize
$^{a}$ $\Delta\mathrm{AIC}$ is the difference in AIC relative to the best-fitting model (Model E).
\end{table*}
		
\underline{\textbf{Model E: Plummer + Gaussian:}}
\begin{equation}
I(r) = \frac{I_{0,p}}{\left[1 + (r/r_{\text{flat}})^2\right]^{\alpha/2}} + I_{0,g} \exp\left[-\frac{(r-r_g)^2}{2\sigma_g^2}\right]
\end{equation}
The parameters $I_{0,p}$, $r_{\text{flat}}$, and $\alpha$ denote the central intensity, characteristic (flattening) radius, and outer power-law index of the Plummer component, respectively. The Gaussian component is characterized by its peak intensity $I_{0,g}$, central radius $r_g$, and width $\sigma_g$, which defines the radial extent of the Gaussian feature. This composite model describes a centrally concentrated Plummer-like structure with an additional radially localized Gaussian component, which may represent a ring-like feature, spiral arm, or an enhanced emission region within the circumstellar environment (see \cite{ca23} for details regarding this model).\\\\
\underline{\textbf{Fitting results:}} The fits were performed on the azimuthally averaged radial profile binned at $0\farcs 1$ intervals, with the beam profile of $2.30^{\prime\prime} \times 1.28^{\prime\prime}$ ($\sim$ 3791 AU). To compare the relative performance of the different models quantitatively, we use the Akaike information criterion (AIC) \citep{al74}. The AIC balances the goodness of fit against model complexity and is defined as:
		
\begin{equation}
\mathrm{AIC} = n \ln\left(\frac{\chi^2}{n}\right) + 2k
\label{eq:AIC}
\end{equation}
where $n$ is the number of data points, $k$ is the number of free parameters, and $\chi^2$ is the sum of squared residuals weighted by the uncertainties. Lower AIC values indicate a better model. The $\Delta\text{AIC}$ for a given model is calculated as the difference between its AIC and the minimum AIC among all models:
\begin{equation}
\Delta\text{AIC}_i = \text{AIC}_i - \text{AIC}_{\text{min}}
\label{eq:deltaAIC}
\end{equation}
Following \citet{bu04}, models with $\Delta\mathrm{AIC} < 2$ are considered to have substantial support, those with $2 < \Delta\mathrm{AIC} < 4$ have strong support, those with $4 < \Delta\mathrm{AIC} < 7$ have considerably less support, and models with $\Delta\mathrm{AIC} > 10$ have essentially no support. The radial intensity profiles and corresponding best-fit models are shown in Figure~\ref{fig:profile_fit}, with the derived fitting parameters summarized in Table~\ref{tab:fit_results}. The individual fits for all five models are displayed in the lower panels of Figure~\ref{fig:profile_fit}.
		
The single power-law fit (Model A) gives $\alpha = 2.20 \pm 0.06$, but this model provides the poorest fit (reduced $\chi^2 = 20.23$) and is strongly disfavored by the model selection criteria, confirming that a single power-law is insufficient to describe the complex structure. The large uncertainties on $I_0$ and $r_0$ further indicate that this model is poorly constrained by the data.
		
The single Plummer model (Model B) yields a characteristic radius $r_{\mathrm{flat}} = 2.58^{\prime\prime}\pm0.07^{\prime\prime}$ ($\sim$3810 AU) and an asymptotic power-law index $\alpha = 4.41 \pm 0.21$ in the outer parts. This represents a significant improvement over the single power-law model, with reduced $\chi^2 = 14.82$.
		
The broken power-law model (Model C) provides a good representation of the data with a clear transition between inner and outer regimes. We find a break radius of $r_{\mathrm{break}} = 2.53^{\prime\prime}\pm0.06^{\prime\prime}$ ($\sim$3800 AU). Inside this radius, the profile is relatively shallow with a power-law index $\alpha_1 = 2.10 \pm 0.37$, while outside it steepens to $\alpha_2 = 3.88 \pm 0.17$. This steepening is characteristic of a density profile that transitions from a flattened core to a more extended envelope where the density falls off more steeply.
		
The double-component models reveal additional structural complexity. The double Plummer fit (Model D) suggests two distinct components with scale radius of $r_1 = 2.31^{\prime\prime}\pm0.80^{\prime\prime}$ ($\sim$3795 AU) and $r_2 = 7.15^{\prime\prime}\pm2.12^{\prime\prime}$ ($\sim$12,730 AU). The inner component has a steep profile ($\alpha_1 = 2.58 \pm 0.72$) while the outer component also shows a steep decline ($\alpha_2 = 3.26 \pm 0.35$). This may trace a compact disk-like structure embedded within a more extended envelope.
		
The Plummer + Gaussian model (Model E) provides the best fit. The best-fit parameters for Model E are $I_{0,\mathrm{p}} = (9.57 \pm 0.27)\times10^{-2}$ K, $r_{\mathrm{flat}} = 2.38^{\prime\prime}\pm0.67^{\prime\prime}$ ($3800$ AU), and $\alpha = 2.86 \pm 0.52$ for the Plummer component, and $I_{0,\mathrm{g}} = (3.71 \pm 0.65)\times10^{-3}$ K, $r_{\mathrm{g}} = 2.90^{\prime\prime}\pm0.56^{\prime\prime}$ ($\sim$5160 AU), and $\sigma_{\mathrm{g}} = 2.08^{\prime\prime} \pm 0.20^{\prime\prime}$ ($\sim$ 3650 AU) for the Gaussian component. Additionally, Model E yields an AIC value of 881.2. Model E is statistically preferred over the other four models. It is favored by $\Delta\mathrm{AIC} = 9.3$ relative to the next best model (Double Plummer), while the remaining models show $\Delta\mathrm{AIC} > 10$, indicating little or no empirical support. The residuals show minimal systematic deviations across the radial range, with a reduced $\chi^2 = 13.64$, confirming that Model E provides the most robust representation of the observed profile among the five models considered. The reduced $\chi^2$ values follow a similar trend, decreasing from 20.23 for Model A to 13.64 for Model E, confirming the progressive improvement in fit quality with increasing model complexity. However, it is worth noting that even the best-fitting model has reduced $\chi^2 > 1$, suggesting either that the uncertainties may be slightly underestimated or that there is additional small-scale structure not captured by any of these analytical models.
		
The fact that the characteristic radius are comparable to or only slightly larger than the synthesized beam size ($2.30^{\prime\prime} \times 1.28^{\prime\prime}$ $\sim$ 3791 AU) indicates that the source is only marginally resolved. Many of the fitted scale radius are of similar magnitude (e.g., $r_{\mathrm{flat}} = 2\farcs38$ in Model E, corresponding to $\sim$3800 au). Therefore, while the profile steepening beyond $\sim$1$\farcs$5--2$\farcs$0 suggests a change in the physical regime, likely marking the transition from the inner envelope/disk region to the outer envelope, the derived power-law slopes should be interpreted with some caution, as they may still be influenced by beam convolution effects. Higher angular resolution observations are needed to fully resolve the inner structure and confirm the intrinsic slope. The presence of a Gaussian component in the best-fit model hints at additional complexity, possibly related to asymmetries in the outflow or infall structures previously reported in NIRS3 \citep{zin15, liu20}.
	
\begin{figure*}
\centering
\includegraphics[width=1.0\textwidth]{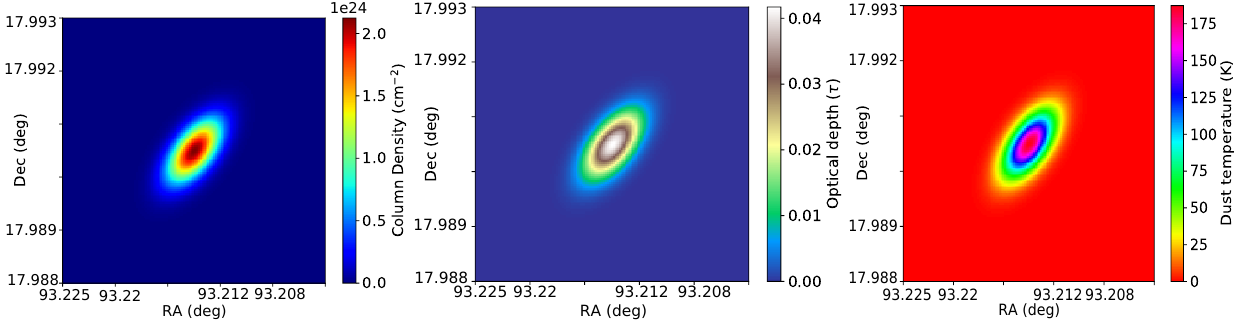}
\caption{Modelled images of the column density of \ce{H2}, dust optical depth, and dust temperature, which were generated with the radiative transfer code RADMC-3D.}
\label{fig:model}
\end{figure*}
	
\begin{table}{}
\centering
\caption{Hydrogen column density and dust optical depth of NIRS3 core.}
\begin{tabular}{cccccccc}
\hline
Wavelength&Column density of \ce{H2}& Dust optical depth\\
(MM)	 &   (cm$^{-2}$) & ($\tau_\nu$) \\                   
\hline
2.32&(1.1$\pm$0.3)$\times$10$^{24}$&1.8$\times$10$^{-2}$ \\
2.29&(1.2$\pm$0.2)$\times$10$^{24}$&1.9$\times$10$^{-2}$ \\
2.12&(1.1$\pm$0.3)$\times$10$^{24}$&2.1$\times$10$^{-2}$ \\
2.10&(1.0$\pm$0.2)$\times$10$^{24}$&2.4$\times$10$^{-2}$ \\
\hline
Average value&(1.1$\pm$0.2)$\times$10$^{24}$&2.1$\times$10$^{-2}$   \\
\hline
\end{tabular}
\label{tab:HI}\\
\end{table}
	
\begin{figure*}
\centering
\includegraphics[width=0.5\textwidth]{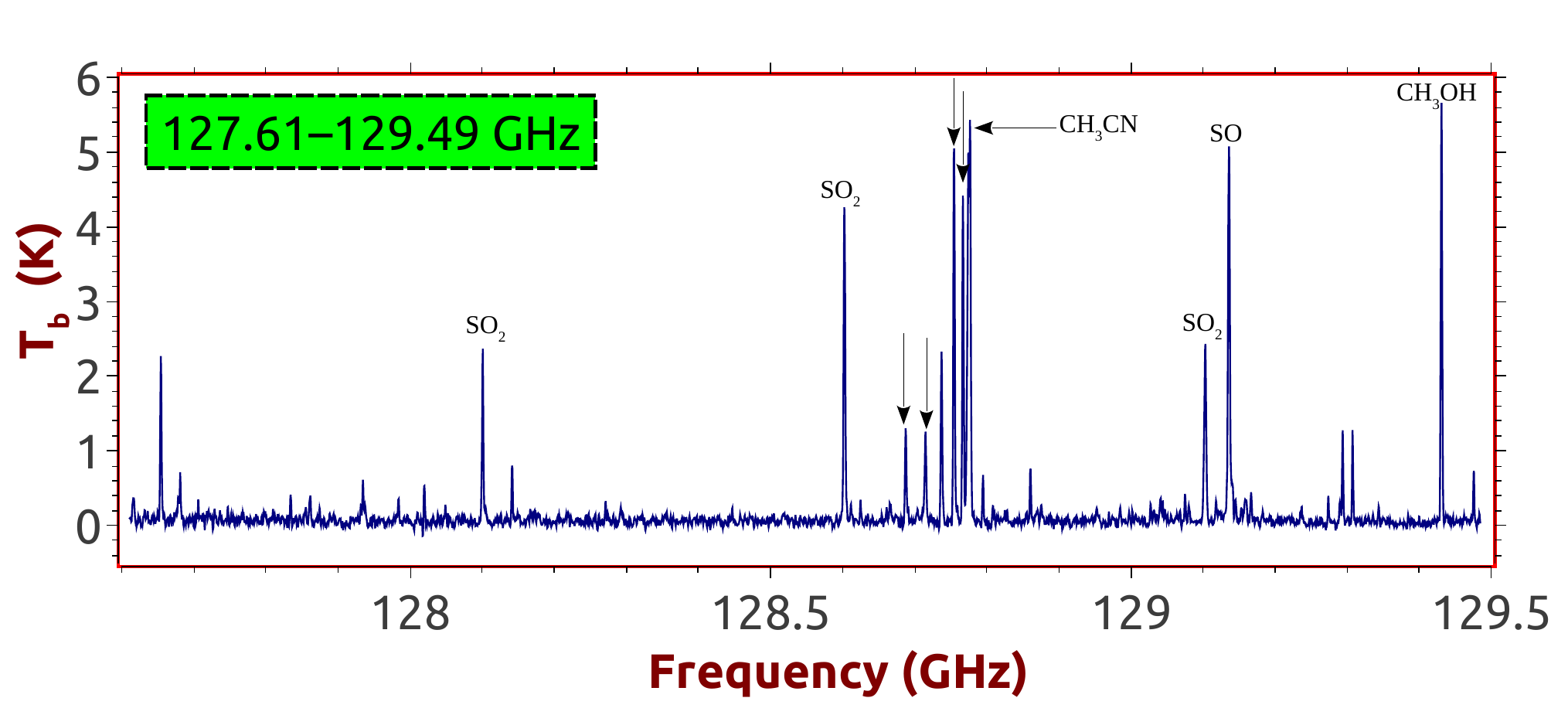}\includegraphics[width=0.5\textwidth]{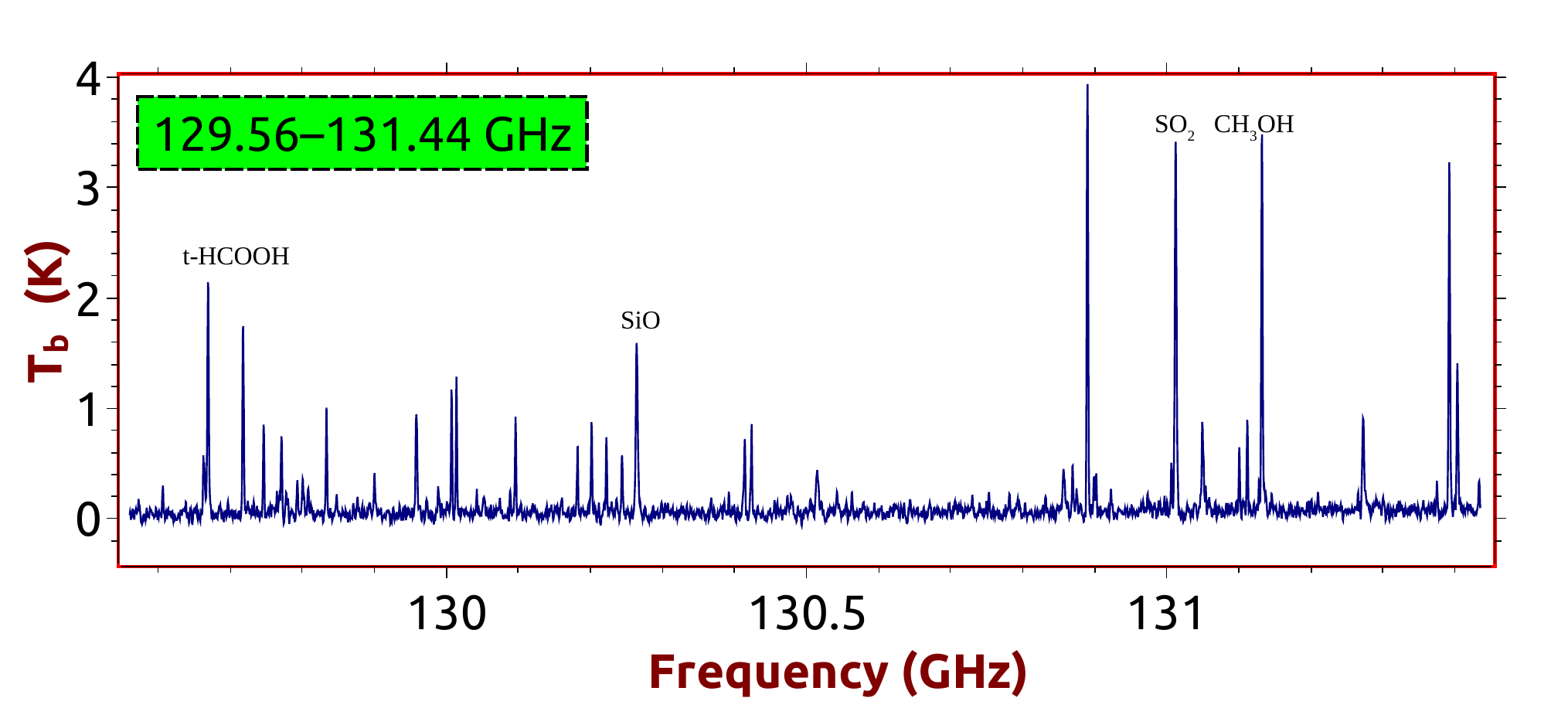}
\includegraphics[width=0.5\textwidth]{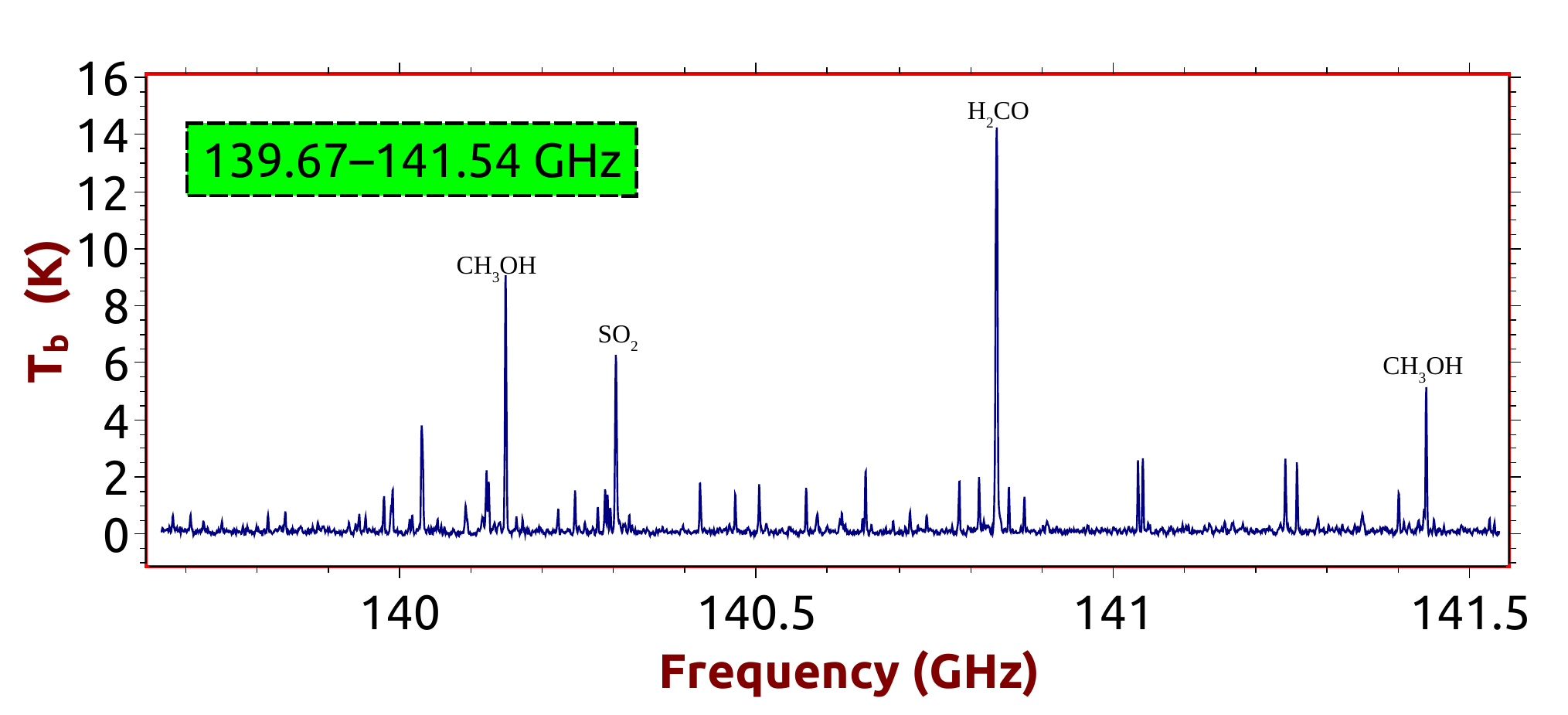}\includegraphics[width=0.5\textwidth]{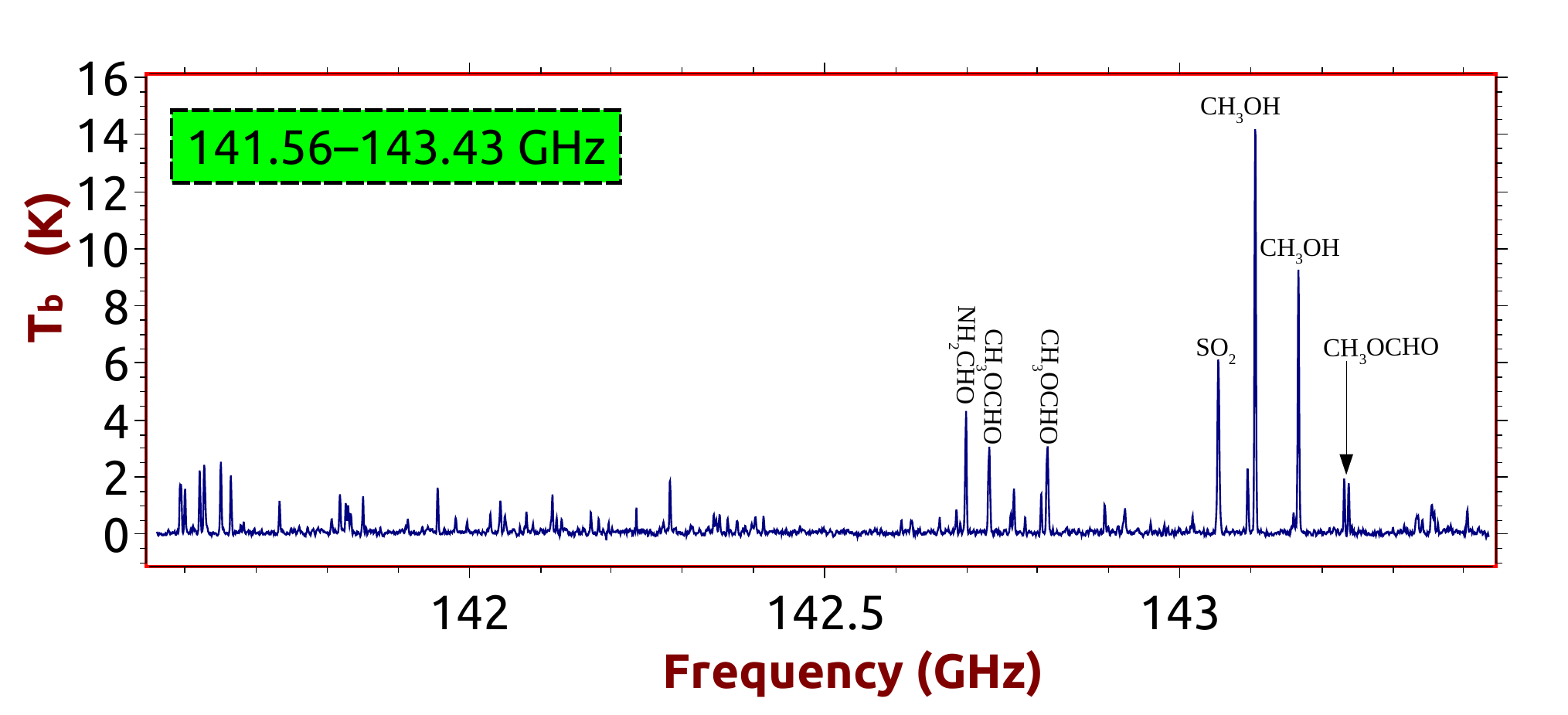}
\caption{Millimeter wavelength molecular emission spectra of NIRS3 from ALMA band 4. The resolution of the spectra is 976.56 kHz.}
\label{fig:Fullspectra}
\end{figure*}
	
\subsubsection{Spectral energy distribution (SED)}
We constructed the spectral energy distribution (SED) of NIRS3 using ALMA bands 3, 4, 6, and 7. The flux densities for bands 3, 6, and 7 were taken from \citet{ce18}, \citet{be23}, and \citet{zin20}, respectively. For band 4, multiple spectral windows are available; however, these share the same calibration and are not independent continuum measurements. Therefore, they are treated as part of a single observational dataset in the SED analysis. In addition, we incorporated flux measurements from AKARI, Herschel, and Spitzer/IRAC covering wavelengths from 3.6 $\mu$m to 500 $\mu$m. All cutout images were obtained from the NASA/IPAC Infrared Science Archive\footnote{\url{https://irsa.ipac.caltech.edu/applications/Gator/}}. Since the angular resolutions of the ALMA observations differ from those of the other instruments, we therefore treated the flux measurements from AKARI, Herschel, and Spitzer/IRAC as upper limits. We then fitted the SED using the radiative transfer model of \cite{rob17} using the Python-based ASTROPY module (version 5.0) \citep{astro22}. For the fitting, the distance to the source was fixed at 1.78 kpc. The details of the radiative transfer fitting procedure applied to the flux measurements are provided in \citet{san14} and \citet{min21}. We employed a Markov Chain Monte Carlo (MCMC) algorithm to fit the radiative model to the observed SED, following the methodology described in \citet{min21}. The resulting SED fit and the corresponding corner plot from the MCMC analysis are shown in Figure~\ref{fig:sed}. The corner diagram was produced using the {\tt corner.py} Python package \citep{fo16}. From the best-fit model, we derived a dust mass of 20.5 $\pm$ 3.0 \(\textup{M}_\odot\), a dust temperature $T_{d}$ = 150 $\pm$ 16 K, and a dust spectral index $\beta$ = 1.7$\pm$0.6 for NIRS3. All derived dust parameters are consistent within uncertainties with previously reported values by \citet{wan11}, \citet{zin15}, and \citet{liu20}.

\begin{table}{}
\scriptsize
\centering
\caption{Priors distribution functions for the Bayesian approach's parameters.}
\begin{tabular}{cccccccc}
\hline
Parameters&Uniform minimum value & Uniform maximum value\\                  
\hline
log$_{10}$(N) (cm$^{-2}$)& 8 & 23 \\
$T_{ex}$ (K)& 4 &300 \\
FWHM (km s$^{-1}$)& 3& 8\\
\hline
\end{tabular}
\label{tab:distri}\\
\end{table}
	
\begin{table*}
\centering
\scriptsize 
\caption{Summary of spectral line parameters of the N-bearing species towards NIRS3. }	%\begin{adjustbox}{width=1.0\textwidth}
\begin{tabular}{ccccccccccccccccc}
\hline 
Molecule&Rest frequency &Transition & $E_{up}$ & $A_{ij}$ &g$_{up}$&S$\mu^{2}$$^{\dagger}$ &FWHM&$\rm{\int T_{b}dV}^\ddagger$& Velocity&Remark\\
		&(GHz) &&(K)&(s$^{-1}$) & &(Debye$^{2}$) &(km s$^{-1}$)&(K km s$^{-1}$)&(km s$^{-1}$) & \\
\hline
\ce{CH3CN}&128.690& 7$_{6}$--6$_{6}$ &281.79&4.72$\times$10$^{-5}$&60&78.73&5.54$\pm$0.18&7.35$\pm$0.25 &5.82$\pm$0.04 &Non blended\\
		  &128.717& 7$_{5}$--6$_{5}$ &203.28&8.73$\times$10$^{-5}$&30&72.68&5.56$\pm$0.19&7.58$\pm$0.37&5.37$\pm$0.06 &Non blended\\
		  &128.739& 7$_{4}$--6$_{4}$ &139.02&1.20$\times$10$^{-4}$&30&99.92&5.51$\pm$0.11&12.94$\pm$0.29&5.92$\pm$0.02 &Non blended\\
		  &128.757& 7$_{3}$--6$_{3}$ &89.02&1.46$\times$10$^{-4}$&60&242.25&5.53$\pm$0.26&25.39$\pm$0.22&5.87$\pm$0.05 &Non blended\\
		  &128.769& 7$_{2}$--6$_{2}$ &53.30&1.64$\times$10$^{-4}$&30&136.26&5.57$\pm$0.29&19.92$\pm$0.35&5.63$\pm$0.07 &Non blended\\
		  &128.776& 7$_{1}$--6$_{1}$ &31.87&1.75$\times$10$^{-4}$&30&145.34&5.55$\pm$0.56&21.31$\pm$0.48&5.23$\pm$0.08 &Non blended\\
		  &128.779& 7$_{0}$--6$_{0}$ &24.72&1.78$\times$10$^{-4}$&30&148.39&5.58$\pm$0.82&21.89$\pm$0.52&5.28$\pm$0.04&Non blended\\
\hline
			
\ce{C2H5CN} & 127.618&14(1,13)--13(1,12)&47.23&1.72$\times$10$^{-4}$&29&206.19&5.45$\pm$0.18&1.51$\pm$0.23&5.23$\pm$0.05 &Non blended\\
		    & 129.795&15(1,15)--14(1,14)&51.09&1.82$\times$10$^{-4}$&31&221.20&5.42$\pm$0.25&1.55$\pm$0.32&5.35$\pm$0.02 &Non blended\\
			& 130.903&15(0,15)--14(0,14)&50.79&1.87$\times$10$^{-4}$&31&221.58&5.46$\pm$0.33&1.48$\pm$0.16&5.63$\pm$0.05 &Non blended\\
			& 142.346&16(2,15)--15(2,14)&62.69&2.37$\times$10$^{-4}$&33&233.27&5.48$\pm$0.65&3.25$\pm$0.38&5.42$\pm$0.08 &Non blended\\
			& ~~143.335$^{*}$&16(8,8)--15(8,7)&129.58&1.85$\times$10$^{-4}$&33&177.89&5.48$\pm$0.85&5.97$\pm$0.48&5.21$\pm$0.02 &Non blended\\
			& ~~143.337$^{*}$&16(7,10)--15(7,9)&112.93&1.99$\times$10$^{-4}$&33&191.75&5.46$\pm$0.55&2.96$\pm$0.36&5.16$\pm$0.06 &Non blended\\
			& ~~143.343$^{*}$&16(9,7)--15(9,6)&148.44&1.68$\times$10$^{-4}$&33&162.11&5.43$\pm$0.15&2.57$\pm$0.38&5.32$\pm$0.08 &Non blended\\
			& ~~143.357$^{*}$&16(6,11)--15(6,10)&98.49&2.12$\times$10$^{-4}$&33&203.81&--&--&5.53$\pm$0.03 &Blended with \\
			&                &                  &     &                     &  &      &  &   &             &$aGg^{\prime}$-(\ce{CH2OH})$_{2}$\\
			& ~~143.360$^{*}$&16(10,6)--15(10,5)&169.51&1.50$\times$10$^{-4}$&33&144.51&--&--&5.52$\pm$0.06 &Blended with \\
			&                &                  &      &                      &  &     &  &   &              &$aGg^{\prime}$-(\ce{CH2OH})$_{2}$\\
			& ~~143.383$^{*}$&16(11,5)--15(11,4)&192.77&1.30$\times$10$^{-4}$&33&125.06&5.41$\pm$0.28&1.51$\pm$0.12&5.51$\pm$0.03 &Non blended\\
			& 143.406&16(5,12)--15(5,11)&86.28&2.23$\times$10$^{-4}$&33&214.01&5.48$\pm$0.42&4.18$\pm$0.36&5.21$\pm$0.05 &Non blended\\
			& 143.407&16(5,11)--15(5,10)&86.28&2.23$\times$10$^{-4}$&33&214.01&5.48$\pm$0.43&4.18$\pm$0.37&5.31$\pm$0.08 &Non blended\\
			& ~~143.410$^{*}$&16(12,4)--15(12,3)&218.22&1.08$\times$10$^{-4}$&33&103.75&5.25$\pm$0.36&3.18$\pm$0.22&5.53$\pm$0.03 &Non blended\\    
			
\hline

\ce{C2H3CN} &131.267&14(0,14)--13(0,13)&47.50&1.85$\times$10$^{-4}$&87&610.55&5.35$\pm$0.14&0.96$\pm$0.02&5.13$\pm$0.06&Non blended\\
			&140.429&15(0,15)--14(0,14)&54.24&2.27$\times$10$^{-4}$&93&654.12&5.38$\pm$0.65&1.21$\pm$0.04&5.18$\pm$0.02&Non blended\\
			&141.945&15(2,14)--14(2,13)&63.22&2.30$\times$10$^{-4}$&93&643.25&5.42$\pm$0.43&1.32$\pm$0.02&5.26$\pm$0.08&Non blended\\
			&~~142.399$^{*}$&15(5,11)--14(5,10)&108.70&2.10$\times$10$^{-4}$&93&582.15&5.32$\pm$0.32 &1.41$\pm$0.05 &5.38$\pm$0.07&Non-blended\\
			&142.401&15(6,10)--14(6,9)&132.44&1.99$\times$10$^{-4}$&93&550.14&5.39$\pm$0.52&1.07$\pm$0.02&5.63$\pm$0.02 &Non-blended\\
			&~~142.419$^{*}$&15(7,8)--14(7,7)&160.44&1.85$\times$10$^{-4}$&93&512.29&5.32$\pm$0.43&1.37$\pm$0.05 &5.52$\pm$0.05 &Non-blended\\
			&142.424&15(4,12)--14(4,11)& 89.27&2.20$\times$10$^{-4}$&93&608.34&5.36$\pm$0.41&1.23$\pm$0.03&5.51$\pm$0.02 &Non blended\\
			&142.426&15(4,11)--14(4,10)&89.27 &2.20$\times$10$^{-4}$&93&608.33&5.53$\pm$0.51&1.01$\pm$0.04&5.36$\pm$0.08 &Non blended\\
			&~~142.447$^{*}$&15(8,7)--14(8,6)&192.71&1.70$\times$10$^{-4}$&93&468.65&5.36$\pm$0.72&1.67$\pm$0.02 &5.86$\pm$0.06 &Non-blended\\
			&142.472&15(3,13)--14(3,12)&74.15&2.28$\times$10$^{-4}$&93&628.73&5.52$\pm$0.63&1.22$\pm$0.04&5.65$\pm$0.02 &Non blended\\
			&~~142.484$^{*}$&15(9,6)--14(9,5)&229.21&1.52$\times$10$^{-4}$&93&419.15&5.21$\pm$0.61&1.46$\pm$0.06&5.16$\pm$0.02 &Non blended\\
			&~~142.527$^{*}$&15(10,5)--14(10,4)&269.92&1.32$\times$10$^{-4}$&93&363.86&5.16$\pm$0.23&1.35$\pm$0.04&5.17$\pm$0.06 &Non blended\\
			&142.573&15(3,12)--14(3,11)&74.16&2.28$\times$10$^{-4}$&93&628.74 &5.35$\pm$0.32 &0.84$\pm$0.03 &5.08$\pm$0.06 &Non blended\\
			&~~142.576$^{*}$&15(11,4)--14(11,3)&314.81&1.10$\times$10$^{-4}$&93&302.71&--&--&5.23$\pm$0.02 &Blended with U line\\
			
\hline
			
\ce{NH2CN} ($\nu = 0$)&139.931&7(2,6)--6(2,5)&84.83&2.56$\times$10$^{-4}$&15&120.26&5.15$\pm$0.36&2.12$\pm$0.26&5.28$\pm$0.03 &Non blended\\
		              &139.940&7(2,5)--6(2,4)&84.83&2.56$\times$10$^{-4}$&15&120.27&5.18$\pm$0.16&1.26$\pm$0.12 &5.23$\pm$0.02 &Non blended \\
		              &139.954&7(0,7)--6(0,6)&26.87&2.79$\times$10$^{-4}$&15&131.03&5.17$\pm$0.25&2.82$\pm$0.14&5.06$\pm$0.08 &Non blended\\
			          &140.877&7(1,6)--6(1,5)&41.54&2.78$\times$10$^{-4}$&45&385.10&5.18$\pm$0.24&5.62$\pm$0.32 &5.12$\pm$0.01 &Non blended\\
\hline          
\ce{NH2CN} ($\nu = 1$)&139.842&7(0,7)--6(0,6)&98.16&2.71$\times$10$^{-4}$&45&383.43&5.12$\pm$0.41 &3.70$\pm$0.48&5.02$\pm$0.06 & Non blended\\
			          &140.710&7(1,6)--6(1,5)&112.46&2.71$\times$10$^{-4}$&15&125.27&5.10$\pm$0.19&1.12$\pm$0.13&5.18$\pm$0.03 &Non blended\\
			
\hline
			
\ce{NH2CHO}&128.103&6(2,4)--5(2,3)&33.38&1.28$\times$10$^{-4}$&13&69.73&5.12$\pm$0.32&12.23$\pm$0.46&5.23$\pm$0.06 &Non blended\\
		   &142.701&7(1,7)--6(1,6)&30.41&1.98$\times$10$^{-4}$&15&89.63&5.13$\pm$0.38&21.60$\pm$0.52&5.18$\pm$0.04 &Non blended \\
			
\hline
			
\ce{HC3N} ($\nu_{7}$ = 2)&128.169&$J$ = 14--13, $l$ = 0&687.83&1.63$\times$10$^{-4}$&29&193.10&5.35$\pm$0.26 &0.72$\pm$0.02&5.17$\pm$0.03 &Non blended\\
			&128.175&$J$ = 14--13, $l$ = 2e&691.10&1.60$\times$10$^{-4}$&29&189.18&5.38$\pm$0.45&1.18$\pm$0.35 &5.12$\pm$0.01 &Non blended \\
			&128.182&$J$ = 14--13, $l$ = 2f&691.10&1.60$\times$10$^{-4}$&29&189.16&5.36$\pm$0.39&0.75$\pm$0.03 &5.18$\pm$0.07 &Non blended\\
			
\hline
\end{tabular}	
%	\end{adjustbox}
\label{tab:MOLECULAR DATA}\\
{{*}}--These transitions are doublets with frequency differences of $\leq$100 kHz. The quantum numbers of the second transition are not shown.\\
$\dagger$--The values of S$\mu^{2}$ are taken from \href{https://splatalogue.online/#/home}{Splatalogue}.\\
$\ddagger$--The values of $\rm{\int T_{b}dV}$ are estimated by fitting the Gaussian model over the non-blended emission lines of detected molecules.\\
\end{table*}

\subsubsection{Estimation of column density of \ce{H2} and dust optical depth}
For optically thin dust continuum emission, the peak flux density ($S_\nu$) can be denoted as,

\begin{equation}
	S_\nu = B_\nu(T_d)\tau_\nu\Omega_{beam}
\end{equation}
In the above equation, $B_\nu(T_d)$ is the Planck function at dust temperature $T_d$, $\tau_\nu$ indicates the dust optical depth, and $\Omega_{beam} = (\pi/4 \ln 2)\times \theta_{major} \times \theta_{minor}$ indicates the solid angle of the synthesized beam \citep{whi92}. The equation of optical depth ($\tau_\nu$) in terms of the mass density of dust can be expressed as,

\begin{equation}
	\tau_\nu =\rho_d\kappa_\nu L
\end{equation}
where $L$ indicates the path length, $\rho_d$ is the mass density of the dust, and $\kappa_{\nu}$ is the mass absorption coefficient. The mass density of the dust can be written in terms of the dust-to-gas mass ratio ($Z$),

\begin{equation}
	\rho_d = Z\mu_H\rho_{H_2}=Z\mu_HN_{H_2}2m_H/L
\end{equation}
where $\rho_{H_2}$ is the mass density of hydrogen, $\mu_H$ is the mean molecular weight per hydrogen molecule, $m_H$ is the mass of hydrogen, and $N_{H_2}$ is the column density of hydrogen. To estimate the column density of \ce{H2}, we used the dust temperature $T_d$ = 150 K (derived from SED), $\mu_H = 1.41$, and $Z = 0.01$ \citep{cox00}. The peak flux densities of NIRS3 at different frequencies are shown in Table~\ref{tab:continuum}. The column density equation of molecular \ce{H2} can be written as,

\begin{equation}
	N_{H_2} = \frac{S_\nu /\Omega}{2\kappa_\nu B_\nu(T_d)Z\mu_H m_H}
\end{equation}
To estimate the mass absorption coefficient ($\kappa_{\nu}$), we adopt the following equation, $\kappa_\nu$ = 0.90($\nu$/230 \textrm{GHz})$^{\beta}\ \textrm{cm}^{2}\ \textrm{g}^{-1}$, where ${k_{230} = 0.90~\textrm{cm}^{2}\ \textrm{g}^{-1}}$ is the dust emissivity at a gas density of $\rm{10^{6}\ cm^{-3}}$ covered by a thin ice mantle at a frequency of 230 GHz \citep{moto19}. We use $\beta$ $\sim$ 1.7, which was estimated from the SED fitting. The values of $\kappa_{\nu}$ are 0.335, 0.343, 0.392, and 0.398. From equation 11, we estimate the column density of molecular \ce{H2} at wavelengths of 2.32 mm, 2.29 mm, 2.12 mm, and 2.10 mm, which are shown in Table~\ref{tab:HI}. We take a mean value to determine the resultant column density of \ce{H2} towards NIRS3. After averaging these four continuum values, we obtain that the final column density of \ce{H2} towards NIRS3 is ($1.1 \pm 0.2)\times 10^{24}$ cm$^{-2}$.

We also calculate dust optical depth ($\tau_\nu$) using the following equation \citep{riv17}:
\begin{equation}
	\tau_{\nu} = -\ln\left(1- \frac{S_{\nu}}{\Omega_{\rm beam}B_{\nu}(T_{\rm d})}\right),
	\label{eq:dust_tau}
\end{equation}
In equation~\ref{eq:dust_tau}, $S_{\nu}$ denotes the peak flux density, $\Omega_{beam} = (\pi/4 \ln 2)\times \theta_{major} \times \theta_{minor}$ is the solid angle of the beam where $\theta_{\rm min}$ and $\theta_{\rm maj}$ are the minor and major axes of the beam, $B_{\nu}$ is the Planck function, and $T_{\rm d}$ is the dust temperature. For the calculations, we used a dust temperature ($T_{d}$) of 150 K, estimated from the SED analysis. The estimated dust optical depths ($\tau_\nu$) of the main core at different wavelengths are shown in Table~\ref{tab:HI}. After averaging, we obtain that the final dust optical depth towards NIRS3 is 2.1$\times$10$^{-2}$. The estimated dust optical depth suggests that the dust in NIRS3 is optically thin, considering the current angular resolution of this observation.

To understand the distribution of the column density of \ce{H2}, dust optical depth, and dust temperature towards NIRS3, we also created modelled images of these physical parameters using the 3-D Monte Carlo radiative transfer code RADMC-3D \citep{dul12}. The modelled images are shown in Figure~\ref{fig:model}. To construct the \ce{H2} column density map, we used the gas density distribution as input and generated a map using the {\tt image} task, which integrates the gas density along the line of sight. The dust temperature map was produced using the {\tt mctherm} task, which performs a thermal Monte Carlo simulation assuming radiative equilibrium between dust and the radiation field. A 2D image of the dust temperature was then generated using the {\tt image} task. The dust optical depth was calculated with the {\tt tausurf} task based on the radiation field and dust properties. The resulting 2D images contain intensity or temperature values and were analyzed to interpret the distributions of column density, dust temperature, and optical depth. Additionally, we constructed a spherical grid and modelled a dust envelope with a 1D radial density power-law distribution, along with temperature and opacity profiles. The detailed procedures and parameter choices adopted in RADMC-3D follow the methods outlined by \citet{jac18} and \citet{du23}. The model parameters for NIRS3 used in these simulations are summarized in Table~\ref{tab:modelsetup} (see Appendix). From the modelled images, we observe that the dust temperature towards the warm inner regions of NIRS3 is more than 150 K. The dust temperature, optical depth, and \ce{H2} column density gradually decrease towards the peripheral areas of NIRS3.

\begin{figure*}
	%\centering
	\includegraphics[width=1.0\textwidth]{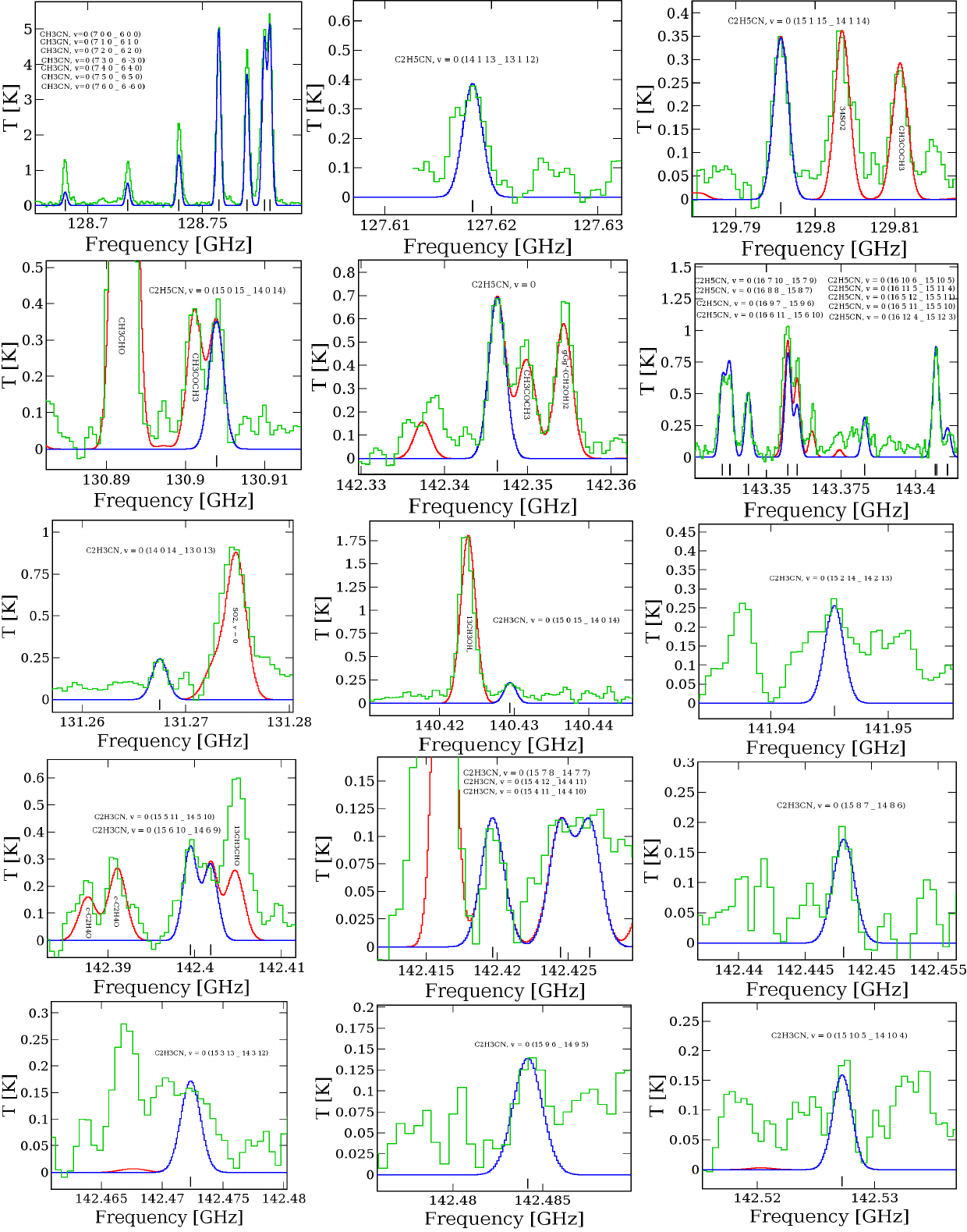}
	\caption{Emission lines of N-bearing molecules \ce{CH3CN}, \ce{C2H5CN}, \ce{C2H3CN}, \ce{NH2CN}, \ce{NH2CHO}, and \ce{HC3N} ($\nu_{7}$ = 2). The green lines indicate the observed molecular spectra of NIRS3, and the blue lines represent the LTE model spectra of detected N-bearing species. Red spectra are the LTE model spectra of all species, including detected N-bearing molecules. The systemic velocity of the spectra is 5.0 km s$^{-1}$.}
	\label{fig:spectra}
\end{figure*}
\begin{figure*}
	%\centering
	\text{{\large Figure~6 continued.}}
	\centering
	\includegraphics[width=1.0\textwidth]{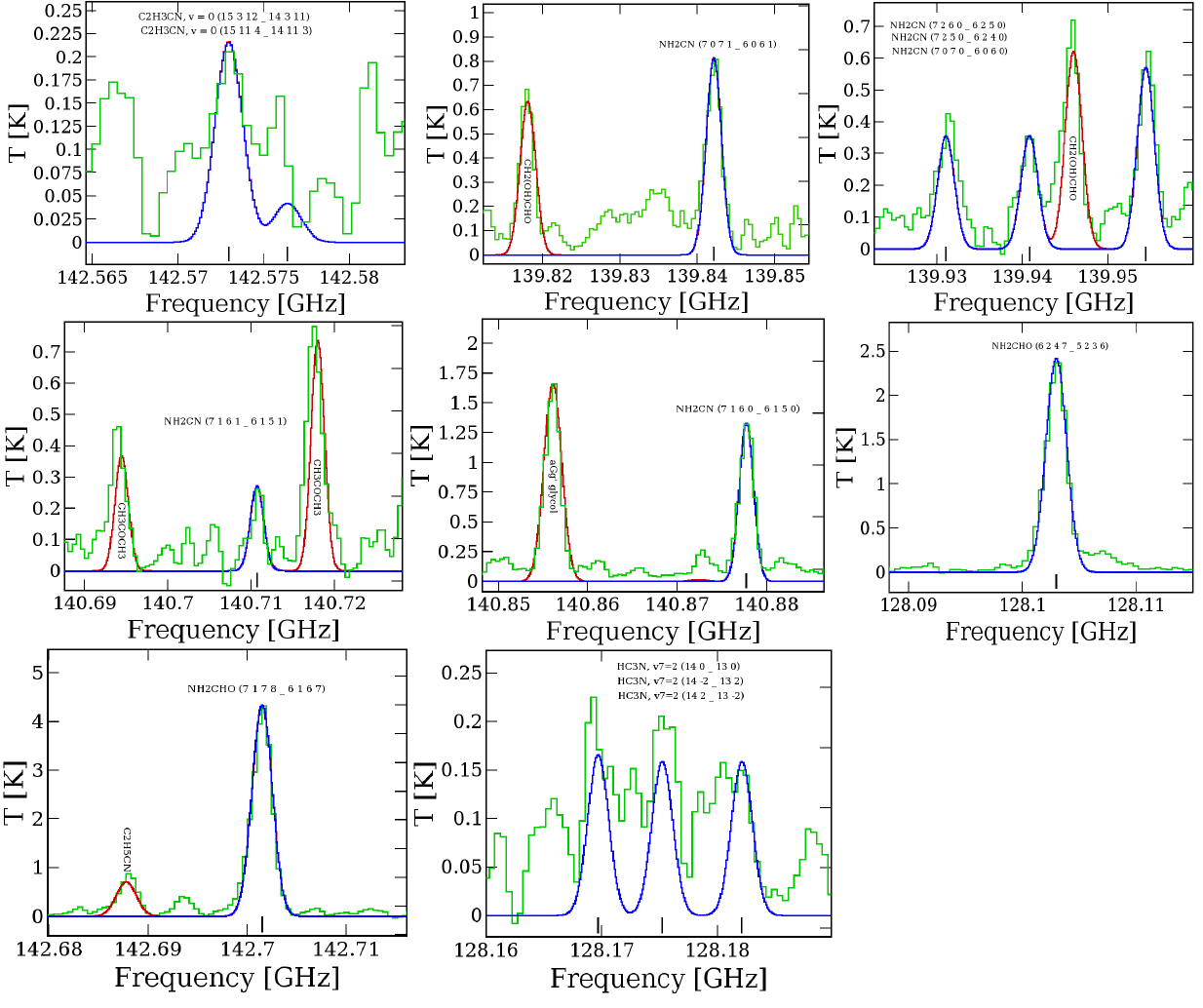}
\end{figure*}

\begin{table*}{}
	%\scriptsize
\centering
\caption{Summary of the LTE-fitted physical parameters of N-bearing species.}
\begin{tabular}{cccccccc}
\hline
Molecule&Number of lines &Column density ($N$)&$T_{ex}$&Abundance&Abundance&Abundance \\
		& &(cm$^{-2}$) & (K)&($\frac{N_{molecule}}{N({H_2})}$)&($\frac{N_{molecule}}{N({CH_{3}OH})}$)&($\frac{N_{molecule}}{N({CH_{3}CN})}$) \\                   
\hline
\ce{CH3CN}&7 & (1.5$\pm$0.7)$\times$10$^{15}$&180$\pm$26&(1.4$\pm$0.7)$\times$10$^{-9}$&(1.6$\pm$0.8)$\times$10$^{-2}$&1\\ 
		
\ce{C2H5CN}&13 &(3.8$\pm$1.2)$\times$10$^{14}$&178$\pm$35&(3.5$\pm$1.3)$\times$10$^{-10}$ &(4.0$\pm$1.3)$\times$10$^{-3}$&(2.5$\pm$1.4)$\times$10$^{-1}$ \\
		
\ce{C2H3CN}&14 &(2.1$\pm$0.6)$\times$10$^{14}$&176$\pm$12&(1.9$\pm$0.7)$\times$10$^{-10}$&(2.2$\pm$0.7)$\times$10$^{-3}$&(1.4$\pm$0.8)$\times$10$^{-1}$\\
		
\ce{NH2CN}&6 &(3.1$\pm$0.9)$\times$10$^{14}$&220$\pm$18&(2.8$\pm$1.0)$\times$10$^{-10}$&(3.3$\pm$1.0)$\times$10$^{-3}$&(2.1$\pm$1.2)$\times$10$^{-1}$\\
		
\ce{NH2CHO}&2 &(2.3$\pm$0.8)$\times$10$^{15}$&210$\pm$22&(2.1$\pm$0.9)$\times$10$^{-9}$  &(2.4$\pm$0.9)$\times$10$^{-2}$&1.5$\pm$0.9  \\
		
\ce{HC3N} ($\nu_{7}$ = 2)&3 &(2.6$\pm$0.5)$\times$10$^{14}$&180$\pm$16&(2.4$\pm$0.7)$\times$10$^{-10}$   &(2.8$\pm$0.6)$\times$10$^{-3}$&(1.8$\pm$0.9)$\times$10$^{-1}$  \\  
		
\hline
\end{tabular}
\label{tab:LTE}\\
\end{table*}

\begin{figure*}
	\centering
	\includegraphics[width=0.5\textwidth]{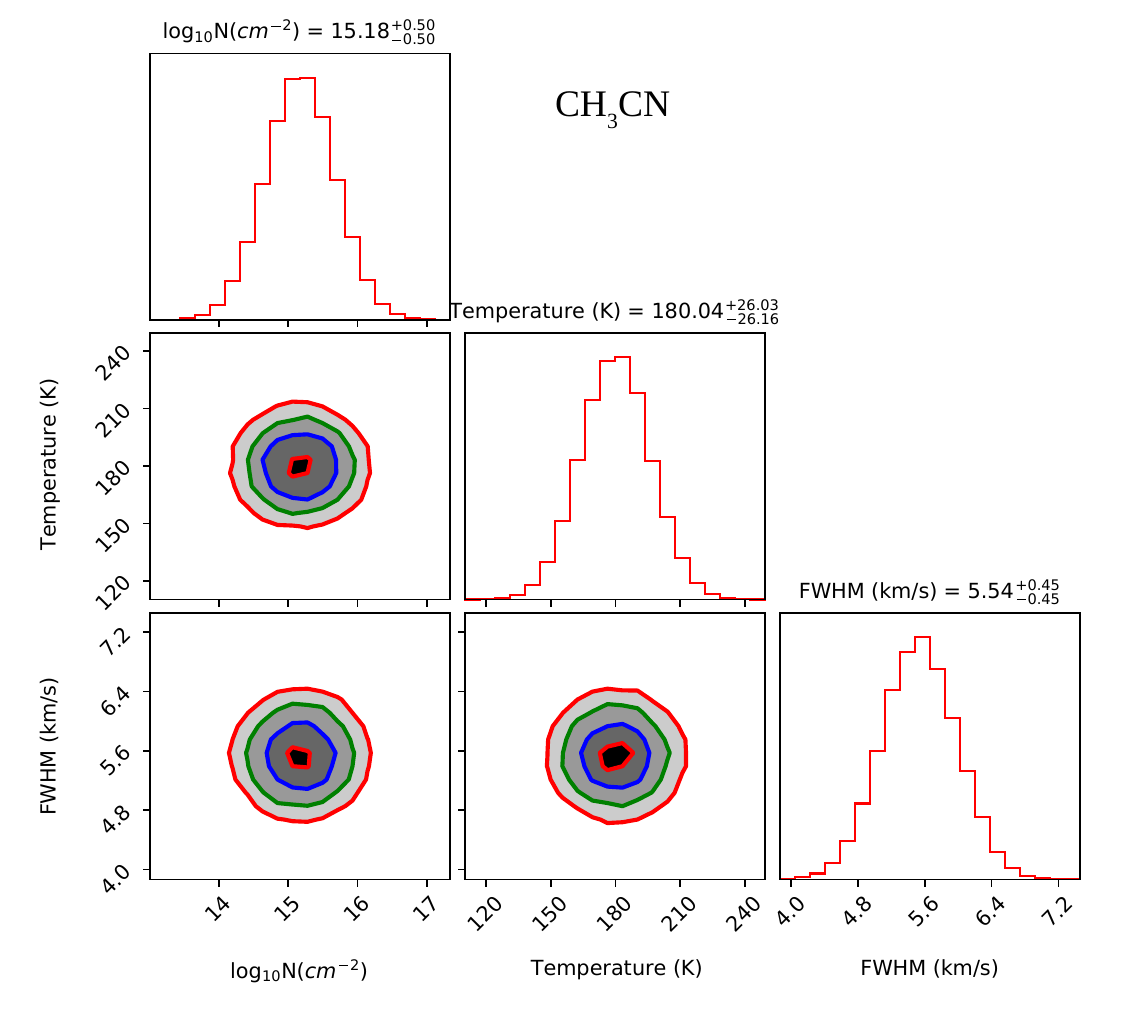}\includegraphics[width=0.5\textwidth]{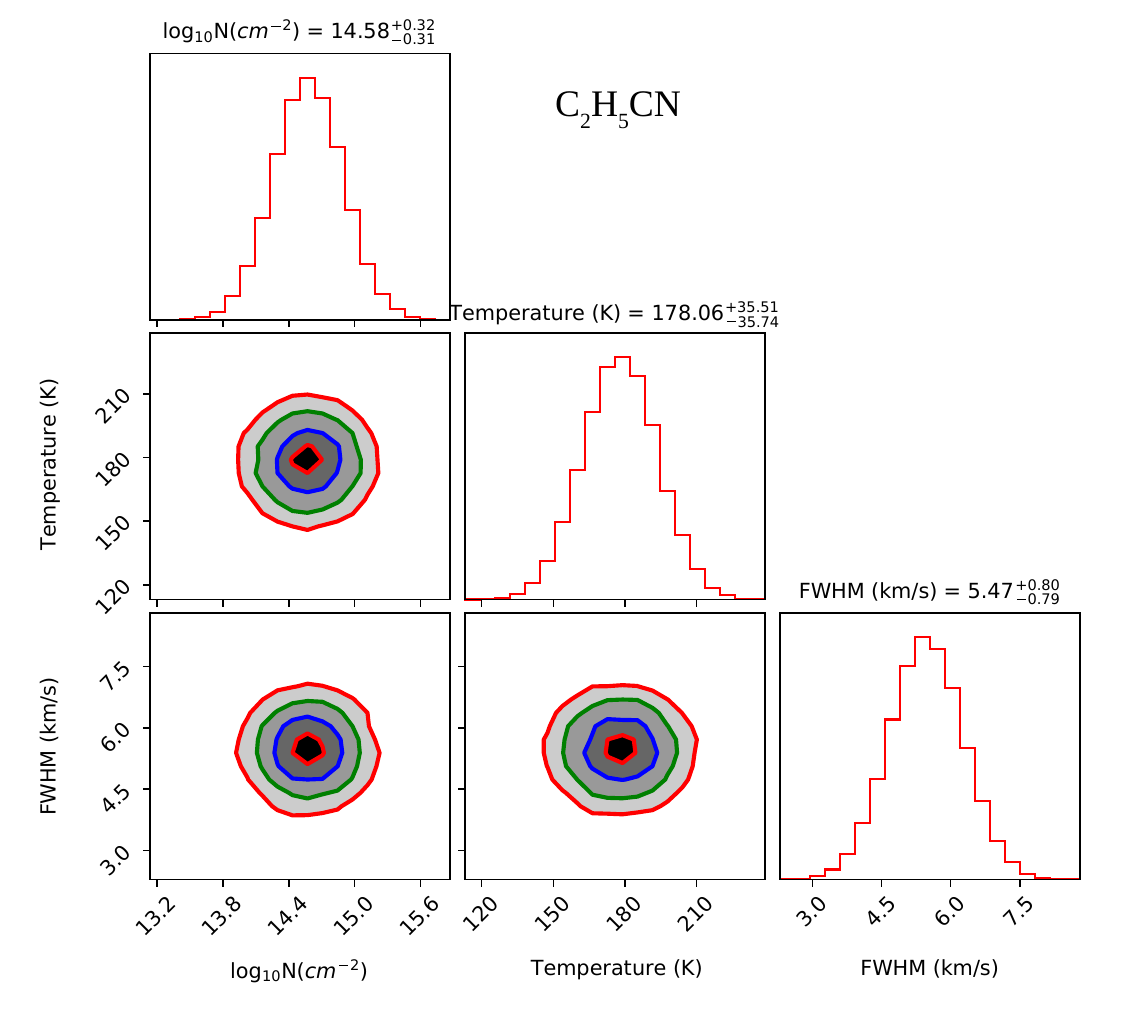}
	\includegraphics[width=0.5\textwidth]{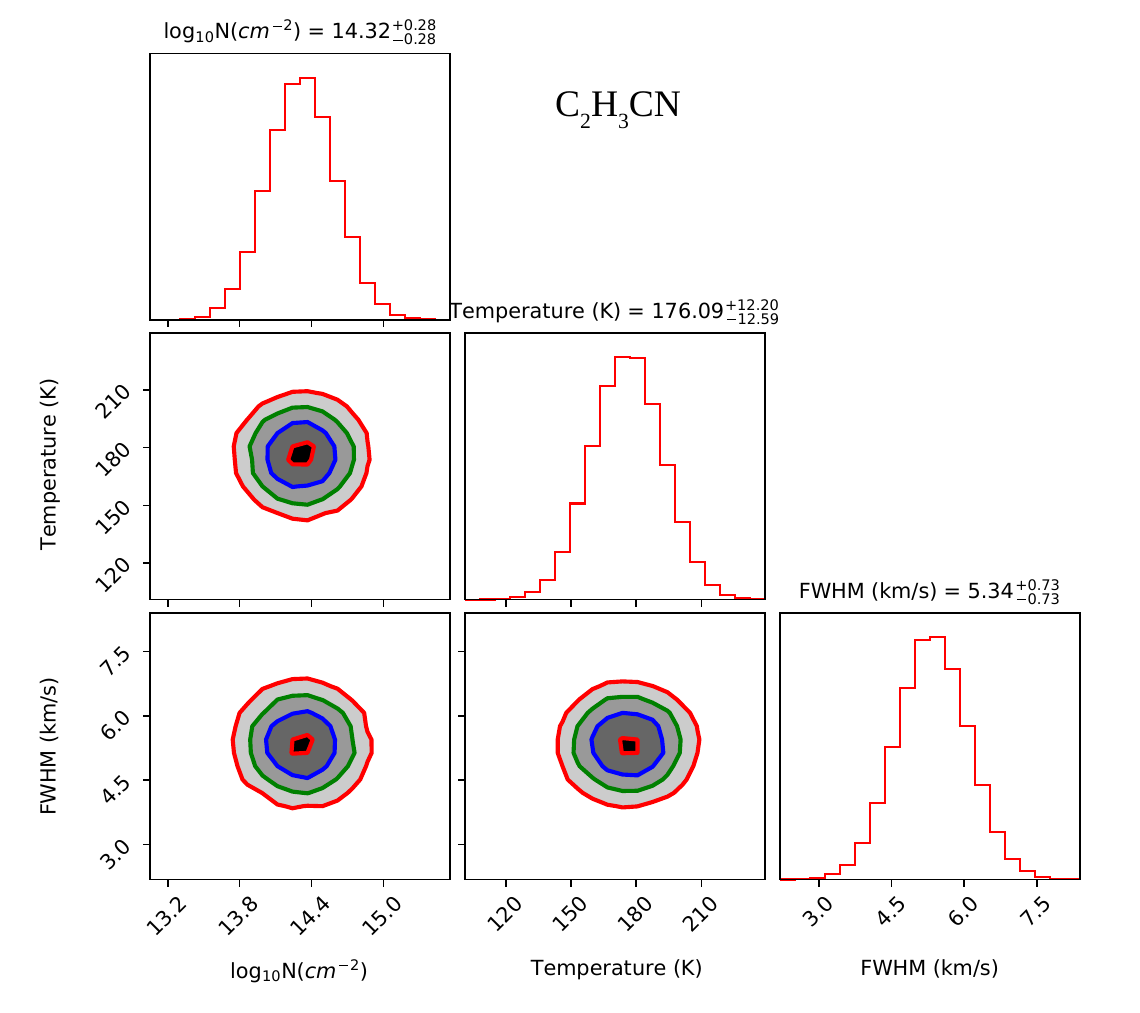}\includegraphics[width=0.5\textwidth]{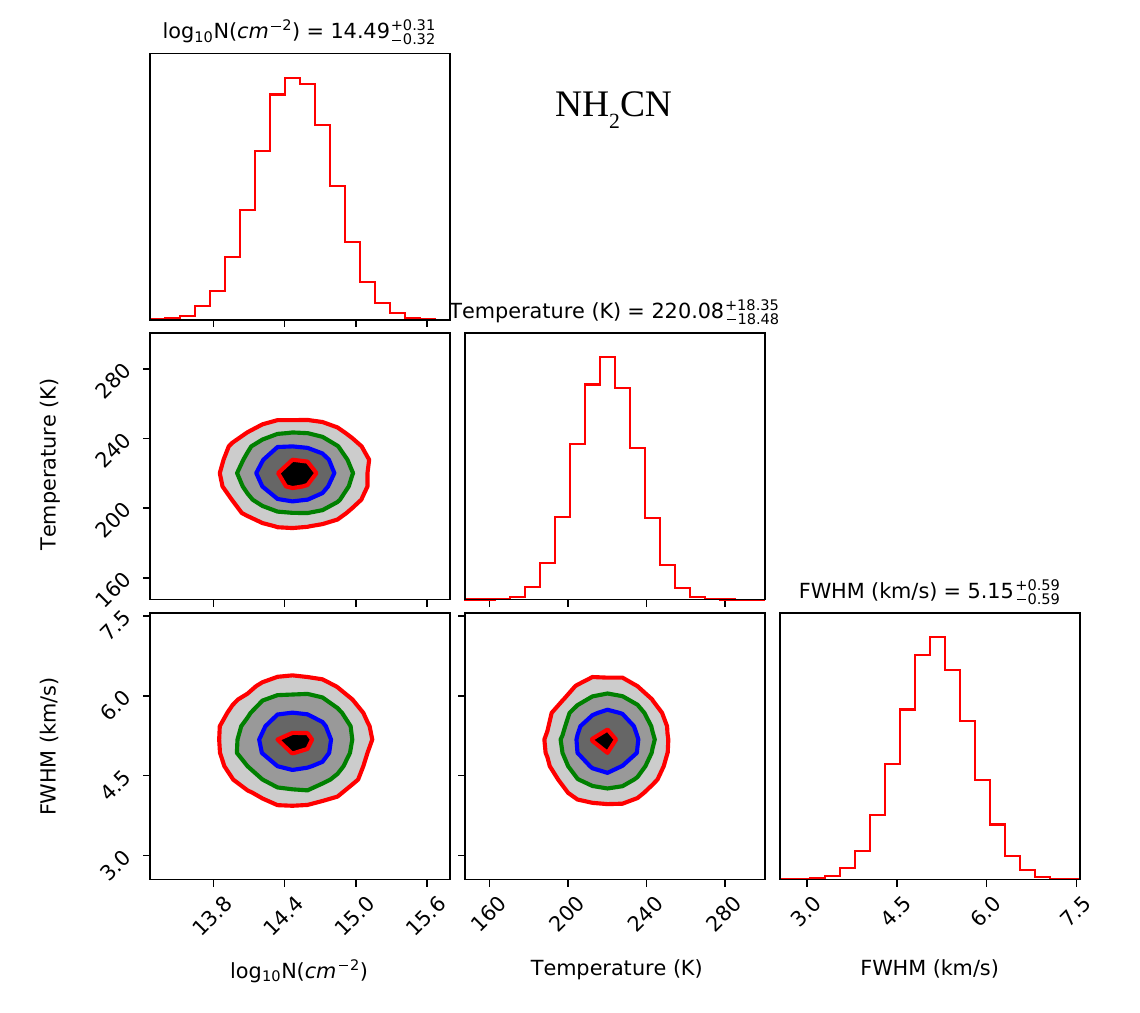}
	\includegraphics[width=0.5\textwidth]{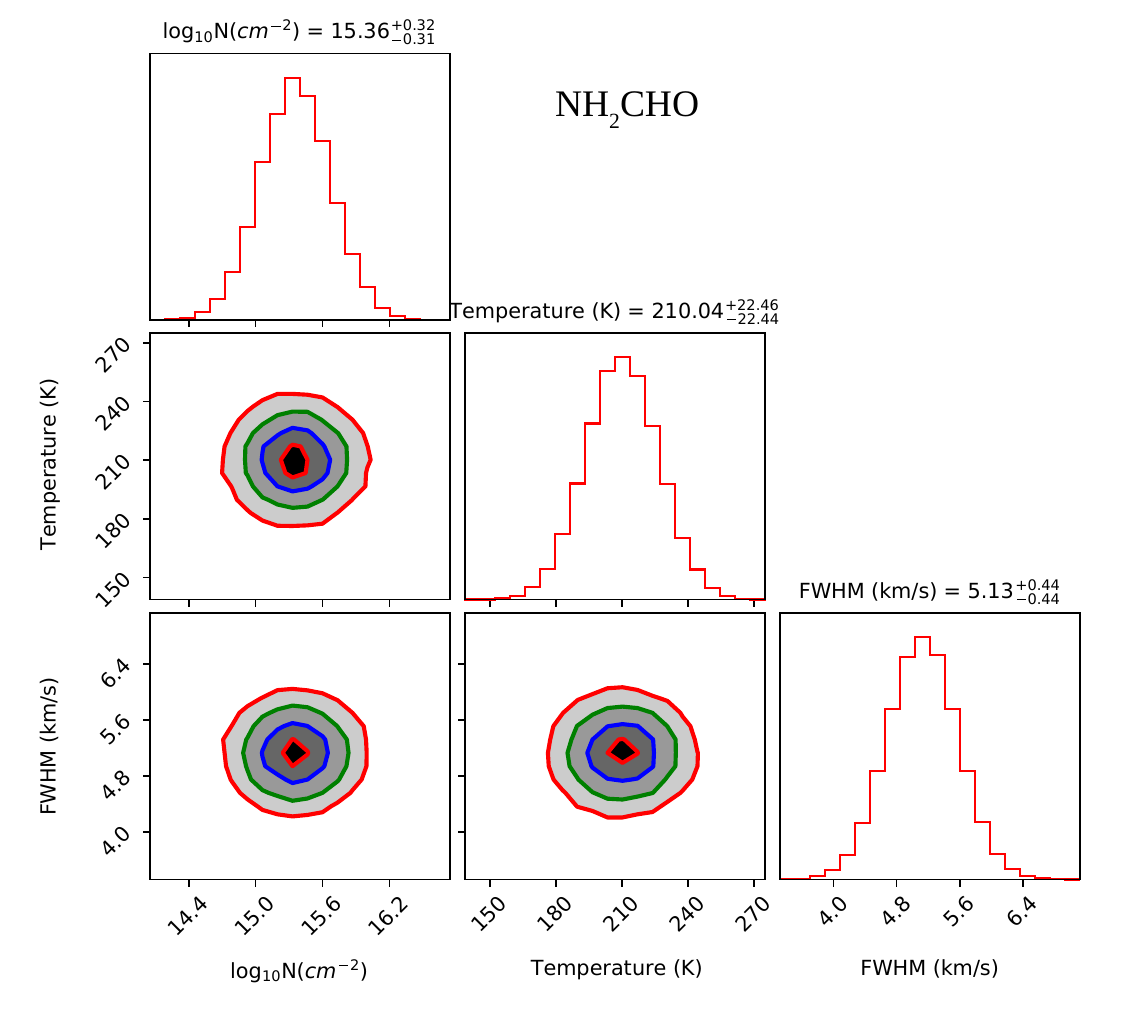}\includegraphics[width=0.5\textwidth]{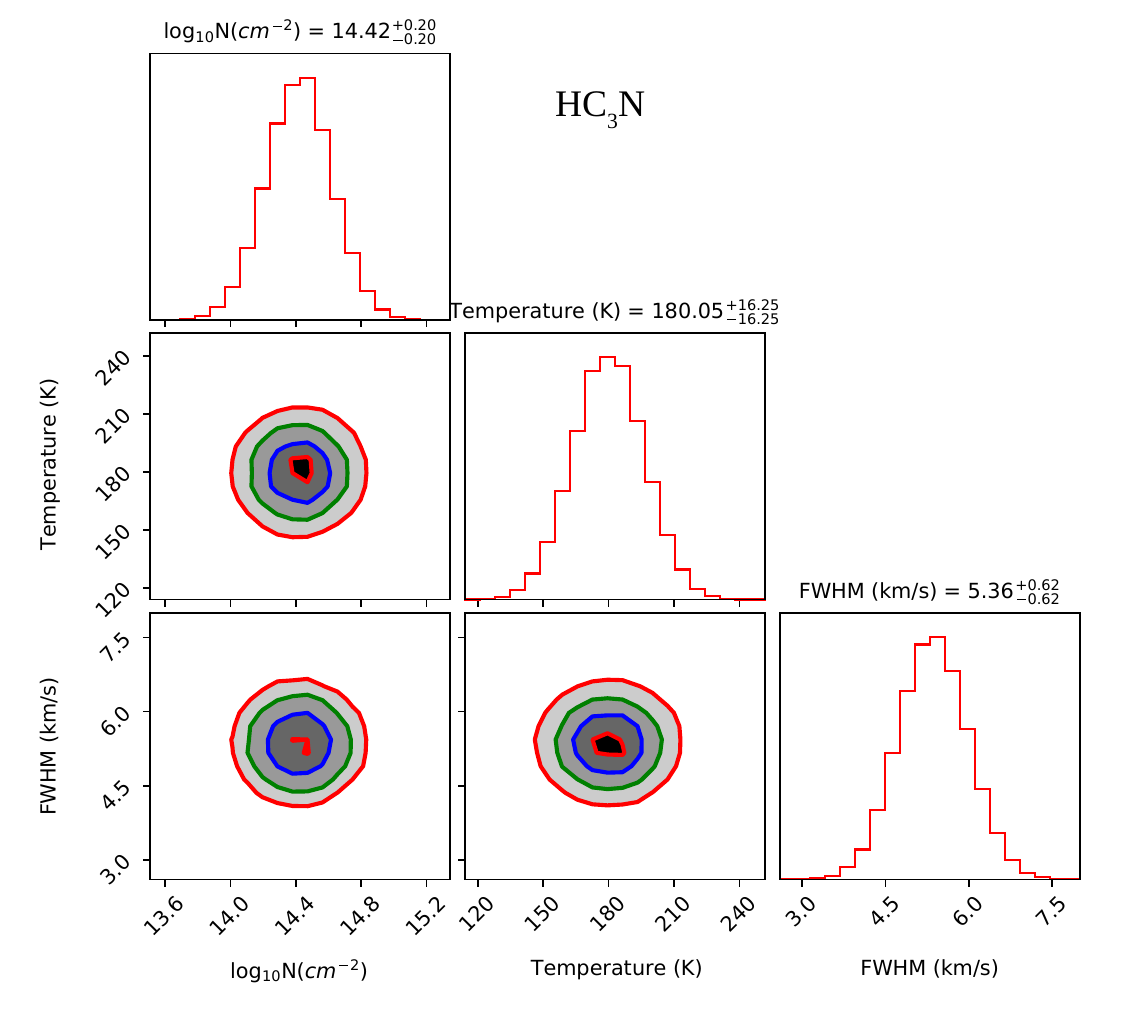}
	\caption{Corner plots showing the covariances of the posterior probability distributions of the column density (log$_{10}$(N) cm$^{-2}$), excitation temperature (K), and FWHM (km s$^{-1}$) of the detected N-bearing molecules.}
	\label{fig:corner}
\end{figure*}

\subsection{Line emission towards NIRS3}
We extracted the molecular spectra from the spectral data cubes using a circular region with a diameter of 2.32$^{\prime\prime}$ (average emitting size) centered on the NIRS3 core. Note that the synthesized beam sizes of the spectral data cubes of NIRS3 are 2.30$^{\prime\prime}$$\times$1.28$^{\prime\prime}$, 2.29$^{\prime\prime}$$\times$1.28$^{\prime\prime}$, 2.07$^{\prime\prime}$$\times$1.19$^{\prime\prime}$, and 2.13$^{\prime\prime}$$\times$1.18$^{\prime\prime}$ at frequency ranges of 127.61--129.49 GHz, 129.56--131.44 GHz, 139.67--141.54 GHz, and 141.56--143.43 GHz, respectively. The extracted molecular spectra of NIRS3 are shown in Figure~\ref{fig:Fullspectra}. The systemic velocity (V$_{LSR}$) of the molecular spectra of NIRS3 is 5.0 km s$^{-1}$ \citep{liu20}. After spectral extraction, we used CASSIS software for molecular identification and spectral analysis using the local thermodynamic equilibrium (LTE) model with an optically thin assumption with the JPL, CDMS, and LSD molecular databases \citep{vas15, pic98, mu05, mot25}. The assumption of LTE is valid in the inner region of NIRS3, where the gas density is approximately 3$\times$10$^{7}$ cm$^{-3}$ \citep{sz18}, comparable to the critical density of dense gas tracers such as C$^{34}$S($J$ = 7-6) ($\sim10^{7}$ cm$^{-3}$) (see \citet{zin20}). Moreover, previous studies have shown that the molecular lines detected toward NIRS3 are optically thin \citep{liu20, be23}, indicating that the molecular energy levels are effectively thermalized under LTE conditions. Following the spectral analysis, we detected six individual N-bearing molecules, such as \ce{CH3CN}, \ce{C2H3CN}, \ce{C2H5CN}, \ce{NH2CN} (both $\nu = 0$ and 1), \ce{NH2CHO}, and \ce{HC3N} ($\nu_{7}$ = 2), respectively. Additionally, we detected several O-bearing molecules, including \ce{CH3OH}, toward NIRS3, and a detailed physical and chemical analysis of these species will be presented in a follow-up study.

To determine the column densities ($N$ in cm$^{-2}$) and excitation temperatures ($T_{ex}$ in K) of the detected N-bearing molecules, we use the LTE model spectra. The brightness temperature ($T_{b}$) of the molecular spectra is estimated by CASSIS according to the following expression:

\begin{equation}
	T_{b} = T_{C}e^{-\tau} +(1-e^{-\tau})(J_{\nu}(T_{ex})- J_{\nu}(CMB))
	\label{eq:tb}
\end{equation}
where $T_{C}$ is the continuum temperature, $\tau$ is the opacity, $J_{\nu}(T)$ $ = (h\nu/k)\times1/(e^{h\nu/k T}-1)$ is the radiation temperature, and CMB is the cosmic microwave background at 2.7 K \citep{vas15}. We used the MCMC algorithm in the CASSIS software for fitting LTE-modelled spectra of molecules over the observed spectra. The MCMC method is initialised by randomly selecting a seed point ($X_0$) in the four-dimensional parameter space. Then, it randomly chooses a nearby point ($X_1$) based on a variable step size, which is recalculated for each iteration. The $\chi^{2}$ value of the new state ($X_1$) is calculated, and if the ratio $p$ = $\chi^{2}$($X_0$)/$\chi^{2}$($X_1$) $>$ 1, the new state is accepted. However, even if $p$ $<$ 1, the new state may still be accepted with a certain probability. If the new state is rejected, the original state ($X_0$) remains, and another nearby point ($X_1$) is randomly selected. By allowing a finite probability of accepting a worse $\chi^{2}$ value, the algorithm avoids converging directly to a local minimum and instead ensures a more thorough exploration of the entire parameter space. During our MCMC analysis, we employed 1000 walkers, uniformly distributed within the specified parameter ranges, which are shown in Table~\ref{tab:distri}, and ran the chains for a burning sequence of 20,000 steps to ensure convergence. The MCMC approach allows us to vary all parameters simultaneously, including column density, excitation temperature, full width at half maximum (FWHM), and V$_{LSR}$ until the best fit is obtained. CASSIS also calculated the partition functions of the detected molecules using the following expression:

\begin{equation}
	Q(T) = \sum_{i} g_{i} \times exp(-E_{i}/kT)
\end{equation}
where $g_{i}$ and $E_{i}$ represent the statistical weight and energy, respectively, of the $i$th energy level. In the LTE model, we used a source size of $\sim$2.32$^{\prime\prime}$ (half-maximum diameter of our synthesized beam size). After LTE spectral analysis, we determine the column density, excitation temperature, and FWHM of the identified molecules. We estimate the fractional abundances of detected molecules by using the column densities of detected molecules and dividing them by the \ce{H2} column density found in Section~3.1.1. We also estimate the abundances of the detected molecules with respect to \ce{CH3OH}. Since several high-intensity transitions of \ce{CH3OH} are detected in the ALMA band 4 data used in this study (see Figure~\ref{fig:methanol}), we performed LTE spectral modelling to determine its physical parameters. The resulting column density and excitation temperature of \ce{CH3OH} towards NIRS3 are $(9.52 \pm 0.62)\times10^{16}$ cm$^{-2}$ and $250 \pm 24$ K, respectively. We find that the column density and excitation temperature derived for \ce{CH3OH} in this work are consistent with those reported in previous studies by \citet{zin15}. The spectral lines and physical parameters of detected N-bearing species are shown in Tables~\ref{tab:MOLECULAR DATA} and \ref{tab:LTE}. The LTE-fitted spectral lines using the MCMC approach and posterior probability distribution of each parameter of detected N-bearing species are shown in Figures~\ref{fig:spectra} and \ref{fig:corner}. The following subsections provide a thorough discussion of the individual characteristics of the identified N-bearing molecules.

\subsubsection{Methyl cyanide (\ce{CH3CN})}
Methyl cyanide (\ce{CH3CN}) is a well-known, strongly prolate, symmetric top molecule. The rotational energy levels (or $J$ transitions) of \ce{CH3CN} are divided into successive $K$ components. The emission lines of \ce{CH3CN} are shifted in frequency with respect to each other due to different $K$ values; this is known as the `$K$ ladder' spectrum. Due to its $K$ ladder, the \ce{CH3CN} molecule can determine the proper temperature of interstellar gas and is hence known as a gas thermometer for interstellar gas. From the spectra of NIRS3, we have detected the rotational emission lines of \ce{CH3CN} with transition $J$ = 7--6 with $K$ = 0--6. A list of molecular lines and their spectroscopic parameters for \ce{CH3CN} was taken from CDMS based on \cite{mu09}. Previously, \citet{wan11} and \citet{liu20} detected the emission lines of \ce{CH3CN} from NIRS3, and this is a re-detection of this molecule in this source using the ALMA band 4. Based on LTE spectral fitting, the column density, excitation temperature, and FWHM of \ce{CH3CN} are ($1.5\pm0.7)\times10^{15}$ cm$^{-2}$, $180\pm26$ K, and 5.5 $\pm$ 0.4 km s$^{-1}$. The fractional abundance of \ce{CH3CN} with respect to \ce{H2} is ($1.4\pm0.7)\times10^{-9}$. Similarly, the fractional abundance of \ce{CH3CN} with respect to \ce{CH3OH} is $(1.6\pm0.8)\times10^{-2}$. After the detection of emission lines of \ce{CH3CN} ($\nu$ = 0), we also searched the emission lines of \ce{CH3CN} ($\nu_{8}$ = 1). After spectral analysis, we notice that all detected emission lines of \ce{CH3CN} ($\nu_{8}$ = 1) are blended with different O-bearing molecules. Assuming an excitation temperature of 180 K, the upper-limit column density of \ce{CH3CN} ($\nu_{8}$ = 1) is $\leq(2.4\pm0.2) \times 10^{14}$ cm$^{-2}$.

\begin{figure*}
	\centering
	\includegraphics[width=0.98\textwidth]{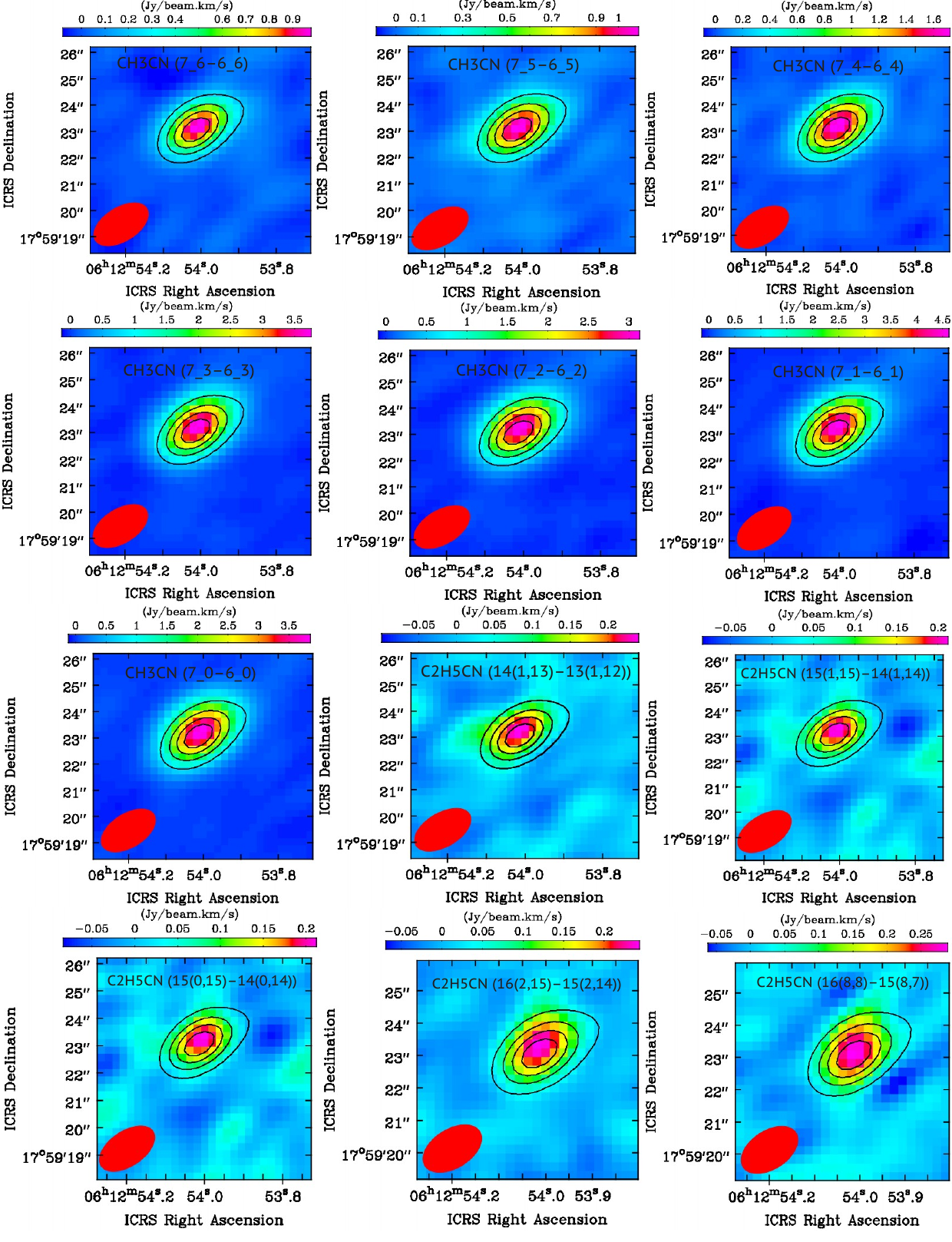}
	\caption{Integrated emission maps (moment zero) of non-blended transitions of \ce{CH3CN}, \ce{C2H5CN}, \ce{C2H3CN}, \ce{NH2CN}, \ce{NH2CHO}, and \ce{HC3N} ($\nu_{7}$ = 2) towards NIRS3. The emission maps of detected N-bearing species are overlaid with the 2.29 mm continuum emission map (black contour). The contour levels are at 20\%, 40\%, 60\%, and 80\% of the peak flux. The red circles represent the synthesised beams of the emission maps.}
	\label{fig:emi}
\end{figure*}
\begin{figure*}
	\text{{\large Figure~8 continued.}}
	\centering
	\includegraphics[width=1.0\textwidth]{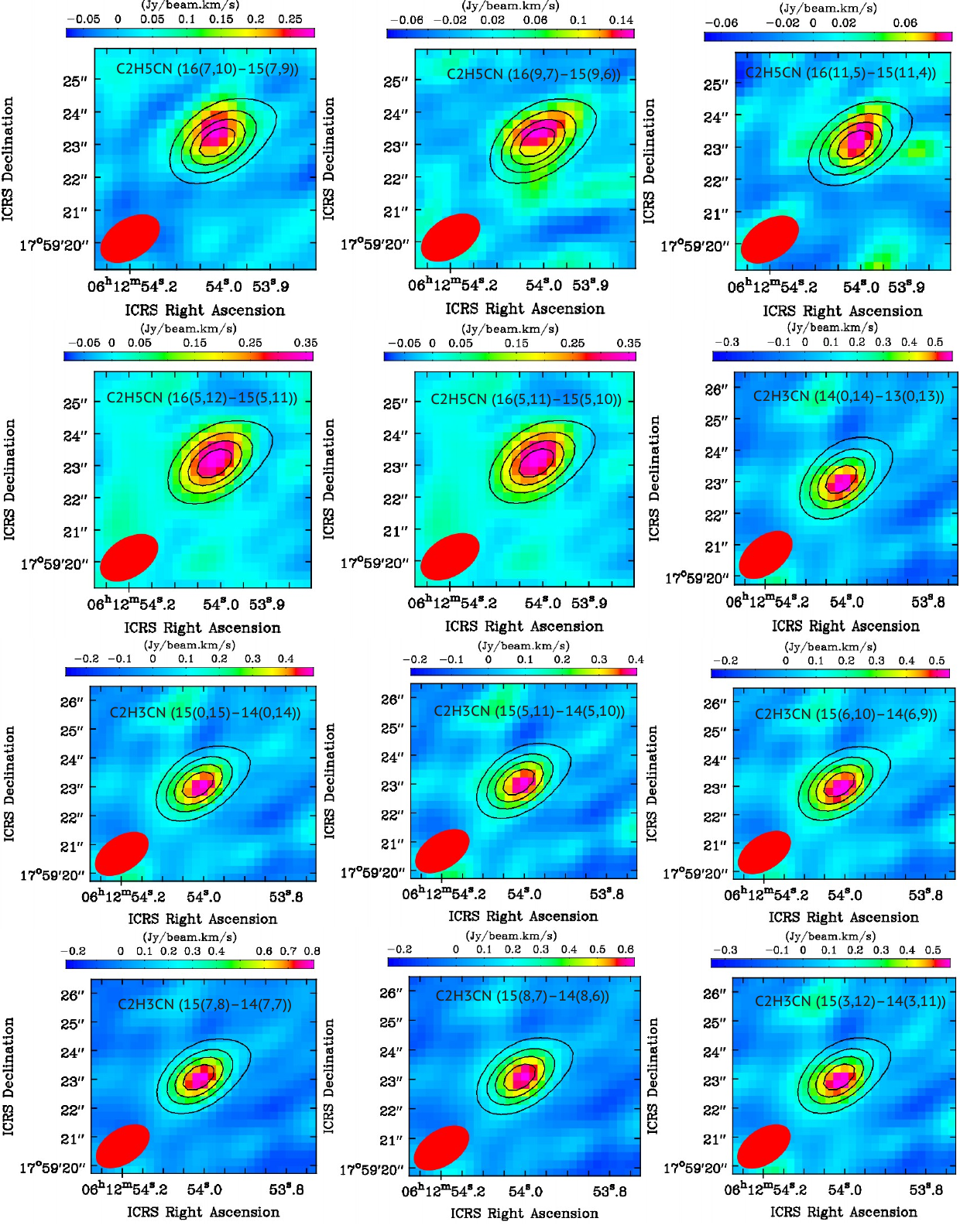}
\end{figure*}
\begin{figure*}
	\text{{\large Figure~8 continued.}}
	\centering
	\includegraphics[width=0.95\textwidth]{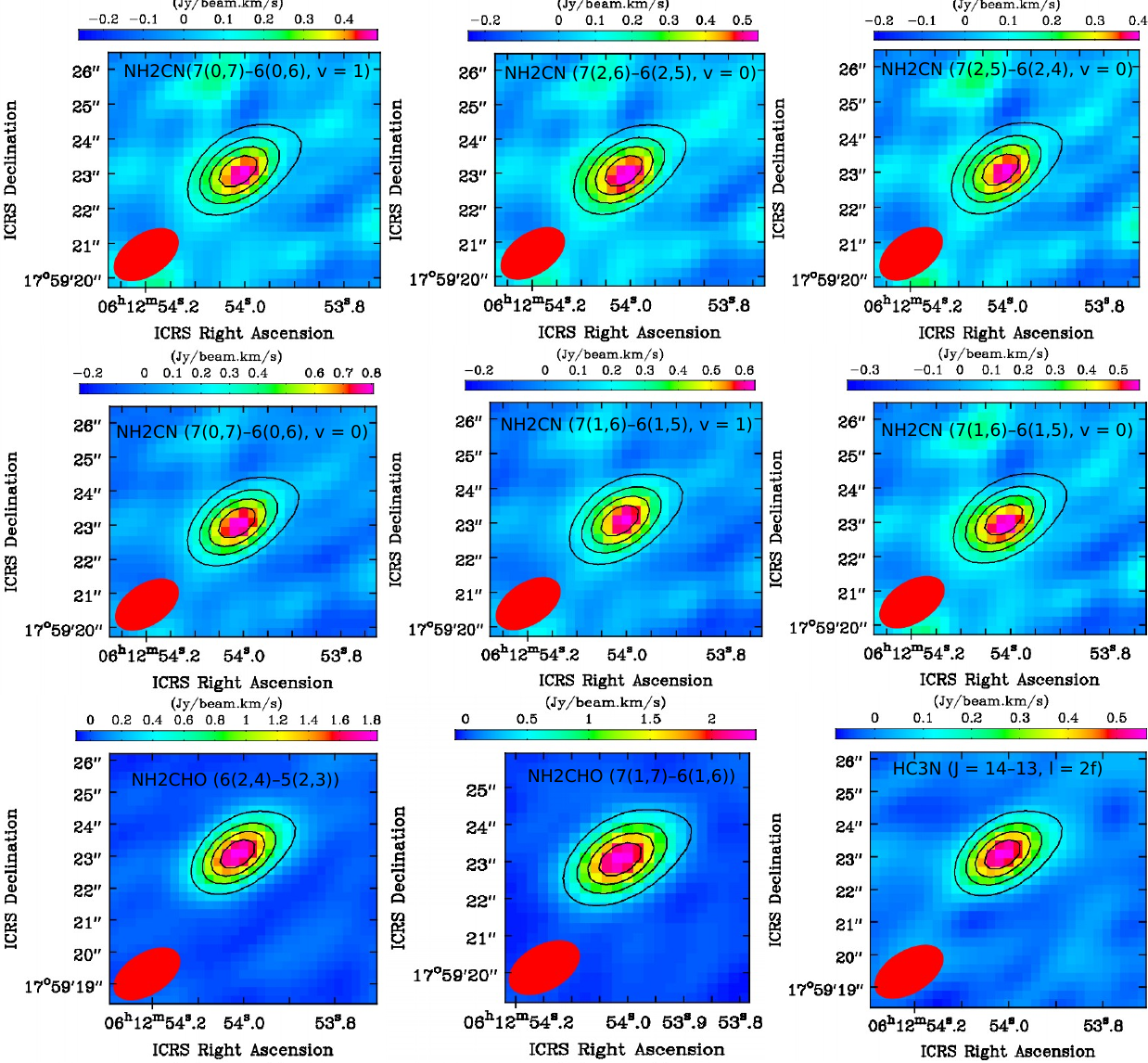}
\end{figure*}

\subsubsection{Ethyl cyanide (\ce{C2H5CN})}	
Ethyl cyanide (\ce{C2H5CN}) is one of the known asymmetric top molecules whose dipole moments are $\mu_{a}$ = 3.85 debye and $\mu_{a}$ = 1.23 debye \citep{he74}. After spectral analysis, we detected a total of 13 transition lines of \ce{C2H5CN} towards NIRS3. To our knowledge, this is the first detection of this molecule in NIRS3. The molecular line lists and laboratory spectroscopic parameters for \ce{C2H5CN} are taken from CDMS based mainly on \citet{br09}. We find that the column density, excitation temperature, and FWHM of \ce{C2H5CN} are $(3.8\pm1.2)\times10^{14}$ cm$^{-2}$, $178\pm35$ K, and $5.5\pm0.8$ km s$^{-1}$, respectively. The fractional abundance of \ce{C2H5CN} with respect to \ce{H2} is (3.5$\pm$1.3)$\times$10$^{-10}$. Similarly, the fractional abundances of \ce{C2H5CN} with respect to \ce{CH3OH} and \ce{CH3CN} are $(4.0\pm1.3)\times10^{-3}$ and $(2.5\pm1.4)\times10^{-1}$. 

\subsubsection{Vinyl cyanide (\ce{C2H3CN})}
Vinyl cyanide (\ce{C2H3CN}) is a planar asymmetric top molecule whose dipole moments are $\mu_{a}$ = 3.82 debye and $\mu_{b}$ = 0.68 debye \citep{st85}. We have detected a total of 14 transition lines of \ce{C2H3CN} from the NIRS3 spectra. We found that all non-blended transitions were detected with a signal-to-noise ratio of 3$\sigma$ ($\sigma\sim0.05$ K)\footnote{The rms ($\sigma$) of the spectra was estimated using the Python function {\tt numpy.std}, which computes the standard deviation across line-free regions of the spectra.}, indicating statistically significant detections. The rest frequencies, transitions, and other spectroscopic parameters for \ce{C2H3CN} are taken from CDMS based on \citet{mu08}. The derived column density, excitation temperature, and FWHM of \ce{C2H3CN} are $(2.1\pm0.6)\times10^{14}$ cm$^{-2}$, $176\pm12$ K, and $5.3\pm0.7$ km s$^{-1}$, respectively. The fractional abundance of \ce{C2H3CN} with respect to \ce{H2} is $(1.9\pm0.7)\times10^{-10}$. The fractional abundances of \ce{C2H3CN} with respect to \ce{CH3OH} and \ce{CH3CN} are $(2.2\pm0.7)\times10^{-3}$ and $(1.4\pm0.8)\times10^{-1}$, respectively.

\begin{table}{}
\scriptsize
\centering
\caption{Emitting regions of N-bearing species towards NIRS3.}
\begin{tabular}{cccccccccccc}
\hline
Molecule&Frequency&Transition& E$_{up}$&Emitting region\\
		&(GHz)& &(K)&($^{\prime\prime}$)\\
\hline
\ce{CH3CN}&128.690& 7$_{6}$--6$_{6}$ &281.79&2.26$\pm$0.26\\
		&128.717& 7$_{5}$--6$_{5}$ &203.28&2.27$\pm$0.31\\
		&128.739& 7$_{4}$--6$_{4}$ &139.02&2.27$\pm$0.28\\
		&128.757& 7$_{3}$--6$_{3}$ &89.02&2.27$\pm$0.32\\
		&128.769& 7$_{2}$--6$_{2}$ &53.30&2.28$\pm$0.24\\
		&128.776& 7$_{1}$--6$_{1}$ &31.87&2.28$\pm$0.34\\
		&128.779& 7$_{0}$--6$_{0}$ &24.72&2.31$\pm$0.22\\
\hline
\ce{C2H5CN}& 127.618&14(1,13)--13(1,12)&47.23&2.31$\pm$0.22\\
		& 129.795&15(1,15)--14(1,14)&51.09&2.29$\pm$0.38\\
		& 130.903&15(0,15)--14(0,14)&50.79&2.30$\pm$0.25\\
		& 142.346&16(2,15)--15(2,14)&62.69&2.27$\pm$0.32\\
		& ~~143.335$^{*}$&16(8,8)--15(8,7)&129.58&2.29$\pm$0.36\\
		& ~~143.337$^{*}$&16(7,10)--15(7,9)&112.93&2.28$\pm$0.29\\
		& ~~143.343$^{*}$&16(9,7)--15(9,6)&148.44&2.30$\pm$0.31\\
		& ~~143.383$^{*}$&16(11,5)--15(11,4)&192.77&2.29$\pm$0.32\\
		& 143.406&16(5,12)--15(5,11)&86.28&2.27$\pm$0.26\\
		& 143.407&16(5,11)--15(5,10)&86.28&2.29$\pm$0.35\\
\hline
\ce{C2H3CN}&131.267&14(0,14)--13(0,13)&47.50&2.29$\pm$0.18\\
		&140.429&15(0,15)--14(0,14)&54.24&2.30$\pm$0.21\\
		&~~142.399$^{*}$&15(5,11)--14(5,10)&108.70&2.26$\pm$0.25\\
		&142.401&15(6,10)--14(6,9)&132.44&2.28$\pm$0.18\\
		&~~142.419$^{*}$&15(7,8)--14(7,7)&160.44&2.27$\pm$0.28\\
		&~~142.447$^{*}$&15(8,7)--14(8,6)&192.71&2.30$\pm$0.21\\
		&142.573&15(3,12)--14(3,11)&74.16&2.29$\pm$0.27\\
\hline
\ce{NH2CN}&139.842&7(0,7)--6(0,6), $v$ = 1&98.16&2.22$\pm$0.32\\
		&139.931&7(2,6)--6(2,5), $v$ = 0&84.83&2.21$\pm$0.28\\
		&139.940&7(2,5)--6(2,4), $v$ = 0&84.83&2.21$\pm$0.21\\
		&139.954&7(0,7)--6(0,6), $v$ = 0&26.87&2.20$\pm$0.18\\
		&140.710&7(1,6)--6(1,5), $v$ = 1&112.46&2.22$\pm$0.25\\
		&140.877&7(1,6)--6(1,5), $v$ = 0&41.54&2.21$\pm$0.30\\	
\hline
\ce{NH2CHO}&128.103&6(2,4)--5(2,3)&33.38&2.25$\pm$0.22\\
		&142.701&7(1,7)--6(1,6)&30.41&2.24$\pm$0.23\\
		\hline
\ce{HC3N} ($\nu_{7}$ = 2)&128.169&$J$ = 14--13, $l$ = 0&687.83&2.27$\pm$0.28\\
		&128.175&$J$ = 14--13, $l$ = 2e&691.10&2.28$\pm$0.36\\
		&128.182&$J$ = 14--13, $l$ = 2f&691.10&2.28$\pm$0.32\\	
\hline
\end{tabular}		
*--These transitions contain double with frequency difference  $\leq$100 kHz. The second transition is not shown.\\
\label{tab:emittingregion}
\end{table}

\begin{figure}
	\centering
	\includegraphics[width=0.48\textwidth]{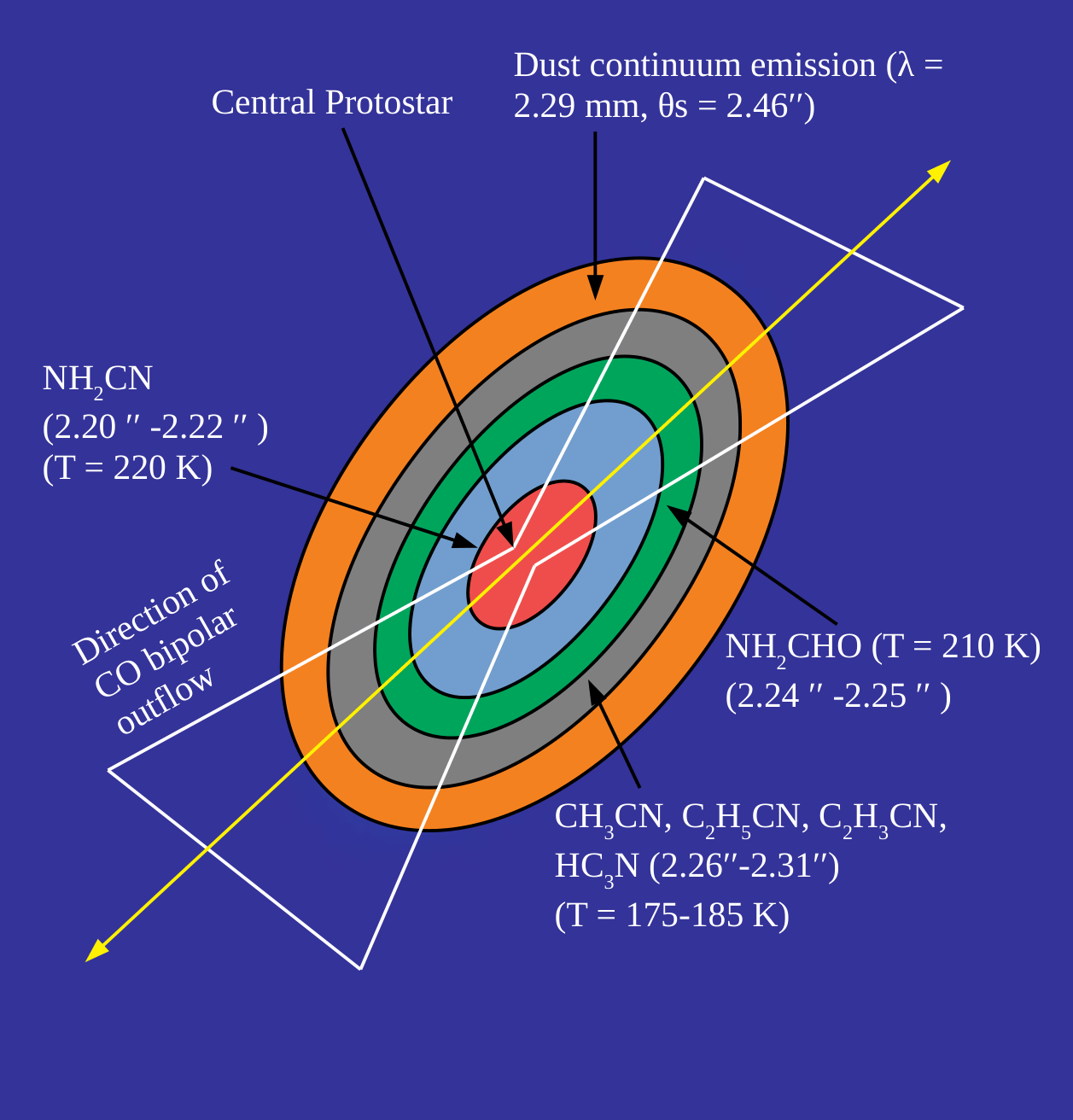}
	\caption{Cartoon illustration (not up to scale) of molecular distribution of detected N-bearing molecules towards NIRS3 based on estimated source sizes and temperatures. The direction of CO ($J$ = 3--2) outflow is obtained from \citet{zin20}.}
	\label{fig:diagram}
\end{figure}

\begin{figure*}
	\centering
	\includegraphics[width=0.95\textwidth]{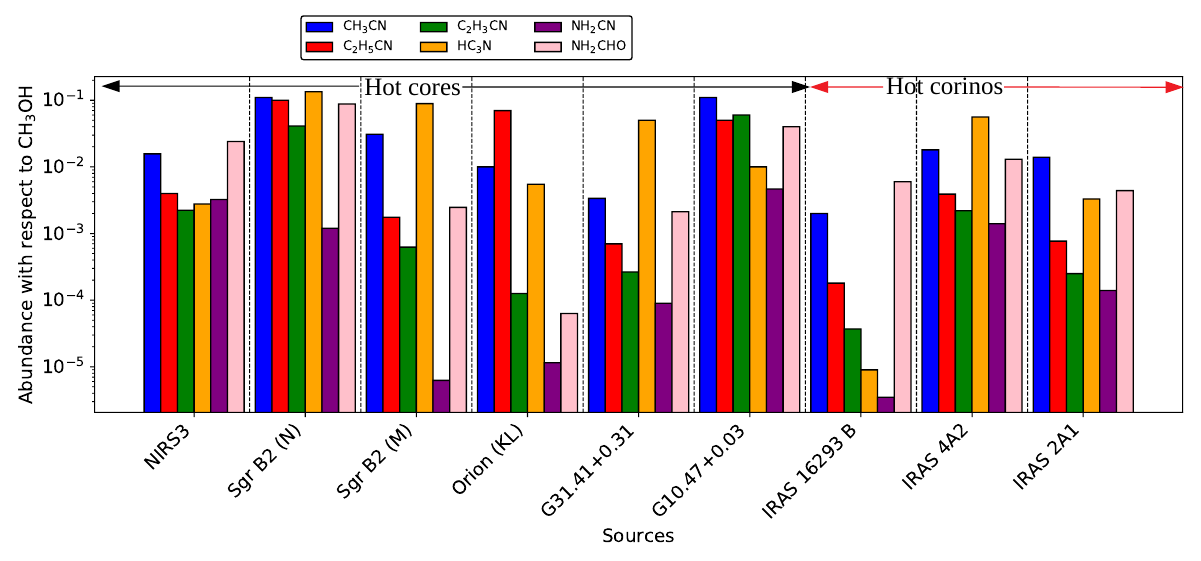}
	\includegraphics[width=0.95\textwidth]{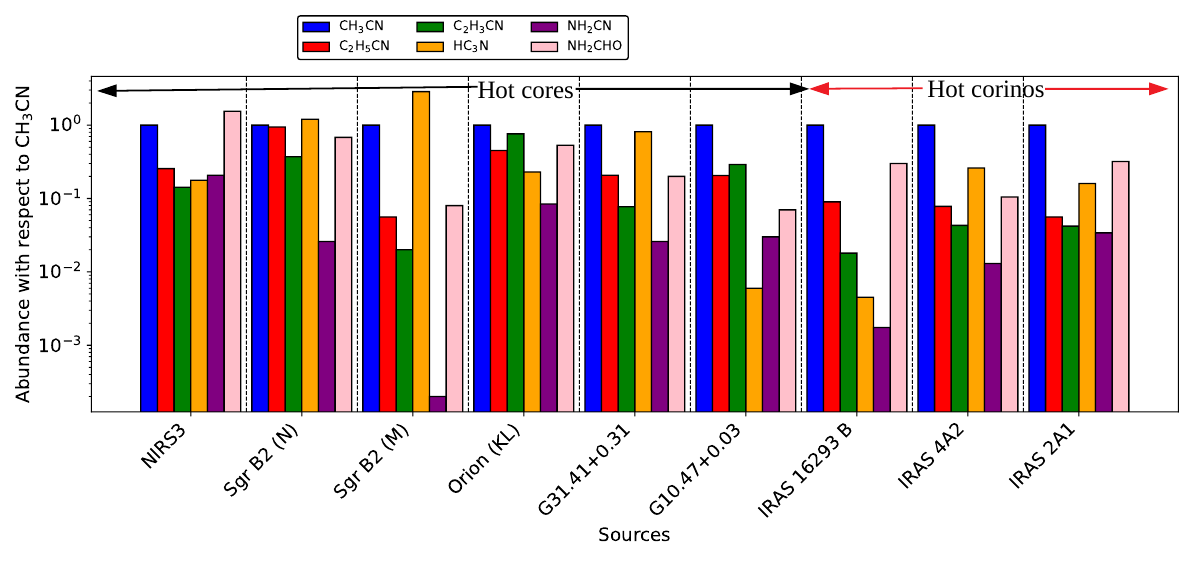}
	\caption{Comparison of the abundances of detected N-bearing molecules relative to \ce{CH3OH} (upper panel) and \ce{CH3CN} (lower panel) towards NIRS3, with different hot cores and hot corinos.}
	\label{fig:bar}
\end{figure*}

\subsubsection{Cyanamide (\ce{NH2CN}) and formamide (\ce{NH2CHO})}
Cyanamide (\ce{NH2CN}) and formamide (\ce{NH2CHO}) are amide-like molecules found in both high- and low-mass star-formation regions \citep{bel13, cou18, man24c}. A previous study showed that both \ce{NH2CN} and \ce{NH2CHO} play key roles in the formation of several biomolecules, including amino acids, in star-forming regions \citep{lig20}. Analysis of the observed spectra depicted a total of 6 and 2 rotational emission lines of \ce{NH2CN} (both $\nu = 0$ and 1) and \ce{NH2CHO} towards NIRS3, respectively, above 4$\sigma$ significance. The molecular line lists and rotational spectroscopic parameters for \ce{NH2CN} and \ce{NH2CHO} are taken from CDMS based on \cite{re86} and \cite{ku71}. This is the first detection of \ce{NH2CN} in this source. Previously, \cite{liu20} also detected the emission lines of \ce{NH2CHO} from NIRS3 for which $E_{up}$ varies between 153 K and 568.5 K, but all detected lines of \ce{NH2CHO} were blended. That might have refrained \cite{liu20} from the estimation of the column density and excitation temperature of \ce{NH2CHO}. We estimated that the column density, excitation temperature, and FWHM of \ce{NH2CN} are $(3.1\pm0.9)\times10^{14}$ cm$^{-2}$, 220 $\pm$ 18 K, and 5.2 $\pm$ 0.6 km s$^{-1}$, respectively. Similarly, the estimated column density, excitation temperature, and FWHM of \ce{NH2CHO} are $(2.3\pm0.8)\times10^{15}$ cm$^{-2}$, 210$\pm$22 K, and 5.1 $\pm$ 0.4 km s$^{-1}$, respectively. The fractional abundances of \ce{NH2CN} and \ce{NH2CHO} with respect to molecular \ce{H2} are $(2.8\pm1.0)\times10^{-10}$ and (2.1 $\pm$ 0.9)$\times$10$^{-9}$, respectively. The fractional abundances of \ce{NH2CN} and \ce{NH2CHO} with respect to \ce{CH3OH} are $(3.3\pm1.0)\times10^{-3}$ and $(2.4\pm0.9)\times10^{-2}$, respectively. The fractional abundances of \ce{NH2CN} and \ce{NH2CHO} with respect to \ce{CH3CN} are $(2.1\pm1.2)\times10^{-1}$ and 1.5 $\pm$ 0.9, respectively. Similarly, the column density ratio of \ce{NH2CN} and \ce{NH2CHO} (hereafter \ce{NH2CN}/\ce{NH2CHO}) for NIRS3 is 0.13 $\pm$ 0.06. The detected two low-energy transitions of \ce{NH2CHO} cannot robustly constrain the proper excitation temperature. A detailed study of \ce{NH2CHO} over a broad spectral range is needed to properly constrain its excitation temperature and improve our understanding of its physical conditions. We also searched the emission lines of \ce{NH2CHO}, $\nu_{12}$ = 1, but we did not detect that molecule in the spectra of NIRS3. Assuming an excitation temperature of 210 K, the upper-limit column density of \ce{NH2CHO}, $\nu_{12}$ = 1 is $\leq(2.35\pm0.82)\times10^{12}$ cm$^{-2}$.

\subsubsection{Cyanoacetylene (\ce{HC3N}, $\nu_{7}$ = 2)}
Cyanoacetylene (\ce{HC3N}) is the shortest cyanopolyyne (HC$_{2n+1}$N, where $n$ = 1, 2, 3, 5, 7, ..) molecule, which has so far mainly been found in the warm inner region of the hot cores and hot corinos (see \cite{tan22} and \citet{ho25}). We detected 3 transition lines of \ce{HC3N} with a higher vibrational state $\nu_{7}$ = 2 in the spectra of NIRS3. The line lists and spectroscopic parameters for \ce{HC3N} ($\nu_{7}$ = 2) are taken from CDMS based on \cite{th00}. The higher vibrational state of \ce{HC3N} ($\nu_{7}$ = 2) exhibits a higher $E_{up}$ $\sim$ 687.83--691.10 K, which can trace the inner hot gas near the central star, i.e., the disk. The estimated column density, excitation temperature, and FWHM of \ce{HC3N} ($\nu_{7}$ = 2) are $(2.6\pm0.5)\times10^{14}$ cm$^{-2}$, 180 $\pm$ 16 K, and 5.4 $\pm$ 0.6 km s$^{-1}$, respectively. The fractional abundance of \ce{HC3N} ($\nu_{7}$ = 2) with respect to \ce{H2} is $(2.4\pm0.7)\times10^{-10}$. Similarly, the fractional abundances of \ce{HC3N} ($\nu_{7}$ = 2) with respect to \ce{CH3OH} and \ce{CH3CN} are $(2.8\pm0.6)\times10^{-3}$ and $(1.8\pm0.9)\times10^{-1}$. The column density, excitation temperature, and abundance of \ce{HC3N} ($\nu_{7}$ = 2) obtained here may be considered conservative estimates, as they are derived from a limited number of spectral lines. These values may vary slightly if a larger set of spectral lines is used for the estimation. Since we have detected only one rotational quantum number ($J$) of \ce{HC3N} ($\nu_{7}$ = 2) alongside various vibrational angular momentum quantum numbers ($l$), those lines can not reliably constrain the appropriate excitation temperature. A comprehensive study of \ce{HC3N} ($\nu$ = 0), \ce{HC3N} ($\nu_{7}$ = 2), and other higher vibrational states across a wide spectral range is required to accurately determine its excitation temperature and enhance our understanding of its physical conditions.

\subsubsection{Study of N-bearing possible glycine precursor molecules}
After the detection of different complex N-bearing molecules, we also searched the emission lines of possible N-bearing glycine (\ce{NH2CH2COOH}) precursor molecules such as aminocetonitrile (\ce{NH2CH2CN}), methylamine (\ce{CH3NH2}), and methanimine (\ce{CH2NH}) towards NIRS3. The molecular line lists and laboratory spectroscopic parameters for \ce{NH2CH2CN} are taken from CDMS based on \cite{pic73} and \cite{mel20}. Similarly, the spectroscopic parameters for \ce{CH3NH2} and \ce{CH2NH} are taken from the JPL database based on \cite{il05} and \cite{kir73}. After spectral analysis using the LTE-modelled spectra, we do not detect the emission lines of \ce{NH2CH2CN} and \ce{CH3NH2} in the spectra of NIRS3. We also observed that all detected emission lines of \ce{CH2NH} are blended with other molecules. Assuming an excitation temperature of 180 K (similar to that of \ce{CH3CN}), we derive the upper-limit column densities of $\leq(3.2\pm0.3)\times10^{13}$ cm$^{-2}$ for \ce{NH2CH2CN}, $\leq(4.5\pm0.2)\times10^{14}$ cm$^{-2}$ for \ce{CH2NH}, and $\leq(1.3\pm0.2)\times10^{14}$ cm$^{-2}$ for \ce{CH3NH2}.

\begin{figure*}
	\centering
	\includegraphics[width=0.5\textwidth]{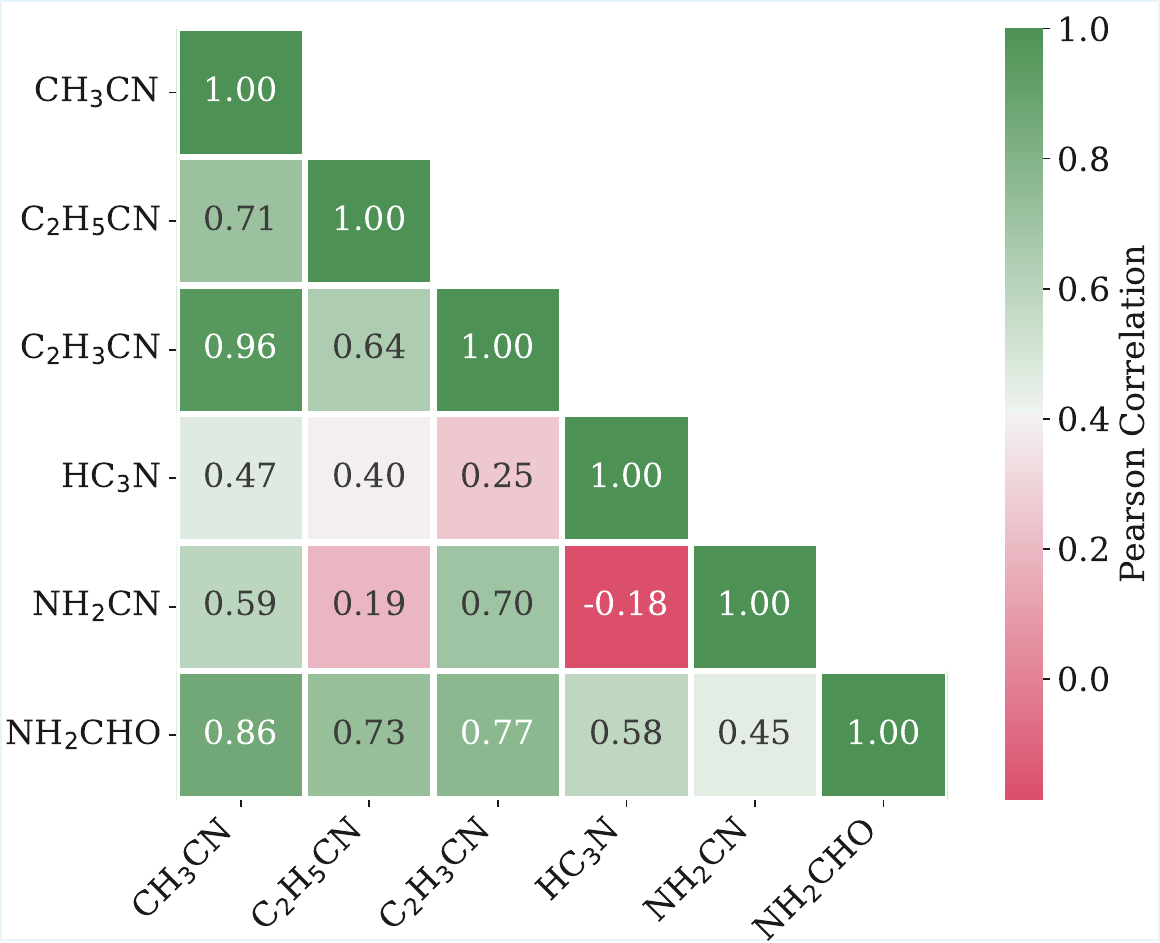}\includegraphics[width=0.5\textwidth]{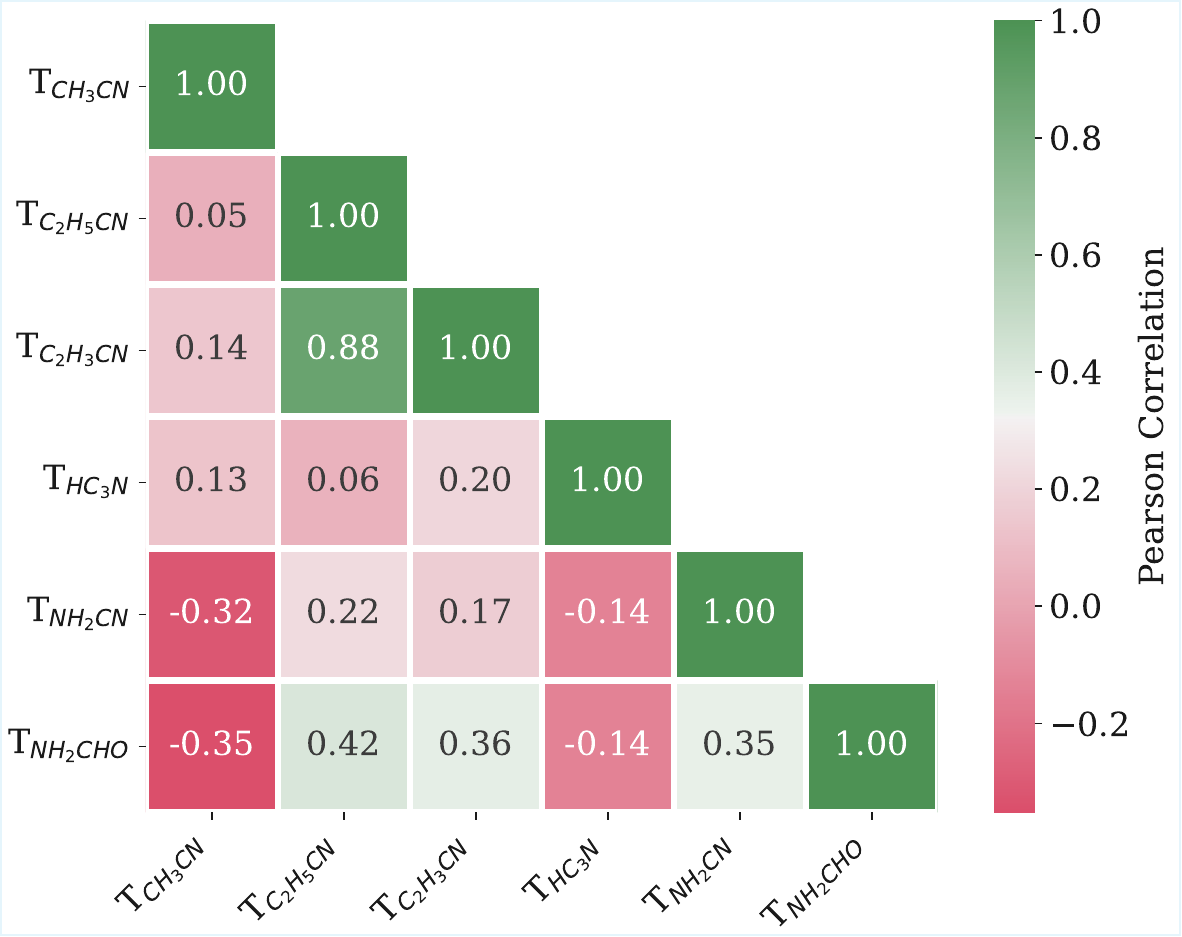}
	\caption{Pearson correlation coefficient heat maps of the abundances relative to \ce{CH3OH} (left panel) and excitation temperatures (right panel) of the identified N-bearing molecules. The colours show the level of correlation, which corresponds to the Pearson coefficient ($r$) shown in each box.}
	\label{fig:Heat}
\end{figure*}

\subsection{Spatial distribution of N-bearing molecules}
We created integrated emission maps (moment zero) of some highly intense emission lines \ce{CH3CN}, \ce{C2H5CN}, \ce{C2H3CN}, \ce{NH2CN}, \ce{NH2CHO}, and \ce{HC3N} ($\nu_{7}$ = 2) using the CASA task IMMOMENTS, as shown in Figure~\ref{fig:emi}.  During the making of the integrated emission maps of the detected N-bearing molecules, we used the channel ranges of the spectral data cubes in the IMMOMENTS task, where the emission lines of the N-bearing molecules were detected. We further fitted the 2D Gaussian over the emission maps using the task IMFIT to determine the source size of the N-bearing molecules. The emitting regions of detected N-bearing molecules are determined using the following equation:

\begin{equation}
	\theta_s=\sqrt{\theta^2_{50}-\theta^2_{\textrm{beam}}}
\end{equation}
where $\theta_{50}=2\sqrt{A/\pi}$ is the diameter of the circle whose area ($A$) encloses the $50\%$ line peak of the emission maps, and $\theta_{\textrm{beam}}$ is the half-power width of the synthesized beam of the emission maps \citep{riv17}. The derived emitting region sizes of different N-bearing molecules are shown in Table~\ref{tab:emittingregion}. 

It appeared that the emitting diameters of the different N-bearing molecular transitions are smaller than the synthesized beam sizes of the integrated emission maps, implying that the detected transitions of the N-bearing molecules are spatially unresolved. The emitting region of \ce{CH3CN} vary from 2.26$^{\prime\prime}$ to 2.31$^{\prime\prime}$, and we observe that the emitting regions of \ce{CH3CN} decrease with an increase of E$_{up}$. This means that \ce{CH3CN} could trace the temperature distribution of NIRS3, but these data can not be used to determine the NIRS3 temperature distributions owing to the low angular resolution. The estimated emitting regions of \ce{C2H5CN} and \ce{C2H3CN} vary from 2.27$^{\prime\prime}$ to 2.31$^{\prime\prime}$ and 2.26$^{\prime\prime}$ to 2.30$^{\prime\prime}$. Similarly, the estimated emitting region of \ce{HC3N} vary from 2.27$^{\prime\prime}$ to 2.28$^{\prime\prime}$. We observed that the emitting regions of \ce{C2H5CN}, \ce{C2H3CN}, and \ce{HC3N} are nearly the same as \ce{CH3CN}, which means that these molecules may be existed in the region of \ce{CH3CN}. The emitting regions of \ce{NH2CN} and \ce{NH2CHO} vary from 2.20$^{\prime\prime}$ to 2.22$^{\prime\prime}$ and 2.24$^{\prime\prime}$ to 2.25$^{\prime\prime}$. We also made a cartoon illustration to understand the molecular distributions of detected N-bearing species based on estimated source sizes and temperatures towards NIRS3, which is shown in Figure~\ref{fig:diagram}. We find that the integrated emission maps of all detected transitions of N-bearing molecules are very compact towards dust continuum emissions. We also note that a potential relationship between the emitting source size ($\theta$) and excitation temperature ($T_{ex}$) has been examined in previous studies (e.g., \citealt{bou22}). However, owing to the limited parameter space and small sample size in our dataset, we refrain from performing a quantitative analysis.

\subsection{Validation of estimated abundance relative to \ce{H2}}
The column density of \ce{H2} is estimated from the dust continuum emissions, and the beam sizes of the continuum and line images are the same. We noticed that the emitting regions of some molecules are the same as the line-emitting regions of the dust continuum. For example, the emitting region of the dust continuum at frequency 141.07 GHz is 2.28$^{\prime\prime}$ (see Table~\ref{tab:continuum}), and the emitting regions of \ce{C2H5CN}, \ce{C2H3CN}, and \ce{HC3N} with transition $J$ = 16(7,10)--15(7,9), $J$ = 15(6,10)---14(6,9), and $J$ = 14--13, $l$ = 2f are 2.28$^{\prime\prime}$ (see Table~\ref{tab:emittingregion}). Therefore, we can say that \ce{H2} and some transitions of different molecules are emitted in the same regions. So, the estimation of the abundance relative to \ce{H2} is valid for this study.

\section{Discussion}
\label{sec:dis}
\subsection{Molecular correlation with other sources}
In this section, we compare the abundances of the detected N-bearing molecules relative to \ce{CH3OH} and \ce{CH3CN} between NIRS3 and other sources. We also examine the molecular correlations of the abundances relative to \ce{CH3OH} and excitation temperature using Pearson correlation coefficient heat maps. For the comparison of chemical abundances, we used only those hot core and hot corino sources where the emission lines of \ce{CH3CN}, \ce{C2H3CN}, \ce{C2H5CN}, \ce{HC3N}, \ce{NH2CN}, and \ce{NH2CHO} were detected. In this study, we did not use the abundance of N-bearing molecules relative to \ce{H2} because of the challenges in accurately determining the column density of \ce{H2} from the dust continuum or the CO column density. Instead, we present the abundance of the detected N-bearing species relative to \ce{CH3OH} and \ce{CH3CN}. We compare the abundances of detected N-bearing species in NIRS3 with several hot core and hot corino sources, such as Sgr B2 (N) \citep{bel13}, Sgr B2 (M) \citep{bel13, bel16}, Orion KL \citep{fri08, lo14, mot12, bell14, pen17, pag17}, G31.41+0.31 \citep{min23, wy99, man24c}, G10.47+0.03 \citep{rof11, han14}, IRAS 16293--2422 B \citep{cou18, cal18}, IRAS 4A2 \citep{lo17, bel20}, and IRAS 2A1 \citep{bel20}. 

A bar diagram of the comparison of the detected N-bearing species towards NIRS3 and other sources is shown in Figure~\ref{fig:bar}. It can be seen that the abundances of N-bearing species vary in different hot cores and hot corinos, which may indicate that their chemical formation environments are different. The Galactic centre hot core Sgr B2 (N) generally shows higher abundances than NIRS3 for most of the detected species by approximately one order of magnitude or more. The molecular abundances in NIRS3 show notable variations when compared with other sources. 
	
Following our comparison shown in the upper panel of Figure~\ref{fig:bar}, we found that the abundance of \ce{CH3CN} relative to \ce{CH3OH} varies from $2\times10^{-3}$ in IRAS 16293--2422 B to $1.1\times10^{-1}$ in Sgr B2 (N) and G10.47+0.03, corresponding to a spread of nearly two orders of magnitude. The abundance of \ce{CH3CN} relative to \ce{CH3OH} in NIRS3 shows an intermediate value of $1.6\times10^{-2}$, comparable to Sgr B2 (M), Orion KL, IRAS 4A2, and IRAS 2A1, but higher by factors of $\sim$5--10 compared to G31.41+0.31 and IRAS 16293--2422 B, and lower by a factor of $\sim$7 compared to Sgr B2 (N). 
	
The abundance of \ce{C2H5CN} relative to \ce{CH3OH} spans almost three orders of magnitude, from $1.8\times10^{-4}$ in IRAS 16293--2422 B to $1.0\times10^{-1}$ in Sgr B2 (N). The abundance of \ce{C2H5CN} relative to \ce{CH3OH} in NIRS3 ($4.0\times10^{-3}$) is comparable to Sgr B2 (M) and IRAS 4A2, higher by factors of $\sim$5--20 than G31.41+0.31, IRAS 16293--2422 B, and IRAS 2A1, but lower by factors of $\sim$10--25 than Orion KL, G10.47+0.03, and Sgr B2 (N). 
	
For \ce{C2H3CN}, the abundance relative to \ce{CH3OH} ranges from $3.7\times10^{-5}$ in IRAS 16293--2422 B to $6.0\times10^{-2}$ in G10.47+0.03, representing a spread of more than three orders of magnitude. The abundance of \ce{C2H3CN} relative to \ce{CH3OH} in NIRS3 ($2.2\times10^{-3}$) is comparable to IRAS 4A2, higher by one to two orders of magnitude than Orion KL, G31.41+0.31, and IRAS 16293--2422 B, but lower by factors of $\sim$20--30 than Sgr B2 (N) and G10.47+0.03. 
	
The abundance of \ce{HC3N} relative to \ce{CH3OH} exhibits one of the widest variations, from $9\times10^{-6}$ in IRAS 16293--2422 B to $1.3\times10^{-1}$ in Sgr B2 (N), spanning over four orders of magnitude. The abundance of \ce{HC3N} relative to \ce{CH3OH} in NIRS3 ($2.8\times10^{-3}$) is comparable to Orion KL and IRAS 2A1, higher than IRAS 16293--2422 B by nearly three orders of magnitude, but lower by factors of $\sim$10--50 than Sgr B2 (N), Sgr B2 (M), G31.41+0.31, G10.47+0.03, and IRAS 4A2. 
	
For \ce{NH2CN}, the abundance relative to \ce{CH3OH} ranges from $3.5\times10^{-6}$ to $4.7\times10^{-3}$, corresponding to nearly three orders of magnitude. The abundance of \ce{NH2CN} relative to \ce{CH3OH} in NIRS3 ($3.3\times10^{-3}$) is among the most enriched sources, exceeded only slightly by G10.47+0.03, and higher by factors of $\sim$10--10$^{3}$ than most other sources. Similarly, the abundance of \ce{NH2CHO} relative to \ce{CH3OH} spans from $6.3\times10^{-5}$ in Orion KL to $8.8\times10^{-2}$ in Sgr B2 (N), covering more than three orders of magnitude. The abundance of \ce{NH2CHO} relative to \ce{CH3OH} in NIRS3 ($2.4\times10^{-2}$) ranks among the highest values, exceeded only by Sgr B2 (N) and G10.47+0.03, and is typically higher by factors of $\sim$5--300 than those observed in other hot cores and hot corinos. Overall, these comparisons show that NIRS3 exhibits moderate to high abundances of N-bearing molecules relative to \ce{CH3OH}. While it does not reach the extreme chemical richness observed in Sgr B2 (N) or G10.47+0.03, it is significantly more enriched than typical hot corino sources.

Along with \ce{CH3OH}, we also compare the observed abundances of detected N-bearing molecules relative to \ce{CH3CN}, which is shown in the lower panel in Figure~\ref{fig:bar}. When abundances are normalized to \ce{CH3CN}, the source-to-source variations generally span 1--3 orders of magnitude, highlighting differences in cyanide chemistry efficiency. The abundance of \ce{C2H5CN} relative to \ce{CH3CN} varies from $5.6\times10^{-2}$ to $9.0\times10^{-2}$ in hot corinos to $9.4\times10^{-1}$ in Sgr B2 (N). The \ce{C2H5CN}/\ce{CH3CN} abundance ratio in NIRS3 $(2.5\times10^{-1})$ is comparable to G31.41+0.31 and G10.47+0.03, lower than Sgr B2 (N) and Orion KL by factors of 2--4, but higher than Sgr B2 (M) and hot corinos by factors of 3--5. For \ce{C2H3CN}/\ce{CH3CN} abundance ratios range from $\sim1.8\times10^{-2}$ to $7.6\times10^{-1}$. The \ce{C2H3CN}/\ce{CH3CN} abundance ratio in NIRS3 $(1.4\times10^{-1})$ lies in the middle of this distribution, lower than Sgr B2 (N), Orion KL, and G10.47+0.03 by factors of 2--5, but higher than Sgr B2 (M) and hot corinos by factors of 3--8. The \ce{HC3N}/\ce{CH3CN} ratio spans nearly three orders of magnitude, from $4\times10^{-3}$ in IRAS 16293--2422 B to 2.9 in Sgr B2 (M). The abundance of \ce{HC3N} relative to \ce{CH3CN} in NIRS3 $(1.8\times10^{-1})$ is lower than that in Sgr B2 (N), Sgr B2 (M), Orion KL, G31.41+0.31, and IRAS 4A2 by factors of $\sim1.3-16$, but higher than that in G10.47+0.03 and IRAS 16293--2422 B by factors of $\sim30$ and $\sim40$, respectively, and is comparable to that in IRAS 2A1. In contrast, \ce{NH2CN}/\ce{CH3CN} and \ce{NH2CHO}/\ce{CH3CN} show particularly strong enhancements in NIRS3. The \ce{NH2CN}/\ce{CH3CN} ratio ranges from $2\times10^{-4}$ to $\sim2.1\times10^{-1}$, with NIRS3 displaying the highest value among all sources. Similarly, \ce{NH2CHO}/\ce{CH3CN} varies from $7.0\times10^{-2}$ to 1.54, and NIRS3 again shows the maximum value, exceeding most sources by factors of 3--20. These results demonstrate that, although the cyanide-bearing species in NIRS3 are not as extreme as in Sgr B2 (N), the source is chemically unique in terms of its exceptionally high relative abundances of \ce{NH2CN} and \ce{NH2CHO}. This strongly suggests that the physical and chemical conditions in NIRS3, such as high temperatures, strong radiation fields, and efficient gas-grain interactions, favour the formation of certain N-bearing species through pathways that are less efficient in other hot cores and hot corinos.

Figure~\ref{fig:Heat} presents the correlation matrix heat map for Pearson's $r$ coefficients of abundances relative to \ce{CH3OH} and the excitation temperatures of the detected N-bearing species. The Pearson $r$ coefficient is estimated using the following equation:
\begin{equation}
r = \frac{\sum_{i=1}^{n} (X_{1,i} - \bar{X_1})(X_{2,i} - \bar{X_2})}{\sqrt{\sum_{i=1}^{n} (X_{1,i} - \bar{X_1})^2 \sum_{i=1}^{n} (X_{2,i} - \bar{X_2})^2}}
\label{eq:pear}
\end{equation}
where $r$ is calculated using the observed abundances of two species, $X_{1}$ and $X_{2}$, with $\bar{X}_{1}$ and $\bar{X}_{2}$ denoting their respective mean values. The correlation matrix heat map shows that some molecules exhibit strong positive correlations ($r \geq 0.7$) with each other. The choice of $r = 0.7$ as the threshold follows previous works, particularly \citet{naz22}, who applied the same criterion in their correlation analysis. We adopt this threshold to identify strong correlations. However, their study is based on a substantially larger sample ($>$50 sources), whereas our analysis includes only 9 sources. As the statistical robustness of a correlation depends sensitively on the number of data points, the same threshold does not correspond to an equivalent level of significance in the two studies. We found that \ce{HC3N} did not strongly correlate ($r$ $\leq$ 0.7) with any other N-bearing species. Similarly, we noticed that \ce{NH2CN} is strongly correlated with \ce{C2H3CN} but not strongly correlated with \ce{NH2CHO}, \ce{CH3CN}, and \ce{C2H5CN}. We also observe a negative correlation ($r$ = --0.18) between \ce{NH2CN} and \ce{HC3N}. There is a strong correlation ($r \geq 0.7$) between all three molecules in the cyanide family, \ce{CH3CN}, \ce{C2H3CN}, and \ce{C2H5CN}, indicating that there might be potential chemical connections between these molecules. Other observational studies also reported a strong correlation between these cyanide families \citep{naz22, ya21}. From the temperature correlation heat map, we observed that only the excitation temperature of \ce{C2H3CN} is strongly correlated with \ce{C2H5CN}, which indicates a strong chemical link between these two molecules of the CN family. We obtain strong negative correlations in the excitation temperature heat map of other molecules because there are too few samples for the simultaneous detection of these molecules.

\subsection{Prebiotic chemistry of detected N-bearing molecules}
Here, we focus on understanding the prebiotic chemistry of the detected N-bearing molecules. To understand the possible formation pathways of the detected N-bearing species, we compared the observed abundances with the existing astrochemical modelling abundances of these species. For the chemical modelling results, we followed the three-phase (gas + grain + icy mantle) warm-up chemical models of \citet{gar13} (hereafter, Model A) and \citet{suz18} (hereafter, Model B). \cite{gar13} used the MAGICKAL astrochemical code for three-phase warm-up chemical modelling in the context of hot cores. 

In the chemical model, \cite{gar13} considered an isothermal collapse stage followed by a static warm-up phase. In the isothermal collapse stage, the gas density increases from n$_{H}$ = $3\times10^{3}$ cm$^{-3}$ to $1\times10^{7}$ cm$^{-3}$, and the dust temperature falls to 8 K from 16 K. The second stage is known as the warm-up phase, where the gas density remains constant at $1\times10^{7}$ cm$^{-3}$ and the temperature increases from 8 to 400 K. \cite{gar13} estimated the modelled gas-phase abundances of the complex molecules over the time scales of $5\times10^{4}$ yr (fast warm-up), $2\times10^{5}$ yr (medium warm-up), and $1\times10^{6}$ yr (slow warm-up). 

Similarly, \cite{suz18} computed a three-phase warm-up chemical model in the context of hot cores and high-mass protostars using the gas-grain chemical kinetics code NAUTILUS. In the first stage (isothermal collapse phase), \citet{suz18} increased the gas density from $3\times10^{3}$ cm$^{-3}$ to $2\times10^{7}$ cm$^{-3}$, and the dust temperature fell to 8 K from 16 K. In the second stage (warm-up phase), the gas density remained fixed at $2\times10^{7}$ cm$^{-3}$. The dust temperature increased from 8 to 400 K. \cite{suz18} estimated the modelled abundances of different complex molecules in the gas-phase, grain surface, and icy mantle over fast ($7.12\times10^{4}$ yr) and slow ($1.43\times10^{6}$ yr) warm-up conditions. \citet{suz18} showed that the fast warm-up condition is most suitable for hot cores. 

Models A and B are suitable for NIRS3 because the gas density and temperature of NIRS3 are $\sim3\times10^{7}$ cm$^{-3}$ and 150 K, respectively, and the source contains a chemically rich hot molecular core \citep{sz18}. The comparison between the observed and modelled (both Model A and B) abundances of the detected N-bearing molecules is shown in Table~\ref{tab:model}. In the warm-up stages, \cite{gar13} estimated the peak modelled abundances in the gas-phase, whereas \cite{suz18} estimated the abundances of molecules in the gas-phase, grain surfaces, and in icy mantles. Since the observed abundances are in the gas-phase, we utilized the modelled gas-phase abundances of N-bearing molecules from \cite{suz18} during the comparison. The chemical formation pathways of the detected N-bearing molecules are discussed in the following subsections.

\begin{table*}{}
\centering
\caption{Comparison between observed and modelled abundances of different N-bearing molecules.}
\begin{tabular}{|c|c|c|c|c|c|c|}
\hline
Molecule & Observed value & \multicolumn{3}{c|}{Model A$^{(a)}$} & \multicolumn{2}{c|}{Model B$^{(b)}$} \\
		 & Abundance  & Fast & Medium & Slow & Abundance \\
\hline  
		
\ce{CH3CN}/\ce{H2} & (1.4$\pm$0.7)$\times$10$^{-9}$  & 4.9$\times$10$^{-9}$  & 2.1$\times$10$^{-9}$  & 1.3$\times$10$^{-9}$  & 6.2$\times$10$^{-9}$  \\
		
\hline
		
\ce{C2H5CN}/\ce{H2} & (3.5$\pm$1.3)$\times$10$^{-10}$  & 5.7$\times$10$^{-9}$  & 7.7$\times$10$^{-8}$  & 3.0$\times$10$^{-8}$  & 4.8$\times$10$^{-8}$  \\
		
\hline
		
\ce{C2H3CN}/\ce{H2} & (1.9$\pm$0.7)$\times$10$^{-10}$ & 1.1$\times$10$^{-8}$  & 5.6$\times$10$^{-9}$  & 5.6$\times$10$^{-9}$  & 3.9$\times$10$^{-9}$  \\
		
\hline
		
\ce{NH2CN}/\ce{H2} & (2.8$\pm$1.0)$\times$10$^{-10}$  & 2.3$\times$10$^{-8}$  & 3.6$\times$10$^{-8}$  & 3.3$\times$10$^{-8}$  & --  \\
		
\hline
		
\ce{NH2CHO}/\ce{H2} & (2.1$\pm$0.9)$\times$10$^{-9}$  & 3.9$\times$10$^{-7}$  & 1.5$\times$10$^{-7}$  & 2.9$\times$10$^{-9}$  & 1.8$\times$10$^{-6}$  \\
		
\hline
		
\ce{HC3N}/\ce{H2} & (2.4$\pm$0.7)$\times$10$^{-10}$  & --  & --  & --  & 2.6$\times$10$^{-9}$  \\
		
\hline
		
\ce{CH3CN}/\ce{CH3OH} & (1.6$\pm$0.8)$\times$10$^{-2}$  & 4.5$\times$10$^{-4}$  & 2.3$\times$10$^{-4}$  & 3.4$\times$10$^{-4}$  & 8.5$\times$10$^{-5}$  \\
		
\hline
		
\ce{C2H5CN}/\ce{CH3OH} & (4.0$\pm$1.3)$\times$10$^{-3}$  & 5.2$\times$10$^{-4}$  & 8.5$\times$10$^{-3}$  & 7.9$\times$10$^{-3}$  & 6.6$\times$10$^{-4}$  \\
		
\hline
		
\ce{C2H3CN}/\ce{CH3OH} & (2.2$\pm$0.7)$\times$10$^{-3}$  & 1.0$\times$10$^{-3}$  & 6.1$\times$10$^{-4}$  & 1.5$\times$10$^{-3}$  & 5.3$\times$10$^{-5}$  \\
		
\hline
		
\ce{NH2CN}/\ce{CH3OH} & (3.3$\pm$1.0)$\times$10$^{-3}$  & 2.1$\times$10$^{-3}$  & 3.9$\times$10$^{-3}$  & 8.7$\times$10$^{-3}$  & --  \\
		
\hline
		
\ce{NH2CHO}/\ce{CH3OH} & (2.4$\pm$0.9)$\times$10$^{-2}$  & 3.5$\times$10$^{-2}$  & 1.6$\times$10$^{-2}$  & 7.6$\times$10$^{-4}$  & 2.5$\times$10$^{-2}$  \\
		
\hline
		
\ce{HC3N}/\ce{CH3OH} & (2.8$\pm$0.6)$\times$10$^{-3}$  & --  & --  & --  & 3.6$\times$10$^{-5}$  \\
		
\hline
		
\ce{C2H5CN}/\ce{CH3CN} & (2.5$\pm$1.4)$\times$10$^{-1}$  & 1.2  & 36.7  & 23.1  & 7.8  \\
		
\hline
		
\ce{C2H3CN}/\ce{CH3CN} & (1.4$\pm$0.8)$\times$10$^{-1}$  & 2.2  & 2.7  & 4.3  & 6.2$\times$10$^{-1}$  \\
		
\hline
		
\ce{NH2CN}/\ce{CH3CN} & (2.1$\pm$1.2)$\times$10$^{-1}$  & 4.7  & 17.1  & 25.4  & --  \\
		
\hline
		
\ce{NH2CHO}/\ce{CH3CN} & 1.5$\pm$0.9  & 79.6  & 71.4  & 2.2  & 290.3  \\
		
\hline
		
\ce{HC3N}/\ce{CH3CN}&(1.8$\pm$0.9)$\times$10$^{-1}$ & -- & -- & -- & 4.1$\times$10$^{-1}$  \\
		
\hline
		
\ce{NH2CN}/\ce{NH2CHO} & 0.13$\pm$0.06  & 5.9$\times$10$^{-2}$  & 2.4$\times$10$^{-1}$  & 11.4  & --  \\
		
\hline
\end{tabular}
\label{tab:model}\\
(a) -- Values are taken from \citet{gar13}.\\
(b) -- The gas-phase abundance values are taken from \citet{suz18}.\\
\end{table*}

\subsubsection{Formation mechanism of \ce{CH3CN}}
Numerous formation pathways for \ce{CH3CN} exist in both the grain surface and gas-phase, yet recent quantum chemical studies have indicated that \ce{CH3CN} predominantly forms on the grain surface of HMCs \citep{gi23}. The potential formation pathways of \ce{CH3CN} on the surface of grains are outlined below: \\\\
\ce{CH3} + CN $\rightarrow$ \ce{CH3CN}~~~~~~~~~~~~~~~~~~~~~~~~~~~~~~~~~~~~~~~~~~~~~(1)\\\\
CN $\stackrel{\rm C}\rightarrow$ \ce{C2N} $\stackrel{\rm H}\rightarrow$\ce{HC2N} $\stackrel{\rm H}\rightarrow$ \ce{CH2CN} $\stackrel{\rm H}\rightarrow$ \ce{CH3CN}~~~~~~~~~~(2)\\\\
Reaction 1 was first proposed by \cite{pra80}, which shows that \ce{CH3CN} is created directly between the reaction of \ce{CH3} and CN on the grain surface, but the rate constants of this reaction are assumed and have not been studied experimentally. The formation of \ce{CH3CN} using reaction 1 on the grain surface is difficult because that reaction might have an activation energy barrier owing to the interaction of radicals with \ce{H2O}-ice molecules \citep{en19, en21, en22}. More importantly, the CN radical reacts with the \ce{H2O} molecules of ice, which shows that recombination reactions involving CN radicals on ice are not possible \citep{rim18}. Therefore, there is a low probability of the formation of \ce{CH3CN} using reaction 1 on the grain surface of the hot cores. 

Reaction 2 shows that \ce{CH3CN} is formed on the grain surface via hydrogenation of the \ce{CH2CN} \citep{gar13, gar17, suz18, gar22}. Similarly, \ce{CH2CN} is formed on the grain surface via the atom addition sequence of carbon (C) and hydrogen (H) with the CN radical \citep{gar13, gar17, suz18}. Previous studies show that the CN radical forms in the star-formation regions after the destruction of HCN in the gas-phase \citep{gar17}. The three-phase warm-up chemical models and quantum chemical studies have shown that reaction 2 is the most efficient for the formation of \ce{CH3CN} on the grain surface of hot cores, particularly Sgr B2 (N) \citep{gar13, gar17, suz18, bon19, gar22, gi23}.

To understand the possible formation pathways of \ce{CH3CN} towards NIRS3, we compared our estimated abundances of \ce{CH3CN} with the modelled abundances. We estimate that the abundance of \ce{CH3CN} with respect to \ce{H2} towards NIRS3 is $(1.4\pm0.7)\times10^{-9}$, which is similar to the slow warm-up modelled abundance of \citet{gar13} (Model A). The gas-phase modelled abundance of \cite{suz18} (Model B) is slightly higher than the observed abundance of \ce{CH3CN} towards NIRS3. The estimated abundance of \ce{CH3CN} with respect to \ce{CH3OH} towards NIRS3 is $(1.6\pm0.8)\times10^{-2}$, which is approximately two and three orders of magnitude higher than Models A and B, respectively. In the slow warm-up conditions, \cite{gar13} showed that reaction 2 is the dominant pathway for the production of \ce{CH3CN} on the grain surface. Since the observed abundance of \ce{CH3CN} with respect to \ce{H2} towards NIRS3 is similar to the slow warm-up modelled value, this indicates that \ce{CH3CN} may be formed via the hydrogenation of \ce{CH2CN} (reaction 2) on the grain surface of NIRS3. \cite{mc89} and \cite{gar22} showed that \ce{CH3CN} is destroyed via the radiative association reaction of \ce{CH3}$^{+}$ and \ce{CH3CN} when the source temperature is above 100 K. We estimate the excitation temperature of \ce{CH3CN} to be 180 K, indicating that the radiative association reaction of \ce{CH3}$^{+}$ and \ce{CH3CN} towards NIRS3 could potentially destroy \ce{CH3CN}, leading to more active cyanide chemistry.

\subsubsection{Formation mechanism of \ce{HC3N}}
There are many chemical reactions available that produce high \ce{HC3N} abundance \citep{bon19, gar22}. Previously, \citet{wak15} discussed \ce{HC3N} in the gas-phase, but the three-phase warm-up chemical models of \citet{tan19} and \citet{wil20} showed that \ce{HC3N} is formed on the grain surface of HMCs. The possible formation pathways of \ce{HC3N} in grain surface are described below: \\\\
\ce{C3N} + H $\longrightarrow$ \ce{HC3N} ~~~~~~~~~~~~~~~~~~~~~~(3)\\\\
\ce{C2H2} + CN $\longrightarrow$ \ce{HC3N} + H ~~~~~~~~~~~(4)\\\\
Reaction 3 obtained from \cite{wil20} shows that \ce{HC3N} is formed via the hydrogenation of \ce{C3N} on the grain surface, but this reaction was not verified using quantum chemical studies. We also noticed that the rate constants of reaction 3 are assumed. So, we are reluctant to use reaction 3 as the probable formation chain for \ce{HC3N}. Reaction 4 shows that \ce{HC3N} is formed via the reaction between \ce{C2H2} and CN on the grain surface of hot cores \citep{fu97, me05, tan19, bon19, wil20}. In hot cores, \ce{C2H2} is formed when \ce{CH4} desorbs into the gas-phase from the dust surface and reacts with C$^{+}$ (\ce{CH4} + C$^{+}$ $\longrightarrow$ \ce{C2H2} + \ce{H2}$^{+}$) \citep{has08}. Similarly, C$^{+}$ is created between the reactions of CO and He$^{+}$ \citep{tan19}. Previously, \cite{fu97} showed that reaction 4 is the most efficient pathway for the formation of \ce{HC3N} on the grain surface using quantum chemical studies. Subsequently, \cite{tan19} and \cite{wil20} used reaction 4 for the three-phase warm-up chemical modelling of \ce{HC3N}. After chemical modelling, \cite{tan19} and \cite{wil20} showed that reaction 4 is the most efficient for the formation of \ce{HC3N} towards hot cores. \cite{tan19} showed that the cyanopolyynes are mainly formed in the hot cores above 100 K via the reaction of C$_{2n}$H$_{2}$ and CN (C$_{2n}$H$_{2}$ + CN $\rightarrow$ HC$_{2n+1}$N, where $n$ = 1, 2, ..). Recently, \cite{man24a} showed that reaction 4 is responsible for the formation of \ce{HC3N} towards IRAS 18089--1732.

From the comparison between observed and modelled abundances of \ce{HC3N}, we noticed that \citet{gar13} (Model A) did not estimate the modelled abundance of \ce{HC3N} while \cite{suz18} (Model B) estimated it. We estimated that the abundance of \ce{HC3N} with respect to \ce{H2} towards NIRS3 is $(2.4\pm0.7)\times10^{-10}$, which is approximately one order of magnitude lower than the Model B. Similarly, the estimated abundances of \ce{HC3N} with respect to \ce{CH3OH} and \ce{CH3CN} towards NIRS3 are $(2.8\pm0.6)\times10^{-3}$ and $(1.8\pm0.9)\times10^{-1}$. We observed that the modelled abundance of \ce{HC3N} relative to \ce{CH3OH} is approximately two orders of magnitude lower than the observed value. We also observed that the modelled abundance of \ce{HC3N} relative to \ce{CH3CN} is very close to the observed value. Since observed abundance relative to \ce{H2} does not match with modelled values, it is challenging to understand the formation pathway of \ce{HC3N} based on Model B.

To understand the potential formation pathways of \ce{HC3N} towards NIRS3, we also examined the chemical model of \cite{wil20}. After chemical modelling, \cite{wil20} found that the modelled abundances of \ce{HC3N} relative to \ce{H2} and \ce{CH3CN} are $2.5\times10^{-10}$ (Model 6) and $1.5\times10^{-1}$ (Model 5), which are nearly similar to the observed abundances.  Throughout the chemical modelling, \cite{wil20} demonstrated that reaction 4 is the dominant pathway for producing \ce{HC3N} on the grain surface. Since the observed abundances of \ce{HC3N} with respect to \ce{H2} and \ce{CH3CN} towards NIRS3 closely match the modelled abundances of \cite{wil20}, this suggests that \ce{HC3N} may have been formed through the reaction between \ce{C2H2} and CN (reaction 4) on the grain surface of NIRS3.

\begin{figure*}
	\centering
	\includegraphics[width=0.98\textwidth]{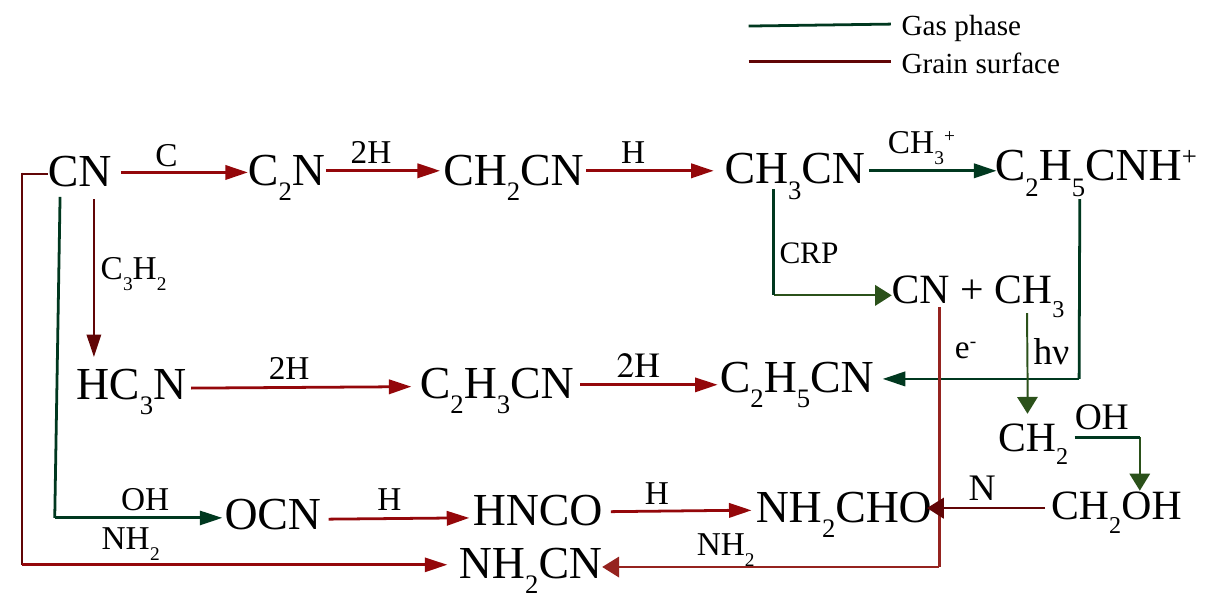}
	\caption{Proposed chemical network to understand the chemical link between detected N-bearing molecules towards NIRS3. The red and green lines show the grain surface and gas-phase pathways. CRP represents the photon produced as a consequence of cosmic ray ionization.}
	\label{fig:network}
\end{figure*}

\subsubsection{Formation mechanism of \ce{C2H5CN} and \ce{C2H3CN}}
We observed that several gas-phase and grain surface reactions of \ce{C2H5CN} and \ce{C2H3CN} are available in the UMIST 2012 \citep{mc13} and KIDA \citep{wak15} astrochemical databases, but previous three-phase warm-up chemical models showed that both molecules are formed in the grain surface of HMCs \citep{gar13, gar17, wil20, gar22}. The formation pathways of \ce{C2H5CN} and \ce{C2H3CN} on grain surface are described below:\\\\
\ce{HC3N} $\stackrel{\rm H}\longrightarrow$ \ce{C2H2CN} $\stackrel{\rm H}\longrightarrow$ \ce{C2H3CN}~~~~~~~~~~~~~~~~~~~~~~(5)\\\\
\ce{C2H3CN} $\stackrel{\rm H}\longrightarrow$ \ce{C2H4CN} $\stackrel{\rm H}\longrightarrow$ \ce{C2H5CN}~~~~~~~~~~~~~~~~~~(6)\\\\
Reaction 5 shows that \ce{C2H3CN} is created on the dust grains via the continuous hydrogenation of \ce{HC3N}, which accretes from the gas-phase. Similarly, reaction 6 indicates that the continuous hydrogenation of \ce{C2H3CN} on the grain surfaces quickly produces \ce{C2H5CN}. Several authors have shown that reactions 5 and 6 are the main formation routes of \ce{C2H5CN} and \ce{C2H3CN}, respectively, on the grain surfaces of hot cores \citep{gar13, gar17, suz18, bon19, gar22}. \cite{gar13} and \cite{gar17} showed that \ce{C2H3CN} is produced in small amounts on the grain surface during the collapse stage via the hydrogenation of \ce{HC3N}. However, hydrogenation continues to produce \ce{C2H5CN} in the collapse phase to maintain a low abundance of \ce{C2H3CN}. It allows \ce{C2H5CN} to maintain a steady abundance ($\sim10^{-10}$) on the grain surface in the warm-up phase \citep{gar13, gar17}. Reactions 5 and 6 show that a chemical linkage may exist between \ce{C2H3CN} and \ce{C2H5CN} on the grain surface because \ce{HC3N} acts as a possible precursor of both \ce{C2H3CN} and \ce{C2H5CN}. Using quantum chemical studies, \cite{sin21} showed that reactions 5 and 6 are the most efficient for the formation mechanisms of \ce{C2H3CN} and \ce{C2H5CN}. \cite{sin21} also showed that \ce{C2H5CN} acts as a possible precursor of the most complex N-bearing molecule propyl cyanide (\ce{C3H7CN}) (\ce{C2H5CN} + \ce{CH2} $\longrightarrow$ \ce{C3H7CN}) towards star-forming regions. 

To understand the possible formation mechanisms of \ce{C2H5CN} and \ce{C2H3CN} towards NIRS3, we compared the derived abundances of those molecules with the modelled abundances. We estimated that the abundances of \ce{C2H5CN} and \ce{C2H3CN} with respect to \ce{H2} towards NIRS3 are $(3.5\pm1.3)\times10^{-10}$ and $(1.9\pm0.7)\times10^{-10}$, which are one and two orders of magnitude lower than those of Models A and B. We notice that the observed abundances of \ce{C2H5CN} and \ce{C2H3CN} with respect to \ce{CH3OH} towards NIRS3 are nearly identical to the slow warm-up modelled abundances of Model A. Similarly, we also found that the observed abundance of \ce{C2H5CN} relative to \ce{CH3CN} towards NIRS3 is lower than that of Models A and B, but the observed abundance of \ce{C2H3CN} relative to \ce{CH3CN} is almost identical to the modelled abundance of Model B. Since the observed abundances of \ce{C2H5CN} and \ce{C2H3CN} relative to \ce{H2} do not match with the modelled values, it is difficult to explain the formation pathways of both molecules based on Models A and B.

To understand the possible formation pathways of \ce{C2H5CN} and \ce{C2H3CN} towards NIRS3, we also examined the chemical model of \cite{wil20}. After chemical modelling, \citet{wil20} found that the modelled gas-phase abundances of \ce{C2H5CN} and \ce{C2H3CN} relative to \ce{H2} are $5.2\times10^{-10}$ (Model 4) and $1.5\times10^{-10}$ (Model 2), which are very similar to the observed abundances. Similarly, the modelled gas-phase abundances of \ce{C2H5CN} and \ce{C2H3CN} relative to \ce{CH3CN} are $2.7\times10^{-1}$ (Model 6) and $1.0\times10^{-3}$ (Model 6), respectively. While the modelled abundance of \ce{C2H5CN} relative to \ce{CH3CN} closely matches the observed value, the modelled value of \ce{C2H3CN} relative to \ce{CH3CN} does not match the observation. During chemical modelling, \cite{wil20} demonstrated that reactions 5 and 6 are the dominant pathways for producing \ce{C2H5CN} and \ce{C2H3CN} on the grain surface. Since the observed abundances of \ce{C2H5CN} and \ce{C2H3CN} with respect to \ce{H2} towards NIRS3 are similar to the modelled abundances of \cite{wil20}, this suggests that \ce{C2H3CN} may form on the dust grains of NIRS3 via the continued hydrogenation of \ce{HC3N} (reaction 5). Likewise, \ce{C2H5CN} may be produced through the continuous hydrogenation of \ce{C2H3CN} (reaction 6) on the grain surfaces of NIRS3.

\subsubsection{Formation mechanism of \ce{NH2CN} and \ce{NH2CHO}}
Many chemical reactions are available to create high \ce{NH2CN} and \ce{NH2CHO} abundances. Previous chemical models showed that \ce{NH2CN} and \ce{NH2CHO} are formed on the grain surface instead of in the gas-phase \citep{gar13, gar22}. The chemical reactions for the formation of \ce{NH2CN} and \ce{NH2CHO} in grain surface are described below: \\\\
\ce{NH2} + CN $\longrightarrow$ \ce{NH2CN}~~~~~~~~~~~~~~~~~~~~~~~~~~~~~~~~~~~~~~~~~~~~~~~~~~~(7)\\\\
OCN $\stackrel{\rm H}\longrightarrow$ \ce{HNCO} $\stackrel{\rm H}\longrightarrow$ \ce{NH2CO} $\stackrel{\rm H}\longrightarrow$ \ce{NH2CHO}~~~~~~~~~~~~~~~(8)\\\\
Reaction 7 shows that \ce{NH2CN} is formed on the grain surface of the hot cores via the reaction between \ce{NH2} and CN. Previous three-phase warm-up chemical models showed that reaction 7 is responsible for the production of \ce{NH2CN} in hot cores and hot corinos \citep{gar13, cou18, gar22}. Recently, \cite{man24c} computed a two-phase (gas + grain) warm-up chemical model in the context of hot core object G31.41+0.31 using the gas-grain chemical model UCLCHEM and showed that reaction 7 is responsible for the formation of \ce{NH2CN} towards G31.41+0.31. Reaction 8 shows that the continued hydrogenation of OCN forms \ce{NH2CHO} on the grain surface, and \cite{bon19} showed that this reaction is the most suitable pathway for the formation of \ce{NH2CHO} on the grain surface of Sgr B2 (N). Several three-phase warm-up chemical models have shown that reaction 8 is the most suitable route for the formation of \ce{NH2CHO} towards hot cores \citep{gar13, suz18, bon19, gar22}.

We compare our estimated abundances of \ce{NH2CN} and \ce{NH2CHO} with the modelled abundances. We observe that Model B \citep{suz18} did not estimate the modelled abundance of \ce{NH2CN}, but Model A \citep{gar13} estimated the modelled abundance of \ce{NH2CN}. We estimated that the abundances of \ce{NH2CN} and \ce{NH2CHO} with respect to \ce{H2} towards NIRS3 are $(2.8\pm1.0)\times10^{-10}$ and $(2.1\pm0.9)\times10^{-9}$. We found that the observed abundance of \ce{NH2CN} is approximately two orders of magnitude lower than the modelled abundance of Model A, but the abundance of \ce{NH2CHO} is nearly similar to the slow warm-up modelled value of Model A. We observed that the derived abundance of \ce{NH2CN} relative to \ce{CH3OH} towards NIRS3 is nearly similar to the medium warm-up modelled value of Model A, but the abundance of \ce{NH2CHO} relative to \ce{CH3OH} is similar to Model B. Similarly, we notice that the observed abundances of \ce{NH2CN} and \ce{NH2CHO} relative to \ce{CH3CN} are lower than that of both Models A and B.  \cite{gar13} showed that reaction 8 is the main formation route for the formation of \ce{NH2CHO} on the grain surface in slow warm-up conditions. After comparison, we found that the observed and modelled abundances of \ce{NH2CHO} relative to \ce{H2} are very close, which indicates \ce{NH2CHO} may have been formed via the hydrogenation of \ce{NH2CO} (reaction 8) on the grain surface of NIRS3. We also noticed that the observed \ce{NH2CN}/\ce{NH2CHO} ratio towards NIRS3 is not similar to the modelled values. Since the observed and modelled abundances of \ce{NH2CN} relative to \ce{H2} are not similar, we do not draw any conclusions about the formation mechanism of \ce{NH2CN} towards NIRS3 using Model A.

We also examined the three-phase warm-up chemical model from \cite{cou18} to understand the potential formation pathway of \ce{NH2CN} towards NIRS3. \cite{cou18} estimated that the modelled gas-phase abundance of \ce{NH2CN} is $3.7\times10^{-10}$. During the chemical modelling, \cite{cou18} used the grain surface reaction between \ce{NH2} and CN (reaction 7) for the formation of \ce{NH2CN}. We noticed that the observed abundance of \ce{NH2CN} with respect to \ce{H2} towards NIRS3 is nearly similar to the modelled abundance of \citet{cou18}. This suggests that \ce{NH2CN} may form through the reaction between \ce{NH2} and CN (reaction 7) on the grain surface of NIRS3.

\subsection{Proposed chemical network between detected N-bearing molecules}
To understand the nitrile chemistry of NIRS3, we created a chemical network between the detected N-bearing molecules using gas-phase and grain surface chemical reactions, which are shown in Figure~\ref{fig:network}. During the making of chemical networks, we mainly used the reactions discussed in Sections 4.2.1, 4.2.2, 4.2.3, and 4.2.4. We also used the additional reactions, which are taken from the UMIST and KIDA astrochemistry reactions databases \citep{mc13, wak15}.  In the chemical network, we observed that \ce{CH3CN} is formed via the continued hydrogenation of \ce{C2N}. Similarly, the continued hydrogenation of \ce{HC3N} forms \ce{C2H3CN} and \ce{C2H5CN}. We also observed that \ce{CH3CN} is destroyed via the radiative association reaction of \ce{CH3}$^{+}$ and \ce{CH3CN} and forms \ce{C2H5CNH}$^{+}$ in the gas-phase. Again, the dissociative recombination of \ce{C2H5CNH}$^{+}$ forms \ce{C2H5CN}. Thus, the N-bearing molecules \ce{CH3CN}, \ce{HC3N}, \ce{C2H3CN}, and \ce{C2H5CN} may be chemically linked via both gas-phase and grain surface chemical routes. Similarly, \ce{CH3CN} may be chemically linked with \ce{NH2CHO} and \ce{NH2CN} in both the gas-phase and grain surface routes. Our molecular correlation heat maps show that \ce{CH3CN}, \ce{C2H5CN}, \ce{C2H3CN}, \ce{NH2CN}, and \ce{NH2CHO} exhibit a tight correlation with each other, and our proposed chemical network also exhibits the chemical link of those molecules. Previously, \cite{gar17} showed that \ce{HC3N}, \ce{C2H5CN}, and \ce{C2H3CN} are chemically linked towards hot cores. Detailed quantum chemical and three-phase warm-up chemical modelling are needed to verify the proposed chemical network and understand the nitrile chemistry in the warm inner regions of the star-formation regions.

\section{Summary and conclusions}
\label{sec:con}
We analyzed ALMA band 4 data of a massive protostar, S255IR NIRS3, and the following conclusions are made. \\

$\bullet$From the dust continuum emission, we estimate that the mass, dust temperature, and dust spectral index are $20.5\pm3.0$ \(\textup{M}_\odot\), $150\pm16$ K, and $1.7\pm0.6$, respectively. Similarly, the column density of molecular hydrogen and dust optical depth towards NIRS3 are $(1.1\pm0.2)\times10^{24}$ cm$^{-2}$ and $2.1\times10^{-2}$. \\

$\bullet$ We found the emission lines of \ce{CH3CN}, \ce{C2H5CN}, \ce{C2H3CN}, \ce{NH2CN}, \ce{NH2CHO}, and \ce{HC3N} ($\nu_{7}$ = 2) towards NIRS3. We derive the column density and excitation temperatures of detected molecules using the LTE-modelled spectra. We observed the excitation temperatures of detected N-bearing vary between 175 K and 220 K, which means all detected N-bearing molecules emit from the warm inner regions ($T$ $>$ 100 K) of NIRS3. We also estimate the fractional abundances of all detected N-bearing species relative to \ce{H2}, \ce{CH3OH}, and \ce{CH3CN}. Since \ce{NH2CN} and \ce{NH2CHO} have \ce{NH2} as a common precursor, the \ce{NH2CN}/\ce{NH2CHO} column density ratio towards NIRS3 is $0.13\pm0.06$.\\

$\bullet$ We compared our estimated abundances of detected N-bearing molecules with existing three-phase warm-up chemical models of \cite{gar13}, \cite{suz18}, \cite{wil20}, and \cite{cou18}. We find that the modelled abundances of \ce{CH3CN} and \ce{NH2CHO} reported by \citet{gar13} are broadly consistent with our observed values, differing by factors of only 1.04 and 0.72, respectively. Likewise, the modelled abundances of \ce{C2H3CN}, \ce{C2H5CN}, and \ce{HC3N} predicted by \citet{wil20} agree well with our observations within factors of 1.28, 0.67, and 0.96. Similarly, the modelled abundance of \ce{NH2CN} from \citet{cou18} matches the observed value within a factor of 0.76. We also discuss the possible formation pathways of the detected N-bearing species based on gas/grain chemistry towards NIRS3. Since the observed abundance of \ce{CH3CN} towards NIRS3 is in good agreement with the modelled value reported by \citet{gar13}, we conclude that \ce{CH3CN} is likely formed via the hydrogenation of \ce{CH2CN} (reaction 2) on the grain surface of NIRS3. Based on the comparison between the observed abundances of \ce{HC3N}, \ce{C2H5CN}, and \ce{C2H3CN} and the model predictions of \citet{wil20}, we infer that \ce{HC3N} may be formed through the reaction between \ce{C2H2} and CN (reaction 4) on the grain surface of NIRS3. Likewise, \ce{C2H3CN} may have been formed via subsequent hydrogenation of \ce{HC3N} (reaction 5), and \ce{C2H5CN} may then be formed through the continued hydrogenation of \ce{C2H3CN} (reaction 6) on the grain surfaces of NIRS3. Additionally, the observed abundances of the amide-like molecules \ce{NH2CN} and \ce{NH2CHO} are in close agreement with the modelled values reported by \citet{cou18} and \citet{gar13}. This agreement suggests that \ce{NH2CN} and \ce{NH2CHO} are likely formed on the grain surfaces of NIRS3 via the reaction between \ce{NH2} and CN (reaction 7) and the hydrogenation of \ce{NH2CO} (reaction 8), respectively. However, a more detailed analysis of dominant chemical pathways in tailored chemical models is required to confirm these proposed interpretations.\\

$\bullet$ We also proposed a chemical network to understand the chemical link between the detected N-bearing species. This chemical link is important to understand the nitrile chemistry towards NIRS3. A detailed quantum chemical model is required to verify this network.

\section*{ACKNOWLEDGEMENTS}{
We appreciate the valuable feedback from the anonymous reviewers, which has helped enhance the manuscript. AM acknowledges the postdoctoral fellowship of the S. N. Bose National Centre for Basic Sciences, Kolkata, India, funded by the Department of Science and Technology (DST), India. AM also acknowledges Somnath Dutta for useful discussions regarding this work and for continued support. AH and TB thank the support of the S. N. Bose National Centre for Basic Sciences under the Department of Science and Technology, Govt. of India. AH also thanks the CSIR-HRDG, Govt. of India, for funding the fellowship. This paper makes use of the following ALMA data: ADS /JAO.ALMA\#2016.A.00008.T. ALMA is a partnership of ESO (representing its member states), NSF (USA), and NINS (Japan), together with NRC (Canada), MOST and ASIAA (Taiwan), and KASI (Republic of Korea), in co-operation with the Republic of Chile. The Joint ALMA Observatory is operated by ESO, AUI/NRAO, and NAOJ.}\\
\section*{Software} CASA \citep{mc07}, ASTROPY \citep{astro22}, corner.py \citep{fo16}, CASSIS \citep{vas15}, RADMC-3D \citep{dul12}.

\section*{DATA AVAILABILITY}
The plots within this paper and other findings of this study are available from the corresponding author on reasonable request. The data used in this paper are available in the ALMA Science Archive (\url{https://almascience.nrao.edu/asax/}), under project code of 2016.A.00008.T.

\section*{Conflicts of interest}{The authors declare no conflict of interest.}

\bibliographystyle{aasjournal}
%\bibliography{./literature.bib,added.bib} % if your bibtex file is called example.bib

\renewcommand{\thefigure}{A\arabic{figure}}  % Force A1, A2... format
\setcounter{figure}{0}                       % Reset figure count if needed
\renewcommand{\thetable}{A\arabic{table}}  % Force table numbers as A1, A2...
\setcounter{table}{0}

\section*{Appendix}
\begin{figure*}
\centering
\includegraphics[width=1.0\textwidth]{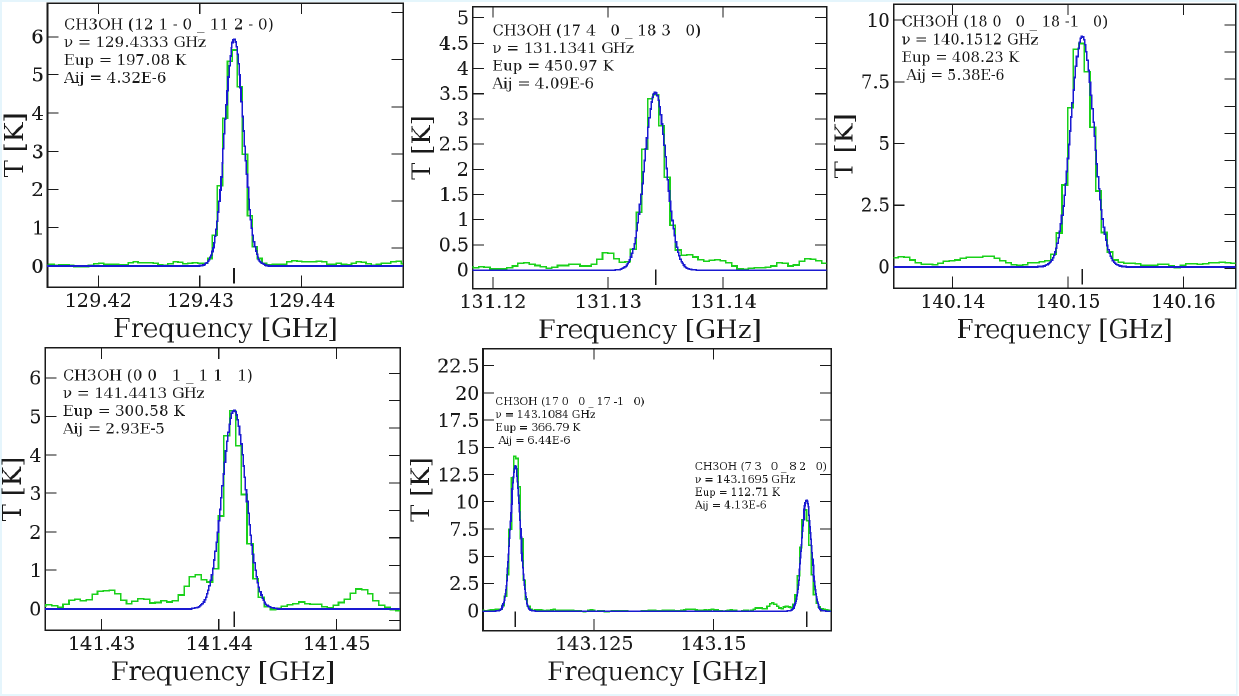}
\caption{Detected rotational emission lines of \ce{CH3OH} towards NIRS3. The green lines indicate the observed molecular spectra of NIRS3, and the blue lines represent the LTE model spectra of \ce{CH3OH}. The optical depths ($\tau$) of the detected \ce{CH3OH} transitions range from 5$\times$10$^{-2}$ to 8$\times$10$^{-2}$.}
\label{fig:methanol}
\end{figure*}

In addition to various N-bearing species, we also detected rotational emission lines of \ce{CH3OH} toward NIRS3. Based on LTE spectral modeling, the column density and excitation temperature of \ce{CH3OH} are $(9.52\pm0.62)\times10^{16}$ cm$^{-2}$ and $250\pm24$ K, respectively. The LTE spectral fit to the \ce{CH3OH} emission lines is presented in Figure~\ref{fig:methanol}. Details of previously detected N-bearing molecules in the ISM and planetary environments are listed in Table~\ref{tab:molecular_sources}. The RADMC model parameters adopted for NIRS3 in the simulations are summarized in Table~\ref{tab:modelsetup}.

\FloatBarrier   % <-- forces all earlier floats to appear before proceeding
%\clearpage 

\begin{table*}[t]
\scriptsize
\centering %{\large	 Appendix}
\centering
\caption{Detailed previous detections of selected N-bearing iCOMs across galactic and solar system sources.} 
\label{tab:molecular_sources}
\begin{adjustbox}{width=1.0\textwidth}
\begin{tabular}{cccccccc}
	\hline
Molecule & Sources & References \\ 
\hline
\ce{CH3CN} &Hot molecular cores, Hot corinos, prestellar cores, & \citet{ol96, bot04, taq15, cal18} \\
		&cold starless cores, protoplanetary disks, and Titan.& \citet{bel20, naz21, mer22, bi22} \\
		&                                &\citet{hs23, min23, min93, vas19, mag23}\\
		&                                &\citet{ob15, il21, ul74, rod01}\\
		&                                &\citet{al19, biv22, I20}\\
\hline 
\ce{C2H3CN}& Hot molecular cores (Sgr B2, Orion KL, G10.47+0.03, G31.41+0.31), & \citet{gar75, lo14, rof11, min23} \\ 
		& hot corino (IRAS 16293--2422 B), carbon-rich star IRC+10216,& \citet{ cal18, ag08, pal17}\\
		& and Titan.& \\
\hline
		
\ce{C2H5CN}& Sgr B2, OMC-1, G10.47+0.03, G31.41+0.31, Orion KL, & \citet{jon77, rof11, man23a, min23} \\ 
		&G34.26+0.15, IRAS 18089--1732, IRAS 16293--2422 B, &  \citet{fri12, mo07, col20, man24a} \\
		&and Titan.& \citet{cal18, cor15}\\
		
\hline
\ce{HC3N}& Hot cores G10.47+0.03, G31.41+0.31, G34.26+0.15, Orion KL, Sgr B2,& \citet{wy99, rof11} \\ 
		&hot corino IRAS 16293--2422 B, high-mass protostar IRAS 18089--1732,&\citet{wy99, mo07, pen17, wil20} \\
		&prestellar core L1544, and Titan.&\citet{jab17, man24a, bi23, cor15} \\
\hline	
		
\ce{NH2CN}& Hot cores G10.47+0.03,  G31.41+0.31,  G358.93--0.03 MM1, Orion KL,  & \citet{rof11, man22, man24c, man23b} \\ 
		&Sgr B2, and and hot corino IRAS 16293--2422 B.&\citet{whi03, pag17, tur75} \\
		&   &\citet{bel13, cou18}\\
		
\hline
		
\ce{NH2CHO}& Hot cores G10.47+0.03, G31.41+0.31, Orion KL , Sgr B2 (N),& \citet{rof11, col21, mot12, hal11} \\ 
		&G358.93--0.03 MM1, high-mass protostar IRAS 18089--1732, &\citet{bel13, man24d, man24a, lig18} \\
		&low-mass protostar IRAS 16293--2422 B, and comets. & \citet{biv14, go15}\\
\hline
\end{tabular}
\end{adjustbox}
\end{table*}

\begin{table*}{}
\caption{Model parameters of NIRS3 used in RADMC-3D}
\begin{adjustbox}{width=1.0\textwidth}
\begin{tabular}{ccccccccccc}
\hline
Parameter Description$^{\color{blue}{\dagger}}$& Values& Units\\
\hline
Radius of envelope ($r_{\mathrm{out}}$)&4.5$\times10^{3}$&AU  \\
Grid start radius 	($r_{\mathrm{in}}$)&20&AU\\ 
Density plateau radius ($r_{\mathrm{plat}}$)&900$^{\color{blue}{\ast}}$& AU\\
Grid cells in the polar range ($n_{\mathrm{\theta}}$)&131&--\\
Grid cells in the azimuthal range ($n_{\mathrm{\phi}}$)&131&--\\
Grid cells in the outer radial region ($r$ $>$ 900 AU) ($n_{\mathrm{r,\:out}}$)&88&-- \\
Grid cells in the inner radial region ($r$ $\leq$ 900 AU) ($n_{\mathrm{r,\:in}}$)&120&--\\
Amount of photons employed in the thermal Monte Carlo process ($n_{\mathrm{photons}}$)&3$\times10^{8}$ &--  \\
Luminosity of NIRS3 ($L_\mathrm{NIRS3}$)&(1--3)$\times$10$^{4}$$^{\color{blue}{\ast}}$  & L$_{\sun}$ \\
Octree refinement radius ($r_{\mathrm{oct}}$)&500& AU \\
Octree refinement number ($n_{\mathrm{octree\:levels}}$)&1 &-- \\
Envelope density ($\rho_{0, \: \mathrm{env}}$)&0.71$^{\color{blue}{\ast}}$& g~cm$^{-3}$ \\
Density power-law exponent ($p_{\mathrm{env}}$) &1.1$^{\color{blue}{\ast}}$  &-- \\
Mass of NIRS3 ($M_\mathrm{NIRS3}$) & 18$^{\color{blue}{\ast}}$ & M$_{\sun}$ \\
Temperature of Star surface ($T_{\mathrm{eff}}$)&3900$^{\color{blue}{\ast}}$& K \\ 
Radius of inner disk ($r_{\mathrm{inner\:disk}}$)& 2 & AU\\
Radius of NIRS3 disk ($r_\mathrm{disk,NIRS3}$) & 150& AU\\ 
Peak flux density ($F_{\mathrm{peak}}$)&117$^{\color{blue}{\ast}}$& mJy beam$^{-1}$ \\
\ce{H2} column density ($N_{\mathrm{H_2}}$) &$\geq$ 1.0$\times$10$^{24}$$^{\color{blue}{\ast}}$ & cm$^{-2}$ \\
Dust optical depth ($\tau$)&$\geq$3$\times$10$^{-2}$$^{\color{blue}{\ast}}$&--\\
Dust temperature ($T_{d}$)&$\geq$100$^{\color{blue}{\ast}}$&K\\		
Mass accretion rate ($\dot{M}_\mathrm{NIRS3}$)& 1.0$\times10^{-4}$$^{\color{blue}{\ast}}$& M$_{\sun}$ yr$^{-1}$ \\
\hline
\end{tabular}
\end{adjustbox}\\
${\color{blue}{\dagger}}$ -- The detailed descriptions of the physical parameters in RADMC-3D are well described in \citet{jac18}.\\
${\color{blue}{\ast}}$ -- Those values of NIRS3 are taken from \citet{liu20} and the present study.
	
\label{tab:modelsetup}
\end{table*}

\end{document}